\documentclass[11pt,a4paper]{article}
\usepackage{jheppub}
\usepackage{orcidlink}
\usepackage{booktabs}
\usepackage{array}
\usepackage{placeins}
\usepackage{microtype}
\usepackage{hyperref}
\usepackage{enumitem}
 \hypersetup{
 breaklinks=true,
 hypertexnames=false,
 pdftitle={Complex Kerr--AdS5 Saddles and Contour-Selected Fisher Zeros},
 pdfauthor={Bum-Hoon Lee, Hocheol Lee, and Somyadip Thakur},
 pdfkeywords={black holes, AdS/CFT correspondence, models of quantum gravity, nonperturbative effects}
}

\newcommand{\R}{\mathbb{R}}
\newcommand{\C}{\mathbb{C}}
\newcommand{\dd}{\mathrm{d}}
\renewcommand{\Re}{\operatorname{Re}}
\renewcommand{\Im}{\operatorname{Im}}
\newcommand{\diag}{\operatorname{diag}}
\newcommand{\sgn}{\operatorname{sgn}}
\newcommand{\abs}[1]{\left|#1\right|}
\newcommand{\rp}{r_+}
\newcommand{\XiA}{\Xi}
\newcommand{\Jth}{\mathcal{J}}
\newcommand{\Kth}{\mathcal{K}}
\newcommand{\lPL}[1]{\lambda_{\mathrm{PL},#1}}
\newcommand{\phiM}{\phi_{\mathrm{M}}}

\title{\boldmath { 
 Complex Phase Structure of Kerr--AdS$_5$ Black Holes: Critical Saddles and Fisher Zeros} }

\author[a,b,c]{Bum-Hoon Lee\,\orcidlink{0009-0008-3322-2087},}
\author[d]{Hocheol Lee\,\orcidlink{0000-0002-4420-5898},}
\author[a]{and Somyadip Thakur\,\orcidlink{0000-0003-2376-5906}}

\affiliation[a]{Center for Quantum Spacetime, Sogang University,
Seoul 04107, Republic of Korea}
\affiliation[b]{Department of Physics, Sogang University,
Seoul 04107, Republic of Korea}
\affiliation[c]{Department of Physics, Shanghai University,
99 Shangda Road, Baoshan District, Shanghai 200444, China}
\affiliation[d]{Department of Physics, Dongguk University,
Seoul 04620, Republic of Korea}

\emailAdd{bhl@sogang.ac.kr}
\emailAdd{insaying@dongguk.edu}
\emailAdd{somyadip@sogang.ac.kr}

\abstract{
We study the complex saddle structure of singly rotating Kerr--AdS$_5$ black holes using a two-variable reduced model in the grand canonical ensemble at fixed angular velocity, $\Omega$. The small and large black hole saddles merge at a critical point where one Takagi singular value of the complex Hessian vanishes, identifying the soft mode associated with the merger.
This merger remains distinct from the Hawking--Page transition throughout the physical domain $|\Omega|<1$. For the reduced integral with the standard measure, we show that thermal AdS dominates uniformly in a complex neighborhood of each real merger point when Newton's constant $G$ is sufficiently small. Consequently, this neighborhood contains no zeros of the partition function, even though the local saddle merger is described by Airy-type behavior. The Fisher zeros instead accumulate near the Hawking--Page coexistence line, with spacing of order $G$, through interference between thermal AdS and the large black hole saddle.
We also show that the real quasi-Euclidean Kerr--AdS$_5$ family satisfies a strict asymptotic Kontsevich--Segal--Witten (KSW) phase bound for $|\Omega|<1$, with the merger point lying inside the allowed domain. In the fixed angular momentum extended ensemble, two such merger points combine into a cusp with a contour-dependent Pearcey approximation. These results distinguish local saddle degeneracies from global phase coexistence and clarify the different roles of Takagi modes, phase transitions, and zeros of the partition function in the complex black hole saddle structure.} 

\keywords{Black Holes; AdS-CFT Correspondence; Models of Quantum Gravity;}

\begin{document}
\maketitle
\flushbottom

\section{Introduction}
\label{sec:introduction}

The Hawking--Page transition between thermal anti-de Sitter space and a large
AdS black hole is the archetypal gravitational phase transition
\cite{HawkingPage:1983}.  The AdS/CFT correspondence relates bulk gravity to
a boundary conformal field theory
\cite{Maldacena:1998,Gubser:1998,Witten:1998ads}; in this setting the
Hawking--Page transition is interpreted as confinement--deconfinement
\cite{Witten:1998thermal}.  The standard thermodynamic treatment compares
real on-shell actions and thereby
identifies the dominant saddle.  That comparison does not, however, exhaust
the information in the gravitational path integral.  A complexified
description must additionally determine the contributing saddles, their
steepest-descent cycles, and the complex zeros generated by their
interference.  Semiclassically, the grand canonical partition function is
\begin{equation}
 Z(\beta,\Omega)
 \simeq
 b_0 Z_0+
 \sum_{\sigma\in{\rm int}} n_\sigma Z_\sigma,
 \label{eq:saddle-sum}
\end{equation}
where $Z_\sigma$ is the unweighted saddle amplitude~\cite{Singhi:2025}.  For an interior Morse
critical point, $n_\sigma\in\mathbb Z$ is the intersection number with the
dual upward cycle.  The thermal saddle lies on the boundary of the reduced
contour; the weight $b_0$ is fixed in Section~\ref{sec:setup}.

The same distinction is familiar in condensed matter and
finite-density QCD.  Lee--Yang and Fisher zeros describe the finite-volume
analytic structure of the partition function; their accumulation determines
how a thermodynamic nonanalyticity emerges in the infinite-volume limit
\cite{YangLee:1952,LeeYang:1952,Fisher:1965}.  This language is especially
useful when the path-integral weight is complex, as in non-Hermitian systems \cite{Bender:2002vv,Li:2024ymo}
or QCD at nonzero baryon chemical potential, where the sign problem prevents
a direct real-probability interpretation.  Recent $PT$-symmetric quantum
models provide explicit microscopic realizations of the Yang--Lee transition
and its $i\phi^3$ continuum description
\cite{ArguelloCruzTarnopolsky:2026}.  Zeros obtained by analytic continuation
in complex chemical potential or temperature are then used to investigate
crossover behavior, first-order lines, and candidate critical
endpoints~\cite{Stephanov:2004,FodorKatz:2004}.  Kerr--AdS$_5$ is a controlled semiclassical setting in which the analogous
question can be answered sharply.  Airy and Pearcey normal forms fix the
local geometry of coalescing saddles, whereas physical zero sequences require
the contour-selected saddle sectors and their relative phases.  Related
black hole applications include Lee--Yang complex phase diagrams and Widom
lines in Euler--Heisenberg AdS thermodynamics~\cite{AwalPhukon:2026}, and a
Picard--Lefschetz and KSW analysis of charged-AdS partition functions in which
contour selection removes formal saddles that do not belong to the physical
integration cycle~\cite{Ailiga:2025osa}.
 A complex black hole solution has the same contour-theoretic
description as a complex critical point in thimble regularization.
Complexification is required to deform the original real contour into
steepest-descent cycles and, in the present problem, to continue the two Kerr
saddles after they coalesce at the fold and leave the real slice.  The
resulting complex solution is a legitimate semiclassical candidate, but it
contributes only when the intersection number
$n_\sigma=\langle\Gamma,\Kth_\sigma\rangle$ is nonzero.  In the reduced
model---not the full gravitational fluctuation problem---its one-loop term is
schematically
$n_\sigma\mu_\sigma e^{\Phi_\sigma+i\phi_{{\rm M},\sigma}}/
\sqrt{\det\Sigma_\sigma}$, where $\Sigma_\sigma$ is the Takagi-value matrix.
The Takagi values fix the Gaussian magnitude, while the oriented thimble
tangent frame fixes its phase.  Neither the existence of the complex saddle
nor its Hessian determines the global contour
coefficient~\cite{DiRenzoEruzzi:2015,DiRenzoEruzzi:2018}.  For gravity, this
contour criterion must still be supplemented by KSW metric allowability on a
specified real spacetime cycle.

Holographically, this complex zero structure refines the thermal-CFT
interpretation of Hawking--Page.  For a CFT on $S^3\times S^1_\beta$,
$Z_{\rm CFT}(\beta,\Omega)=\operatorname{Tr}e^{-\beta(H-\Omega J)}$, and
thermal AdS and Kerr--AdS represent the confined and rotating deconfined
sectors.  Since $G_5/L^3\sim N_c^{-2}$, the $O(G_5)$ Fisher zero spacing found
below becomes $O(N_c^{-2})$; the zeros approach the real transition in the
planar limit.  The small/large black hole fold therefore need not be a
thermal-CFT singularity.  The selected zero sequence is controlled instead
by thermal AdS/large Kerr interference.  Thimble selection and KSW
allowability address which complex geometries may enter the string/gravity
path integral, whereas the Takagi values give local growth and decay rates
of the linearized Picard--Lefschetz flow.  Although both are exponential
rates, Takagi values should not be identified with probe-particle Lyapunov
exponents in four-dimensional Kerr--Newman--AdS~\cite{Gwak:2022xje}, nor
with real-time CFT chaos.
    
Singly rotating Kerr--AdS$_5$ provides a particularly sharp realization of
this hierarchy.  As the thermodynamic sources are varied, the small and
large black hole saddles can merge at a spinodal, analytically continue into
a complex-conjugate pair, and compete with thermal AdS for dominance.  These
phenomena must be distinguished.  The Hessian identifies the breakdown of the
local Gaussian approximation; differences of saddle actions locate candidate
Stokes and anti-Stokes curves; and the intersection numbers determine which
saddles actually enter the partition function.  Picard--Lefschetz theory
\cite{Witten:2010,DunneUnsal:2016} supplies the framework that relates, but
does not conflate, these data.  Hence neither a vanishing thermodynamic
Hessian nor equality of real Gibbs free energies alone determines the
physical complex zero distribution.

The global thimble decomposition for the singly rotating Kerr--AdS$_5$
mini-superspace was obtained in Ref.~\cite{Singhi:2025}.  We take its contour
selection as input and complement it with an analysis of the local complex
Hessian, the uniform critical behavior, and the saddle interference that
produces physical zeros.  For the Hessian analysis we use the Takagi
construction established in Lefschetz-thimble field theory and apply it to the
Kerr--AdS thermodynamic mini-superspace.  The calculation is performed in the
same two-coordinate truncation, parametrized by the horizon radius and
rotation parameter.  It is therefore an analysis of the reduced saddle
geometry, rather than a replacement for the full gravitational fluctuation
problem.
 Our analysis leads to several related conclusions. We first derive the full \(2\times2\) complex symmetric Hessian of the reduced Kerr--AdS\(_5\) action and obtain an exact factorization of its determinant. The associated Takagi problem determines the local spectrum governing the antilinear thimble flow. At the rotating fold, one Takagi value vanishes and singles out the mixed radial rotational soft mode, while the orthogonal mode remains nondegenerate.

The resulting fold, or spinodal, curve is everywhere distinct from the Hawking--Page anti-Stokes line in the physical grand canonical regime. The two meet only at the ultraspinning boundary. In particular, the Hawking--Page transition at finite \(|\Omega|<1\) is not the endpoint of the small/large black hole merger.

Locally, the fold is described by the usual Airy uniform approximation. This local degeneracy, however, does not by itself generate the relevant sequence of zeros of the partition function. In the Stokes-regulated decomposition of Ref.~\cite{Singhi:2025}, the small black hole contributions enter with opposite signs and cancel at leading order in the conjugation-symmetric real-\(G_N\) limit. The reduced Fisher zeros therefore arise instead from interference between thermal AdS and the large black hole saddle, and accumulate toward the Hawking--Page line with spacing of order \(G_N\).

We also construct candidate Stokes and anti-Stokes networks in the Schwarzschild and rotating sectors and analyze the Kontsevich--Segal--Witten allowability condition independently of thermodynamic dominance. For the real quasi-Euclidean Kerr--AdS\(_5\) family, the existence of a strict asymptotic phase margin is equivalent to \(|\Omega|<1\), and this margin remains nonzero at the fold.

Finally, in the fixed-\(J\) extended ensemble, two canonical folds merge at the small/large black hole critical point. The resulting local singularity is a cusp and is described by a Pearcey uniform approximation. Determining the full canonical Lee--Yang/Fisher zero pattern nevertheless requires the analytically continued integration contour together with the corresponding one-loop phases.
The central conclusion is therefore not merely that Kerr--AdS possesses
several notable curves in parameter space, but that these curves answer
different questions.  A fold is a local degeneracy of the saddle map and
requires an Airy rather than a Gaussian approximation~\cite{Chester:1957}.
An anti-Stokes line is an equal-magnitude condition between saddles.  A Stokes
wall concerns phase alignment and can change a thimble decomposition only
when the relevant connecting flow exists.  Actual zeros of the partition function
require all of these ingredients together with the contour coefficients and
one-loop phases.
\paragraph{Organization of the paper}

Section~\ref{sec:setup} introduces the thermodynamic integral and its local
Picard--Lefschetz geometry.  Section~\ref{sec:phase} compares the
Hawking--Page and fold curves.  Section~\ref{sec:exponents} gives the Kerr
stability spectrum, fold uniformization, Fisher zeros, complex phase
networks, and KSW test.  Section~\ref{sec:extended} analyzes the distinct
fixed-$J$ cusp.  Section~\ref{sec:discussion} summarizes the results and
outlines open problems.

\section{Thermodynamic setup and local Picard--Lefschetz geometry}
\label{sec:setup}
This section formulates the finite-dimensional saddle problem used below.
We first collect the Kerr--AdS$_5$ thermodynamic data and construct the
reduced grand canonical exponent whose stationary points reproduce the
black hole saddles.  We then distinguish dynamical Lyapunov exponents from
local Picard--Lefschetz flow rates and separate the resulting Takagi data from
the global contour data.  Finally, we obtain the full $2\times2$ Kerr Hessian
within this truncation and display its determinant factorization.  These are
local data of the reduced saddle problem, not the spectrum or one-loop
determinant of the full gravitational fluctuation operator.
\subsection{Kerr--AdS\texorpdfstring{$_5$}{5} thermodynamics and reduced saddle}

We set the AdS radius to unity and write $G\equiv G_5>0$.  For a real black
hole, $\rp>0$ is the outer horizon radius and $a$ is the rotation parameter.
After eliminating the mass parameter through $\Delta_r(\rp)=0$, the
Lorentzian singly rotating Kerr--AdS$_5$ metric may be written
as~\cite{HawkingHunterTaylor:1999,GibbonsPerryPope:2005}
\begin{align}
 \dd s^2={}&
 -\frac{\Delta_r}{\rho^2}
 \left(\dd t-\frac{a\sin^2\theta}{\XiA}\dd\varphi\right)^2
 +\frac{\rho^2}{\Delta_r}\dd r^2
 +\frac{\rho^2}{\Delta_\theta}\dd\theta^2
 \nonumber\\
 &+\frac{\Delta_\theta\sin^2\theta}{\rho^2}
 \left(a\,\dd t-\frac{r^2+a^2}{\XiA}\dd\varphi\right)^2
 +r^2\cos^2\theta\,\dd\psi^2,
 \label{eq:metric}
\end{align}
where
\begin{equation}
 \rho^2=r^2+a^2\cos^2\theta,
 \qquad
 \XiA=1-a^2,
 \qquad
 \Delta_\theta=1-a^2\cos^2\theta,
\end{equation}
and
\begin{equation}
 \Delta_r=(r^2+a^2)(1+r^2)-(\rp^2+a^2)(1+\rp^2).
\end{equation}
The real Kerr family requires $|a|<1$.  We restrict the grand canonical
sources to $\beta>0$ and $|\Omega|<1$, defined in Eq.~\eqref{eq:thermo-map}, where $\Omega$ is measured relative to
the nonrotating frame at infinity.  This qualification is essential because
the Boyer--Lindquist azimuth in Eq.~\eqref{eq:metric} rotates at infinity.
For real $a$, the continuation $t=-i\tau$ is a complex
\emph{quasi-Euclidean} metric.  Smoothness at $r=\rp$, together with the
twisted thermal identification in a nonrotating azimuth $\varphi$,
\begin{equation}
 (\tau,\varphi)\sim(\tau+\beta,\varphi+i\beta\Omega),
 \label{eq:twisted-identification}
\end{equation}
fixes
\begin{equation}
 \boxed{
 \beta=\frac{2\pi(\rp^2+a^2)}{\rp(2\rp^2+a^2+1)},
 \qquad
 \Omega=\frac{a(1+\rp^2)}{\rp^2+a^2}.
 }
 \label{eq:thermo-map}
\end{equation}
The horizon angular velocity in the displayed Boyer--Lindquist azimuth is
$a\XiA/u=\Omega-a$; the thermodynamic $\Omega$ includes the shift to the
nonrotating frame at infinity.
For a real saddle with $|a|<1$, the source condition $|\Omega|<1$ is
equivalent on shell to $\rp^2>|a|$.

It is convenient to define
\begin{equation}
 u=\rp^2+a^2,
 \qquad
 v=1+\rp^2,
 \qquad
 s=2\rp^2+a^2+1=u+v,
 \qquad
 w=3-a^2-2a\Omega.
 \label{eq:shorthand}
\end{equation}
The energy and angular momentum measured in the nonrotating frame at
infinity, and the entropy, are
\begin{align}
 E&=\frac{\pi(3-a^2)uv}{8G\XiA^2},
 \label{eq:E}\\
 J&=\frac{\pi a uv}{4G\XiA^2},
 \label{eq:J}\\
 S&=\frac{\pi^2\rp u}{2G\XiA}.
 \label{eq:S}
\end{align}
With $T=\beta^{-1}$, these quantities satisfy
$\dd E=T\,\dd S+\Omega\,\dd J$.

The corresponding grand canonical saddle problem is described, following
Ref.~\cite{Singhi:2025}, by a two-coordinate
mini-superspace truncation in $(\rp,a)$,
\begin{equation}
 Z(\beta,\Omega)
 \simeq
 \int_{\Gamma}\dd\rp\,\dd a\;\mu(\rp,a)\,
 e^{\Phi(\rp,a;\beta,\Omega)},
 \label{eq:minisuperspace}
\end{equation}
with
\begin{equation}
 \boxed{
 \Phi(\rp,a;\beta,\Omega)
 =\frac{\pi}{8G}\frac{u}{\XiA^2}
 \left[4\pi\rp\XiA-\beta v w\right].
 }
 \label{eq:Phi}
\end{equation}
The original real cycle is
$\Gamma_{\mathbb R}=[0,\infty)_{\rp}\times(-1,1)_a$, and $\Gamma$ denotes
its complex deformation.  The effective measure $\mu$ and the modes omitted
by the truncation are not determined here; Eq.~\eqref{eq:Phi} retains only
the leading $O(G^{-1})$ exponent.

By construction, $\Phi=S-\beta(E-\Omega J)$.  At a black hole saddle it is
minus the background-subtracted grand canonical action, analytically
continued for a complex saddle.  We set
$E_{\mathrm{th}}=I_{\mathrm{th}}=\Phi_{\mathrm{th}}=0$.  Counterterm
renormalization may retain the AdS$_5$ Casimir constant, without changing
the relative saddle weights.  Thermal AdS is the boundary critical point
$(\rp,a)=(0,0)$ and is not governed by the black hole thermodynamic map.

For regular black hole critical points with $\rp>0$, the stationary
conditions are
\begin{equation}
 \partial_{\rp}\Phi=0,
 \qquad
 \partial_a\Phi=0,
\end{equation}
which reproduce the thermodynamic map~\eqref{eq:thermo-map}:
\begin{equation}
 \boxed{
 \beta=\frac{2\pi u}{\rp s},
 \qquad
 \Omega=\frac{av}{u}.
 }
 \label{eq:saddle-equations}
\end{equation}
On shell,
\begin{equation}
 \boxed{
 I_E^{\mathrm{BH}}
 =-\Phi_{\mathrm{BH}}
 =\frac{\pi\beta u(1-\rp^2)}{8G\XiA}.
 }
 \label{eq:on-shell-action}
\end{equation}
This is the standard background-subtracted Kerr--AdS$_5$ action; in
particular, its factor of $\XiA^{-1}$ follows from
Eqs.~\eqref{eq:E}--\eqref{eq:S}.

\subsection{Local Picard--Lefschetz geometry and contour data}
\label{sec:PL}

The term ``stability exponent'' is used for several inequivalent objects.
For a real dynamical system $\dot X^i=F^i(X)$, the asymptotic Lyapunov
exponent of a tangent vector is
\begin{equation}
 \lambda_{\rm dyn}
 =
 \limsup_{t\to\infty}\frac{1}{t}
 \log\frac{\|\delta X(t)\|}{\|\delta X(0)\|}.
 \label{eq:dynamical-Lyapunov}
\end{equation}
Near a fixed point it is controlled by the Jacobian
$M^i{}_j=\partial_jF^i$.  When $t$ is Lorentzian time, this is a statement
about physical dynamics.

Picard--Lefschetz flow is different.  Set $\mathcal I=-\Phi$, so that the
integrand in Eq.~\eqref{eq:minisuperspace} is $e^{-\mathcal I}$.  On the
real slice, the upward gradient flow and its linearization are
\begin{equation}
 \frac{\dd q^i}{\dd\tau_{\rm PL}}
 =\frac{\partial\mathcal I}{\partial q^i},
 \qquad
 \frac{\dd}{\dd\tau_{\rm PL}}\delta q^i
 =-H_{ij}\,\delta q^j,
 \qquad
 q^i=(\rp,a),
 H_{ij}=\partial_i\partial_j\Phi.
 \label{eq:real-gradient-linearization}
\end{equation}
The signed rates of this real linearized flow are the eigenvalues of $-H$,
equivalently minus the ordinary eigenvalues of $H$.  The parameter
$\tau_{\rm PL}$ is an auxiliary flow time, not Lorentzian time; these rates do
not measure orbital instability, scrambling, or an out-of-time-order
correlator.

After analytic continuation to $z=(z^1,z^2)\in\C^2$, the holomorphic Hessian
at a saddle is complex symmetric but generally non-Hermitian.  With the flat
Hermitian metric in the chosen collective coordinates, the upward flow and
its linearization are
\begin{equation}
 \frac{\dd z^i}{\dd\tau_{\rm PL}}
 =
 -\overline{\frac{\partial\Phi}{\partial z^i}},
 \qquad
 \frac{\dd\eta^i}{\dd\tau_{\rm PL}}
 =-\overline{(H_\sigma)_{ij}\eta^j}.
 \label{eq:complex-gradient-linearization-preview}
\end{equation}
This antilinear system is governed by the Takagi values (equivalently, the
singular values) of $H_\sigma$, as shown below.  With the chosen flat metric
they reduce to the absolute
values of the real Hessian eigenvalues at a real saddle, but not at a
genuinely complex saddle.

For real $(\beta,\Omega,G)$, the exponent obeys
$\Phi(\bar z;\beta,\Omega)=\overline{\Phi(z;\beta,\Omega)}$.  Hence a
nonreal solution generically occurs with its conjugate:
\begin{equation}
 z_{\bar\sigma}=\bar z_\sigma,
 \qquad
 \Phi_{\bar\sigma}=\overline{\Phi_\sigma},
 \qquad
 H_{\bar\sigma}=\overline{H_\sigma},
 \qquad
 \lambda_{\rm PL,i}^{(\bar\sigma)}
 =\lambda_{\rm PL,i}^{(\sigma)}.
 \label{eq:conjugate-saddle-data}
\end{equation}
The pair has equal exponential magnitude and identical local rates, while
the imaginary parts of their saddle exponents have opposite signs.  If the
cycle and measure are conjugation invariant, the thimble orientations are
chosen compatibly, and $n_{\bar\sigma}=n_\sigma\neq0$, then the local
prefactors obey $\mathcal A_{\bar\sigma}=\overline{\mathcal A_\sigma}$ and the
leading contributions combine as
\begin{equation}
 n_\sigma Z_\sigma+n_{\bar\sigma}Z_{\bar\sigma}
 \sim
 2|n_\sigma\mathcal A_\sigma|e^{\Re\Phi_\sigma}
 \cos\!\left(\Im\Phi_\sigma+
 \arg(n_\sigma\mathcal A_\sigma)\right),
 \label{eq:conjugate-pair-preview}
\end{equation}
where $\mathcal A_\sigma$ contains the measure, oriented Gaussian factor,
and any consistently retained fluctuation determinants, but not
$n_\sigma$.  Equal local rates do not determine thimble membership.  A
degenerate critical point is characterized by
\begin{equation}
 \det H_\sigma=0
 \quad\Longleftrightarrow\quad
 \lambda_{\rm PL,min}=0.
 \label{eq:complex-coalescence-preview}
\end{equation}
Equation~\eqref{eq:complex-coalescence-preview} alone establishes rank loss,
not branch coalescence.  When there is exactly one zero mode, the cubic
derivative along it is nonzero, and the source variation unfolds the
degeneracy transversely, the critical point is a fold: two real saddles merge
and continue as a complex-conjugate pair.  The quadratic approximation then
fails along the single soft direction, producing the Airy scaling developed
in Subsection~\ref{sec:airy}~\cite{Chester:1957}.  Stokes and anti-Stokes
conditions, which compare distinct saddles, are introduced below.
Although the numerical rates depend on the collective-coordinate metric and
normalization, the Hessian rank loss is invariant under a nonsingular change
of collective coordinates at the critical point.  Throughout we use the
flat metric in $(\rp,a)$.

The local construction can be stated directly for a holomorphic action
$\mathcal{I}(z)$.  With the flat Hermitian metric chosen above, the upward
Picard--Lefschetz flow is
\begin{equation}
 \frac{\dd z^i}{\dd\tau_{\rm PL}}
 =\overline{\frac{\partial\mathcal{I}}{\partial z^i}}.
 \label{eq:PL-flow}
\end{equation}
Along this flow, $\Re\mathcal{I}$ increases monotonically while
$\Im\mathcal{I}$ is constant.  In the present problem one may take
$\mathcal{I}=-\Phi$; because the integrand is $e^{-\mathcal I}$, this
``upward'' flow of $\Re\mathcal I$ is steepest descent for the integrand.  At
a critical point $z_\sigma$ we define
\begin{equation}
 (H_\sigma)_{ij}=\left.\partial_i\partial_j\Phi\right|_{z=z_\sigma}.
 \label{eq:H-def}
\end{equation}
The Hessian of $\mathcal{I}$ is $-H_\sigma$, which has the same singular
values as $H_\sigma$.

Linearizing around $z_\sigma$ gives an antilinear system,
\begin{equation}
 \frac{\dd\eta^i}{\dd\tau_{\rm PL}}
 =-\overline{(H_\sigma)_{ij}\eta^j}.
 \label{eq:linear-flow}
\end{equation}
Consequently, the ordinary complex eigenvalues of $H_\sigma$ are not the local
thimble-flow exponents.  The relevant quantities are the Takagi singular
values $\sigma_i\geq0$.  If $u_i$ denotes a column of the unitary Takagi
matrix $U$, they are defined by
\begin{equation}
 H_\sigma\overline{u_i}=\sigma_i u_i,
 \qquad
 H_\sigma=U\,\diag(\sigma_1,\ldots,\sigma_n)\,U^T.
 \label{eq:Takagi}
\end{equation}
We denote them by
\begin{equation}
 \boxed{
 \lPL{i}=\sigma_i(H_\sigma).
 }
 \label{eq:PL-exponents}
\end{equation}

This is the standard local Takagi construction used in Lefschetz-thimble
calculations.  In lattice field theory, the Takagi vectors furnish a basis
for the tangent space of the thimble at a critical point, while the
non-negative Takagi values enter the semiclassical Gaussian contribution
through $(\det\Sigma_\sigma)^{-1/2}$~\cite{DiRenzoEruzzi:2018}.  We apply the
same construction to the Kerr--AdS thermodynamic mini-superspace.  Once the
Kerr saddles leave the real slice, their ordinary Hessian eigenvalues no
longer give the Picard--Lefschetz rates, whereas the Takagi values remain real
and non-negative.  The smallest value identifies the mixed
radial rotational soft mode and vanishes on the black hole fold.

Writing $H_\sigma=A+iB$ and $\eta=x+iy$, with $A$ and $B$ real symmetric, the
linearized flow is exactly the real doubled system
\begin{equation}
 \frac{\dd}{\dd\tau_{\rm PL}}
 \begin{pmatrix}x\\y\end{pmatrix}
 =
 \begin{pmatrix}-A&B\\B&A\end{pmatrix}
 \begin{pmatrix}x\\y\end{pmatrix}.
 \label{eq:doubled}
\end{equation}
Its eigenvalues occur in pairs $\pm\sigma_i$.

For a generic $2\times2$ complex symmetric matrix $H$, the upper and lower
signs below give $\lPL{1}$ and $\lPL{2}$, respectively:
\begin{align}
 \lPL{1,2}^2
 &=\frac{T_H\pm\sqrt{T_H^2-4D_H^2}}{2},
 \label{eq:Takagi-2by2}\\
 T_H&=|H_{11}|^2+|H_{22}|^2+2|H_{12}|^2,
 \qquad
 D_H=|\det H|,
\end{align}
where we order $\lPL{1}\geq\lPL{2}\geq0$.

The Takagi data determine only the local geometry.  Global thimble membership
is instead fixed by the integration cycle and its intersections with the dual
cycles.  Away from Stokes walls, let $\Jth_\sigma$ and $\Kth_\sigma$ be the
descent and dual ascent cycles of an interior Morse critical point
\cite{Witten:2010,DunneUnsal:2016}.  With compatible orientations,

\begin{equation}
 \langle\Jth_\sigma,\Kth_\rho\rangle=\delta_{\sigma\rho}.
 \label{eq:dual-cycles}
\end{equation}
Separating the thermal boundary saddle from the interior saddles, we write
\begin{equation}
 \Gamma=b_0\Jth_0+
 \sum_{\sigma\in{\rm int}}n_\sigma\Jth_\sigma,
 \qquad
 n_\sigma=\langle\Gamma,\Kth_\sigma\rangle\in\mathbb Z.
 \label{eq:thimble-decomposition}
\end{equation}
Here $\Jth_0$ follows the convention of Ref.~\cite{Singhi:2025}.  It is the
full thermal thimble, whose real representative in the high-$\beta$ chamber
spans $\rp\in\mathbb R$ and $-1<a<1$.  The original cycle has $\rp\geq0$,
and Ref.~\cite{Singhi:2025} assigns $\Jth_0$ the weight
\begin{equation}
 b_0=\frac12.
 \label{eq:thermal-thimble-weight}
\end{equation}
This is a normalization of the radially extended thermal cycle, not the
intersection number $\langle\Gamma,\Kth_0\rangle$ and not a factor from the
reduced Gaussian determinant.  At quadratic order it agrees with the
half-Gaussian on $\rp\geq0$; beyond quadratic order no such equality follows,
since $\Phi$ is not even under $\rp\to-\rp$.

When the black hole saddles are real, the real-$G$ contour of
Ref.~\cite{Singhi:2025} lies on a Stokes surface.  Accordingly, the notation
$\frac12\Jth_0+\Jth_+$ used below is only shorthand for the effective saddle
sum obtained from the conjugation-symmetric $\Im G\to0^\pm$ prescription.  It
is not a thimble decomposition on the real-$G$ Stokes surface.  The interior
coefficients and the thermal weight are contour data and cannot be obtained
from a local Hessian.

Away from Stokes walls the integer coefficients $n_\sigma$ are locally
constant.  A jump requires a connecting flow and equality of the conserved
flow phases,
\begin{equation}
 \Im\!\left(\Phi_\sigma-\Phi_\rho\right)=0
 \label{eq:Stokes-condition}
\end{equation}
on compatible branches of the action.  For a multivalued action, equality of
the exponent phases may instead be expressed modulo $2\pi$, but that weaker
statement does not guarantee a connecting trajectory.  The additional
connecting-flow condition is essential.  With our terminology, this is
different from an anti-Stokes locus,
\begin{equation}
 \Re\!\left(\Phi_\sigma-\Phi_\rho\right)=0,
 \label{eq:anti-Stokes-condition}
\end{equation}
which defines equal exponential magnitude whether or not either saddle
contributes.  It represents a physical competition only when both contour
coefficients are nonzero.  Thus an anti-Stokes crossing can exchange leading
semiclassical dominance without changing the intersection numbers, whereas
a Stokes crossing can change the thimble decomposition.  At leading
exponential order, the Hawking--Page curve below is an anti-Stokes line
between thermal AdS and the large black hole; its existence alone does not
determine the black hole contour coefficient.

The role of the local Hessian can be made explicit.  Near a nondegenerate
critical point,
\begin{equation}
 \Phi(z)=\Phi_\sigma+\frac{1}{2}\eta^T H_\sigma\eta
 +\mathcal{O}(\eta^3),
 \qquad
 \eta=z-z_\sigma.
 \label{eq:local-expansion}
\end{equation}
Write the Takagi factorization as
$H_\sigma=U_\sigma\Sigma_\sigma U_\sigma^T$, with
$\Sigma_\sigma=\diag(\lPL{1},\lPL{2})$.  A convenient oriented tangent plane
to the local steepest-descent cycle is parameterized by
\begin{equation}
 \eta=i\,\overline{U_\sigma}\,t,
 \qquad t\in\R^2,
 \label{eq:Takagi-descent-coordinates}
\end{equation}
for which
\begin{equation}
 \frac{1}{2}\eta^T H_\sigma\eta
 =-\frac{1}{2}\left(\lPL{1}t_1^2+\lPL{2}t_2^2\right).
 \label{eq:Takagi-Gaussian}
\end{equation}
For an interior saddle at which the holomorphically continued measure is
smooth and nonzero, Gaussian integration gives
\begin{equation}
 Z_\sigma
 \sim
 \mathcal A_\sigma e^{\Phi_\sigma},
 \qquad
 \mathcal A_\sigma
 =\mu_\sigma
 \det\!\left(i\overline{U_\sigma}\right)
 \frac{2\pi}{\sqrt{\lPL{1}\lPL{2}}}
 =
 \mu_\sigma\frac{2\pi}{\sqrt{\det(-H_\sigma)}}.
 \label{eq:local-contribution}
\end{equation}
We use $c_0=b_0$ for the thermal saddle and
$c_\sigma=n_\sigma$ for an interior saddle; the term in the contour sum is
$c_\sigma Z_\sigma$.  The square-root branch in
Eq.~\eqref{eq:local-contribution} is fixed by the orientation of
$\Jth_\sigma$.  For an interior saddle, reversing that orientation changes
both $Z_\sigma$ and $n_\sigma$, leaving their product unchanged.  Since the
reduced integral is two-dimensional, $\det(-H_\sigma)=\det H_\sigma$.  Its
magnitude is
\begin{equation}
 \abs{\det H}^{-1/2}
 =(\lPL{1}\lPL{2})^{-1/2},
 \label{eq:one-loop-magnitude}
\end{equation}
and the local determinant phase is
\begin{equation}
 \boxed{
 \phiM=-\frac{1}{2}\arg\det H\pmod\pi.
 }
 \label{eq:Maslov}
\end{equation}
Thus $\phiM$ is the local Gaussian phase, not a criterion for thimble
membership.  The phase of $c_\sigma Z_\sigma$ also contains
$\Im\Phi_\sigma$, the measure phase, the contour weight, and phases from the
omitted fluctuation modes.  Under a holomorphic change of collective
coordinates, only the complete local amplitude is invariant.

It is useful to collect these logical roles:
\begin{equation}
 \underbrace{\Phi_\sigma}_{\substack{\text{exponential weight}\\
                         \text{and relative phase}}},
 \qquad
 \underbrace{H_\sigma}_{\substack{\text{local rates, Gaussian magnitude,}\\
                         \text{and determinant phase}}},
 \qquad
 \underbrace{c_\sigma}_{\substack{\text{global contour}\\
                         \text{weight}}}.
 \label{eq:local-global-dictionary}
\end{equation}
For two nonzero terms, the Gaussian sum vanishes if and only if
\begin{align}
 |c_\sigma\mathcal A_\sigma|e^{\Re\Phi_\sigma}
 &=|c_\rho\mathcal A_\rho|e^{\Re\Phi_\rho},
 \label{eq:two-saddle-magnitude}\\
 \arg\!\left(
 \frac{c_\sigma\mathcal A_\sigma}
 {c_\rho\mathcal A_\rho}\right)
 +\Im\!\left(\Phi_\sigma-\Phi_\rho\right)
 &=(2k+1)\pi,\qquad k\in\mathbb Z.
 \label{eq:two-saddle-phase}
\end{align}
Equality of the real parts of two exponents, a Hessian zero, or a determinant
phase by itself is therefore insufficient to establish a partition-function
zero.  Finally, when one Takagi value vanishes,
Eq.~\eqref{eq:local-contribution} ceases to be valid: the apparent Gaussian
divergence signals a degenerate critical point, and the soft direction must
be retained beyond quadratic order.  For a simple fold, with one hard mode
and a nonzero soft cubic, the resulting Airy uniform approximation is
developed in Subsection~\ref{sec:airy}.

\subsection{Reduced Hessian and black hole fold}
\label{sec:hessian}

We first differentiate at fixed $(\beta,\Omega)$ and only then impose the
saddle equations~\eqref{eq:saddle-equations}.  With indices $1,2$
corresponding to $(\rp,a)$, the reduced Hessian entries are
\begin{align}
 H_{11}
 &=\frac{\pi^2N_1}{2G\rp\XiA s},
 \label{eq:H11}\\
 H_{22}
 &=\frac{\pi^2vN_2}{4G\rp\XiA^3s},
 \label{eq:H22}\\
 H_{12}=H_{21}
 &=\frac{2\pi^2av(1-\rp^2)}{G\XiA^2s},
 \label{eq:H12}
\end{align}
where
\begin{align}
 N_1&=-6\rp^4-3\rp^2a^2+3\rp^2-a^4-a^2,
 \label{eq:N1}\\
 N_2&=4(\rp^2-2\rp^2a^2+a^4)s
 \nonumber\\
 &\hspace{2.5em}
 -\XiA(10\rp^4+2\rp^2a^2-4a^4+6\rp^2-6a^2).
 \label{eq:N2}
\end{align}
Equivalently,
\begin{equation}
 H=\frac{\pi^2}{2G\rp\XiA s}
 \begin{pmatrix}
  N_1 & \dfrac{4\rp av(1-\rp^2)}{\XiA}\\[8pt]
  \dfrac{4\rp av(1-\rp^2)}{\XiA}
  &\dfrac{vN_2}{2\XiA^2}
 \end{pmatrix}.
 \label{eq:full-H}
\end{equation}

A direct determinant calculation gives
\begin{equation}
 \det H
 =\frac{\pi^4v}{8G^2\rp^2\XiA^4s^2}\,\mathcal{D}(\rp,a),
 \label{eq:detH-D}
\end{equation}
with
\begin{equation}
 \mathcal{D}=N_1N_2-32\rp^2a^2v(1-\rp^2)^2.
 \label{eq:D-unfactorized}
\end{equation}
The key simplification is the exact factorization
\begin{equation}
 \boxed{
 \mathcal{D}
 =2(a^2+3)(\rp^2+a^2)^2(1+\rp^2)
 \left(2\rp^2-a^2-1\right).
 }
 \label{eq:D-factorized}
\end{equation}
Hence
\begin{equation}
 \boxed{
 \det H
 =\frac{\pi^4(a^2+3)u^2v^2}{4G^2\rp^2\XiA^4s^2}
 \left(2\rp^2-a^2-1\right).
 }
 \label{eq:detH-factorized}
\end{equation}
For real black hole saddles with $\rp>0$, $|a|<1$, and $G>0$, every prefactor
in Eq.~\eqref{eq:detH-factorized} is strictly positive.  Therefore the sign
and the zero of the reduced determinant are controlled entirely by
$2\rp^2-a^2-1$.

Thermal AdS is the boundary critical point
$p_0=(\rp,a)=(0,0)$ of the reduced model, rather than a solution of the
black hole saddle map.  Expanding Eq.~\eqref{eq:Phi} at fixed
$(\beta,\Omega)$ gives
\begin{equation}
 H_0=-\frac{3\pi\beta}{4G}\,\mathbb{I}_2,
 \qquad
 \lPL{1}^{(0)}=\lPL{2}^{(0)}=\frac{3\pi\beta}{4G}.
 \label{eq:thermal-H}
\end{equation}
This expression assumes real $\beta,G>0$; under complex continuation the
Takagi value is $|3\pi\beta/(4G)|$.  It is a rate of the reduced exponent,
not an eigenvalue of the full thermal AdS fluctuation operator.  We define
$\mathcal A_0$ to be the Gaussian factor on the full radially extended
$\Jth_0$,
\begin{equation}
 \mathcal A_0
 =\mu_0\frac{2\pi}{\sqrt{\det(-H_0)}}.
 \label{eq:thermal-full-Gaussian}
\end{equation}
It contains no factor $1/2$.  The thermal term in the contour sum is
$b_0\mathcal A_0e^{\Phi_0}=\frac12\mathcal A_0$.

On the Schwarzschild--AdS$_5$ slice $a=0$,
\begin{equation}
 H_{11}^{\mathrm{Schw}}
 =\frac{3\pi^2\rp(1-2\rp^2)}{2G(2\rp^2+1)},
 \qquad
 H_{22}^{\mathrm{Schw}}
 =-\frac{\pi^2\rp(1+\rp^2)^2}{2G(2\rp^2+1)},
 \qquad
 H_{12}=0.
 \label{eq:H-Schwarzschild}
\end{equation}
Thus the reduced black hole map has a fold at $\rp=1/\sqrt{2}$, where the
radial Hessian eigenvalue vanishes while the rotational eigenvalue remains
nonzero.

\section{Grand canonical phase structure: Hawking--Page transition and
black hole fold}
\label{sec:phase}
For real sources $\beta>0$ and $|\Omega|<1$, two distinct conditions organize
the grand canonical saddle diagram.  The Hawking--Page curve follows from
equality of the leading thermal AdS and black hole exponents.  The black hole
fold is instead the singular locus of the map
$(\rp,a)\mapsto(\beta,\Omega)$, where the small and large Kerr saddles
coalesce.  The first condition is anti-Stokes; the second is a degeneracy of
the reduced saddle equations.  Their physical roles should not be
identified.  This section concerns the real Kerr family and its classical
actions.  The thimble coefficients are global contour data and are discussed
in Subsection~\ref{sec:airy}.

The equations are invariant under $(a,\Omega)\mapsto(-a,-\Omega)$.  We quote
formulas for the full range $|\Omega|<1$, while the figures display the
$\Omega\geq0$ half of the diagram.


\subsection{Physical domain, Hawking--Page curve, and black hole fold}

For a real Kerr--AdS black hole, $\rp$ is the outer horizon radius and hence
$\rp>0$.  The same restriction follows directly from the source map.  For a
regular real black hole saddle, $u=\rp^2+a^2>0$ and
$s=2\rp^2+a^2+1>0$, so Eq.~\eqref{eq:saddle-equations} gives
\begin{equation}
 \operatorname{sgn}\beta=\operatorname{sgn}\rp.
 \label{eq:beta-r-sign}
\end{equation}
Thus a real solution with $\rp<0$ lies on the negative-$\beta$ sheet and is
not a second black hole at the same physical sources.  Under
$\rp\mapsto-\rp$ at fixed real $a$, the quantities $E$, $J$, and $\Omega$
are unchanged, whereas the oriented Euclidean period and the entropy change
sign.  The labels $\rp^-$ and $\rp^+$ for the small and large solutions
therefore refer to the smaller and larger \emph{positive} roots.

The negative radial half-line has a separate use in the thimble construction.
The full thermal thimble $\Jth_0$ is defined by extending the flow through the
boundary saddle at $\rp=0$; its coefficient $b_0=1/2$ accounts for the
original half-line contour.  This extension does not add negative-$\rp$
black holes to the real phase diagram.  After complexification, $\rp$ is a
complex coordinate and no ordering by sign is imposed.

Within this physical domain, the Hawking--Page curve is obtained by comparing
the black hole exponent with thermal AdS.  In the subtraction scheme used
here, the classical thermal AdS exponent is $\Phi_{\mathrm{th}}=0$.  On the
physical black hole branch,
Eq.~\eqref{eq:on-shell-action} gives
\begin{equation}
 \Phi_{\mathrm{BH}}=0
 \quad\Longleftrightarrow\quad
 \rp=1,
 \qquad \rp>0.
 \label{eq:HP-r}
\end{equation}
The algebraic root $\rp=-1$ belongs to the negative-$\beta$ sheet and is not
part of the physical ensemble.  Substituting $\rp=1$ into the source map
gives
\begin{equation}
 \beta_{\mathrm{HP}}(a)
 =\frac{2\pi(1+a^2)}{3+a^2},
 \qquad
 \Omega_{\mathrm{HP}}(a)
 =\frac{2a}{1+a^2}.
 \label{eq:HP-parametric}
\end{equation}
Eliminating $a$ yields
\begin{equation}
 \boxed{
 \beta_{\mathrm{HP}}(\Omega)
 =\frac{2\pi\left(2-\sqrt{1-\Omega^2}\right)}{3+\Omega^2},
 \qquad |\Omega|<1.
 }
 \label{eq:HP-Omega}
\end{equation}
This is the leading Hawking--Page anti-Stokes curve.  When the thermal and
large black hole sectors both occur on the contour, their leading
semiclassical dominance is exchanged across it.  It is not a Stokes wall:
equality of the actions does not determine the thimble coefficients.  The
thermal factor $b_0=1/2$ and the Gaussian determinants affect the subleading
finite-$G$ comparison, not the classical curve in
Eq.~\eqref{eq:HP-Omega}.

At leading order in $G^{-1}$, thermal AdS has zero entropy.  The entropy jump
is therefore the black hole entropy at $\rp=1$,
\begin{equation}
 \Delta S
 =\frac{\pi^2(1+a^2)}{2G(1-a^2)},
 \label{eq:entropy-jump}
\end{equation}
and the grand canonical latent heat is
\begin{equation}
 \boxed{
 L_\Omega\equiv T_{\mathrm{HP}}\Delta S
 =\Delta(E-\Omega J)
 =\frac{\pi(3+a^2)}{4G(1-a^2)}.
 }
 \label{eq:latent-heat}
\end{equation}
For $a\neq0$, $L_\Omega$ is not the jump in $E$ alone; rather,
$\Delta E=L_\Omega+\Omega\Delta J$.  At $a=0$,
$\Delta S=\pi^2/(2G)$ and $L_\Omega=3\pi/(4G)$.

The second distinguished locus is the black hole fold.  For a regular real
black hole, the determinant factorization in
Eq.~\eqref{eq:detH-factorized} and the Jacobian of the source map give
\begin{equation}
 \det H=0
 \quad\Longleftrightarrow\quad
 2\rp^2-a^2-1=0.
 \label{eq:fold-condition}
\end{equation}
At fixed $\Omega$, this is the turning point at which the two positive-radius
black hole solutions coalesce.  The positive physical root is
\begin{equation}
 \boxed{
 \rp^{\mathrm{sp}}(a)=\sqrt{\frac{1+a^2}{2}}.
 }
 \label{eq:r-spinodal}
\end{equation}
The negative algebraic root lies on the negative-$\beta$ sheet excluded
above.  Substitution into Eq.~\eqref{eq:saddle-equations} gives
\begin{equation}
 \boxed{
 \beta_{\mathrm{sp}}(a)
 =\frac{\pi(1+3a^2)}{\sqrt{2}(1+a^2)^{3/2}},
 \qquad
 \Omega_{\mathrm{sp}}(a)
 =\frac{a(a^2+3)}{1+3a^2}.
 }
 \label{eq:spinodal-parametric}
\end{equation}
For $0\leq a<1$, the curve runs from
\begin{equation}
 (\Omega,\beta)=(0,\pi/\sqrt{2})
 \quad\text{to}\quad
 (\Omega,\beta)\longrightarrow(1,\pi)
 \quad\text{as}\quad a\to1^-.
\end{equation}
The $a<0$ branch is its reflection under
$(a,\Omega)\mapsto(-a,-\Omega)$.

The heat capacity at fixed angular velocity can be written compactly as
\begin{equation}
 \boxed{
 C_\Omega
 =T\left(\frac{\partial S}{\partial T}\right)_\Omega
 =\frac{\pi^2\rp(3\rp^2-a^2)s}
 {2G\XiA\left(2\rp^2-a^2-1\right)}.
 }
 \label{eq:C-Omega}
\end{equation}
In the physical domain $\rp^2>|a|$, all factors in the numerator are
positive, so the sign of $C_\Omega$ is the sign of
$2\rp^2-a^2-1$.  The heat capacity diverges at the spinodal and changes
sign between the small and large branches.  Subsection~\ref{sec:airy} verifies
that the soft cubic and transverse unfolding are nonzero, so a generic point
on this degeneracy curve is a simple fold with a local Airy uniformization.
That local statement does not determine whether the Airy sector occurs in
the physical partition function.

\subsection{Ordering of the phase boundaries and fixed-potential branches}

The Hawking--Page point has $\rp=1$, while every physical fold point obeys
\begin{equation}
 (\rp^{\mathrm{sp}})^2=\frac{1+a^2}{2}<1
 \qquad (|a|<1).
\end{equation}
Because the values of $a$ at the two loci differ when $\Omega$ is fixed, the
ordering also requires the monotonicity of the fixed-$\Omega$ branch.  Direct
differentiation of Eq.~\eqref{eq:saddle-equations} gives
\begin{equation}
 \left(\frac{\partial T}{\partial\rp}\right)_{\Omega}
 =\frac{2\rp^2-a^2-1}{2\pi(\rp^2-a^2)}.
 \label{eq:T-fixed-Omega-slope}
\end{equation}
The denominator is positive because $\rp^2>|a|>a^2$ for $0<|a|<1$
(with the $a=0$ case immediate).  Hence $T$ increases with $\rp$ on the
large branch.  Since $\rp^{\mathrm{sp}}<1$, the large branch crosses the
Hawking--Page radius before it reaches the fold as $\beta$ is increased.
Therefore
\begin{equation}
 \boxed{
 \beta_{\mathrm{HP}}(\Omega)
 <\beta_{\mathrm{sp}}(\Omega),
 \qquad |\Omega|<1.
 }
 \label{eq:curve-ordering}
\end{equation}
The two curves meet only at the two excluded boundary limits
\begin{equation}
 (\Omega,a)\to(1,1)
 \quad\text{or}\quad
 (\Omega,a)\to(-1,-1),
 \qquad
 \rp\to1,
 \qquad
 \beta\to\pi.
 \label{eq:boundary-meeting}
\end{equation}
There is therefore no critical endpoint in the open grand canonical domain
$|\Omega|<1$.

The on-shell Euclidean action at the fold is
\begin{equation}
 I_E^{\mathrm{sp}}
 =\frac{\pi\beta_{\mathrm{sp}}(1+3a^2)}{32G}>0,
 \qquad
 \mathcal G_{\mathrm{sp}}
 =\frac{I_E^{\mathrm{sp}}}{\beta_{\mathrm{sp}}}
 =\frac{\pi(1+3a^2)}{32G}>0.
 \label{eq:spinodal-action}
\end{equation}
Thus, at leading exponential order, the degenerate black hole saddle lies
above thermal AdS, for which $\mathcal G_{\mathrm{th}}=0$.  The fold is the
spinodal endpoint of the metastable large black hole branch, where it meets
the unstable small branch.  It is not a line of phase coexistence.

Having established this ordering, we now describe the two real branches at
fixed angular potential.  For fixed $|\Omega|<1$, the physical solution of
the angular-potential constraint is most conveniently written as
\begin{equation}
 a(\rp;\Omega)
 =
 \frac{2\Omega\rp^2}
 {1+\rp^2+\sqrt{(1+\rp^2)^2-4\Omega^2\rp^2}}.
 \label{eq:a-fixed-Omega}
\end{equation}
This form is regular at $\Omega=0$, where it gives $a=0$.  The second
algebraic root of the quadratic constraint lies outside $|a|<1$ and does not
belong to the physical Kerr family.  Since
$(1+\rp^2)^2-4\Omega^2\rp^2>0$ for $|\Omega|<1$, Eq.~\eqref{eq:a-fixed-Omega}
gives a unique physical $a$ for every $\rp>0$.

Substitution into Eq.~\eqref{eq:thermo-map} gives $T(\rp;\Omega)$ with a
single minimum, or equivalently $\beta(\rp;\Omega)$ with a single maximum,
at the fold.  For $0<\beta<\beta_{\mathrm{sp}}$ there are two positive-radius
solutions.  The smaller one has $C_\Omega<0$ and the larger one has
$C_\Omega>0$.  At $\beta=\beta_{\mathrm{sp}}$ they coalesce; for
$\beta>\beta_{\mathrm{sp}}$ there is no real small/large pair in $|a|<1$.
These are two branches of one saddle map, not two stable phases separated by
a liquid--gas transition.

\begin{figure}[tbp]
 \centering
 \IfFileExists{kerr_ads_small_large_branches.pdf}{%
   \includegraphics[width=0.92\textwidth]
   {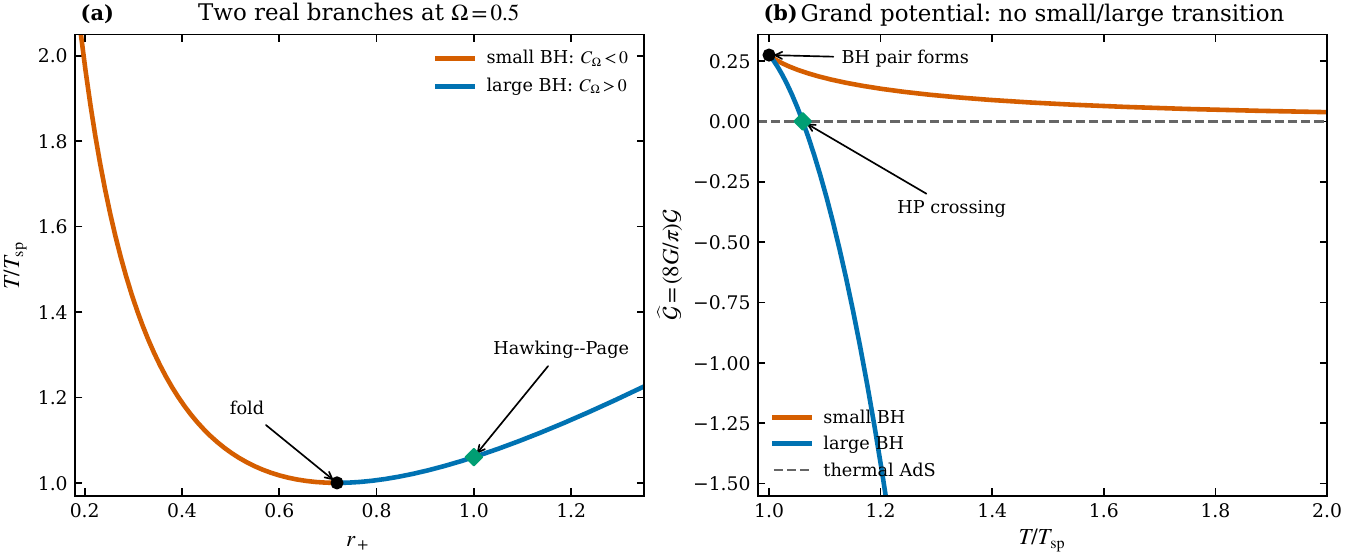}%
 }{%
   \fbox{\parbox[c][5.0cm][c]{0.86\textwidth}{\centering
   Missing figure \texttt{kerr\_ads\_small\_large\_branches.pdf}.}}%
 }
 \caption{Small and large black hole branches in the grand canonical
 ensemble at the representative value $\Omega=1/2$.  (a) The two real
 branches coalesce at the minimum temperature $T_{\rm sp}$; the diamond
 marks the Hawking--Page point on the large branch.  (b) Their common fold
 value satisfies $\mathcal G_{\rm sp}>\mathcal G_{\rm th}=0$.  Only the
 large branch crosses $\mathcal G=0$.  When the thermal and large black hole
 sectors are both present, this crossing is the Hawking--Page transition.
 The plot does not determine the thimble coefficient of the small saddle.}
 \label{fig:small-large-branches}
\end{figure}

\begin{figure}[!ht]
 \centering
 \includegraphics[width=\textwidth]{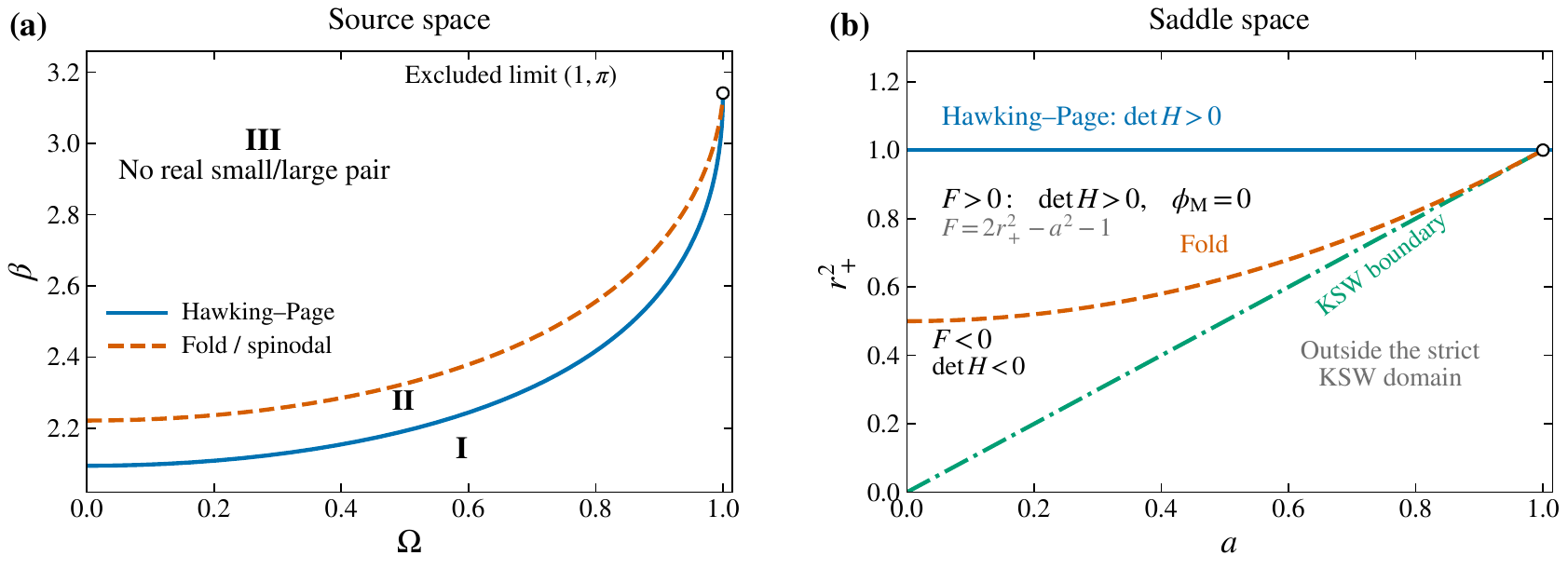}
 \caption{Phase boundaries and determinant phase chambers of singly rotating
 Kerr--AdS$_5$. (a) The solid blue Hawking--Page line lies strictly below
 the dashed orange fold for $0\leq\Omega<1$. Regions I, II, and III have
 $\beta<\beta_{\mathrm{HP}}$,
 $\beta_{\mathrm{HP}}<\beta<\beta_{\mathrm{sp}}$, and
 $\beta>\beta_{\mathrm{sp}}$, respectively. The open circle is the excluded
 limit $(1,\pi)$. (b) In saddle space, the corresponding loci are
 $r_+^2=1$ and $r_+^2=(1+a^2)/2$; the dash-dotted green KSW boundary is
 $r_+^2=a$, with a strict asymptotic KSW margin above it. For signed $a$, the boundary is $r_+^2=|a|$. The fold lies on the
 allowable side. The sign of $F=2r_+^2-a^2-1$ fixes the sign of
 $\det H$ and distinguishes the two local determinant phase chambers;
 their square-root phases differ by $\pi/2$ modulo $\pi$, with the overall
 convention fixed by the thimble orientation. The phase is undefined on
 the fold.}
 \label{fig:phase-structure}
\end{figure}

The distinction is particularly transparent in the on-shell grand potential
\begin{equation}
 \mathcal G=E-TS-\Omega J=-\frac{\Phi_{\rm BH}}{\beta}.
 \label{eq:grand-potential}
\end{equation}
The branches coalesce at the common positive value
$\mathcal G_{\mathrm{sp}}$.  The large branch crosses $\mathcal G=0$ at
$\rp=1$; when the thermal and large black hole sectors are both selected,
this is the Hawking--Page transition.  The small branch remains at $\rp<1$
and hence has $\mathcal G>0$.  Whether it contributes at all is fixed by the
thimble prescription, not by this thermodynamic plot.
Figure~\ref{fig:small-large-branches} illustrates the two branches, while
Table~\ref{tab:phase-regimes} and Figure~\ref{fig:phase-structure}(a) organize
them in source space. In region I, $\beta<\beta_{\mathrm{HP}}$, the large
black hole has lower classical action than thermal AdS. In region II,
$\beta_{\mathrm{HP}}<\beta<\beta_{\mathrm{sp}}$, thermal AdS has lower
action and the large black hole is metastable. Both regions contain the
unstable small saddle. At the fold the two black hole branches merge;
region III, $\beta>\beta_{\mathrm{sp}}$, has no real small/large pair.
The action ordering becomes a statement about semiclassical dominance only
when the corresponding saddle sectors occur on the chosen contour.
The open circle at $(\Omega,\beta)=(1,\pi)$ denotes an excluded boundary
limit, not a critical endpoint in $|\Omega|<1$.

Figure~\ref{fig:phase-structure}(b) separates these thermodynamic and local
conditions from metric allowability. Using $r_+^2$ as the vertical coordinate,
the Hawking--Page line is $r_+^2=1$, the fold is
$r_+^2=(1+a^2)/2$, and the domain with a strict asymptotic KSW margin for $a\geq0$ lies above
$r_+^2=a$. In particular,
\begin{equation}
 (r_+^{\mathrm{sp}})^2-a=\frac{(1-a)^2}{2}>0,
 \qquad 0\leq a<1,
 \label{eq:figure-fold-KSW-gap}
\end{equation}
so the fold remains inside this domain even though its reduced
Hessian is degenerate. The sign of $F=2r_+^2-a^2-1$ labels the two
nondegenerate determinant phase chambers, as discussed in the next section.
Thus the three curves encode different conditions: equality of saddle
actions, loss of local Hessian rank, and the boundary of strict metric
allowability. Equality on the KSW boundary saturates the asymptotic margin;
 the distinction from pointwise allowability at finite radius is explained
 in Subsection~\ref{sec:KSW}. None of these conditions alone fixes a thimble
 coefficient.

At fixed \( |\Omega|<1 \), the Hawking--Page point and the spinodal point characterize two distinct features of the thermodynamic branch structure. The Hawking--Page inverse temperature \(\beta_{\mathrm{HP}}\) is determined by equality of the classical actions of thermal AdS and the large Kerr--AdS saddle, whereas the spinodal inverse temperature \(\beta_{\mathrm{sp}}\) marks the fold at which the small and large black hole branches merge,
\[
\left(\frac{\partial \beta}{\partial r_+}\right)_{\Omega}=0 .
\]
For the Kerr--AdS\(_5\) family considered here,
\[
\beta_{\mathrm{HP}}<\beta_{\mathrm{sp}},
\qquad\text{equivalently}\qquad
T_{\mathrm{HP}}>T_{\mathrm{sp}} .
\]
Hence, in the interval
\(\beta_{\mathrm{HP}}<\beta<\beta_{\mathrm{sp}}\), the large black hole saddle remains locally stable but is globally subdominant to thermal AdS. At \(\beta=\beta_{\mathrm{sp}}\) the two real Kerr saddles coalesce, and for \(\beta>\beta_{\mathrm{sp}}\) they persist only through their complex continuation.
\begin{table}[htbp]
 \centering
 \small
 \renewcommand{\arraystretch}{1.22}
 \begin{tabular}{@{}
 >{\raggedright\arraybackslash}p{0.16\textwidth}
 >{\raggedright\arraybackslash}p{0.18\textwidth}
 >{\raggedright\arraybackslash}p{0.38\textwidth}
 >{\raggedright\arraybackslash}p{0.18\textwidth}@{}}
 \toprule
 Regime & Inverse-temperature range & Small/large Kerr pair
 & Local conclusion\\
 \midrule
 Large black hole region
 & $0<\beta<\beta_{\mathrm{HP}}$
 & Two positive-radius real saddles; $C_\Omega<0$ on the small branch and
 $C_\Omega>0$ on the large branch
 & The large black hole has the lowest classical action\\
 Thermal window
 & $\beta_{\mathrm{HP}}<\beta<\beta_{\mathrm{sp}}$
 & Two positive-radius real saddles; the large branch is metastable and the
 small branch has $C_\Omega<0$
 & Thermal AdS has the lower classical action\\
 Beyond the fold
 & $\beta>\beta_{\mathrm{sp}}$
 & No real pair in $|a|<1$; the two branches continue to complex saddles
 & No contour-independent dominance statement\\
 \bottomrule
 \end{tabular}
 \caption{Classical structure of the small/large Kerr pair at fixed
 $|\Omega|<1$.  The table compares the actions of real saddles in the first
 two rows.  Beyond the fold, dominance cannot be inferred without the
 thimble coefficients; the contour prescription used in
 Subsection~\ref{sec:airy} selects the thermal sector.}
 \label{tab:phase-regimes}
\end{table}

\FloatBarrier

\section{Local flow rates, fold asymptotics, and complex saddles}
\label{sec:exponents}
We first collect the local Takagi spectra and reduced Gaussian phases of the
Schwarzschild and rotating Kerr saddles.  We then construct the fold
uniformization, distinguish the lateral saddle prescription from its
uniform completion, and prove zero exclusion near the real fold for the
integral with a flat measure. The Fisher zeros studied here arise from
thermal AdS/large black hole competition near Hawking--Page.  Finally, we
distinguish the Stokes geometry of the complexified thermodynamic saddle map
from KSW allowability of the spacetime metric.

\subsection{Local Takagi spectra and reduced Gaussian data}

\subsubsection{Schwarzschild limit}

Because the Schwarzschild Hessian is diagonal and real, its Takagi values are
the absolute values of its diagonal entries:
\begin{equation}
 \boxed{
 \lPL{1,2}^{\mathrm{Schw}}
 =\left\{
 \frac{3\pi^2\rp\abs{1-2\rp^2}}{2G(2\rp^2+1)},
 \frac{\pi^2\rp(1+\rp^2)^2}{2G(2\rp^2+1)}
 \right\},
 }
 \label{eq:PL-Schwarzschild}
\end{equation}
with the pair ordered by magnitude when necessary.  At the fold,
\begin{equation}
 \lPL{\min}=0,
 \qquad
 \lPL{\max}=\frac{9\pi^2\sqrt{2}}{32G}.
 \label{eq:PL-at-fold}
\end{equation}
Only one local direction becomes soft.

At the Schwarzschild Hawking--Page point $\rp=1$,
\begin{equation}
 H_{11}=-\frac{\pi^2}{2G},
 \qquad
 H_{22}=-\frac{2\pi^2}{3G},
 \label{eq:H-at-HP}
\end{equation}
so the saddle is nondegenerate.  This explicitly demonstrates that the
Hawking--Page transition is not a local fold of the $(\rp,a)$ reduced
exponent.  This statement concerns the thermodynamic mini-superspace and
does not assert nondegeneracy of the full gravitational fluctuation operator.
Table~\ref{tab:Schwarzschild-rates} collects this result together with the
thermal, fold, and representative small and large black hole rates.

\begin{table}[htbp]
 \centering
 \renewcommand{\arraystretch}{1.2}
 \begin{tabular}{lcccc}
 \toprule
 Configuration & $\rp$ & $\beta_\star$
 & $\widehat\lambda_{\max}$ & $\widehat\lambda_{\min}$\\
 \midrule
 Thermal saddle
 & --- & any & $3\beta/(2\pi)$ & $3\beta/(2\pi)$\\
 Schwarzschild Hawking--Page
 & $1$ & $2\pi/3$ & $4/3$ & $1$\\
 Large Schwarzschild check
 & $2$ & $4\pi/9$ & $50/9$ & $14/3$\\
 Small Schwarzschild check
 & $1/2$ & $2\pi/3$ & $25/48$ & $1/2$\\
 Schwarzschild fold
 & $1/\sqrt{2}$ & $\pi/\sqrt{2}$ & $9\sqrt{2}/16$ & $0$\\
 \bottomrule
 \end{tabular}
 \caption{Thermal and Schwarzschild Picard--Lefschetz rates, with
 $\widehat\lambda=(2G/\pi^2)\lambda_{\mathrm{PL}}$.  The normalization and
 ordering follow Eqs.~\eqref{eq:thermal-H} and
 \eqref{eq:PL-Schwarzschild}.}
 \label{tab:Schwarzschild-rates}
\end{table}

\subsubsection{Rotating Kerr saddles and special loci}

For a real rotating saddle, $H$ is real symmetric and its Takagi values are
the absolute values of its two real eigenvalues.  For a complex saddle, one
must instead use Eq.~\eqref{eq:Takagi-2by2}; ordinary complex eigenvalues do
not give the thimble-flow rates.

To display the complete real Kerr result, define
\begin{equation}
 \widehat H\equiv\frac{2G}{\pi^2}H
 =\begin{pmatrix}\widehat h_{11}&\widehat h_{12}\\
 \widehat h_{12}&\widehat h_{22}\end{pmatrix},
 \label{eq:Hhat-definition}
\end{equation}
where
\begin{equation}
 \widehat h_{11}=\frac{N_1}{\rp\XiA s},
 \qquad
 \widehat h_{22}=\frac{vN_2}{2\rp\XiA^3s},
 \qquad
 \widehat h_{12}=\frac{4av(1-\rp^2)}{\XiA^2s}.
 \label{eq:Hhat-entries}
\end{equation}
The two ordinary eigenvalues and the ordered Takagi rates are therefore
\begin{align}
 \widehat e_\pm
 &=\frac{\widehat t\pm\widehat\Delta}{2},
 &
 \widehat t&=\widehat h_{11}+\widehat h_{22},
 \nonumber\\
 \widehat\Delta
 &=\sqrt{(\widehat h_{11}-\widehat h_{22})^2
 +4\widehat h_{12}^2},
 \label{eq:full-Kerr-eigenvalues}\\
 \boxed{
 \begin{gathered}
 \widehat\lambda_{\max}
 =\max\{\abs{\widehat e_+},\abs{\widehat e_-}\},\\
 \widehat\lambda_{\min}
 =\min\{\abs{\widehat e_+},\abs{\widehat e_-}\}
 \end{gathered}
 }
 \label{eq:full-Kerr-Takagi}
\end{align}
The physical rates are
$\lambda_{\mathrm{PL}}=(\pi^2/2G)\widehat\lambda$.  Equivalently, the
linearized doubled PL-flow matrix has spectrum
\begin{equation}
 \operatorname{spec}\mathcal L_{\mathrm{PL}}
 =\left\{
 \pm\frac{\pi^2}{2G}\widehat\lambda_{\max},
 \ \pm\frac{\pi^2}{2G}\widehat\lambda_{\min}
 \right\}.
 \label{eq:full-Kerr-flow-spectrum}
\end{equation}
Equations~\eqref{eq:Hhat-entries}--\eqref{eq:full-Kerr-flow-spectrum}
give the local Hessian and gradient-flow data at every real singly rotating
Kerr saddle, not only on the Schwarzschild slice.

The product rule
\begin{equation}
 \widehat\lambda_{\max}\widehat\lambda_{\min}
 =|\det\widehat H|
 =\left(\frac{2G}{\pi^2}\right)^2|\det H|
 \label{eq:product-rule}
\end{equation}
shows immediately that the smaller exponent vanishes on the fold while the
larger exponent remains finite for a generic coalescence.

\paragraph{Hawking--Page curve.}

At $\rp=1$, the mixed entry vanishes and the complete rotating Hessian
diagonalizes in the $(\delta\rp,\delta a)$ coordinates:
\begin{equation}
 \widehat H_{\mathrm{HP}}
 =-\operatorname{diag}\!\left(
 \frac{1+a^2}{1-a^2},
 \frac{4(1+a^2)}{(1-a^2)^2(a^2+3)}
 \right).
 \label{eq:Kerr-HP-Hessian}
\end{equation}
Since the second entry has the larger magnitude for $|a|<1$,
\begin{equation}
 \boxed{
 \widehat\lambda_{\min}^{\mathrm{HP}}
 =\frac{1+a^2}{1-a^2},
 \qquad
 \widehat\lambda_{\max}^{\mathrm{HP}}
 =\frac{4(1+a^2)}
 {(1-a^2)^2(a^2+3)}.
 }
 \label{eq:Kerr-HP-rates}
\end{equation}
Both reduced rates are strictly positive: the full rotating
Hawking--Page curve is nondegenerate in the $(\rp,a)$ mini-superspace, not
merely at its Schwarzschild endpoint.

\paragraph{Fold and soft direction.}

On the fold $\rp^2=(1+a^2)/2$, the smaller rate vanishes and the nonzero
eigenvalue equals the trace.  Direct substitution gives
\begin{equation}
 \boxed{
 \widehat\lambda_{\min}^{\mathrm{sp}}=0,
 \qquad
 \widehat\lambda_{\max}^{\mathrm{sp}}
 =\frac{\sqrt{2}(1+3a^2)(9+11a^2)}
 {16(1-a^2)(1+a^2)^{3/2}}.
 }
 \label{eq:Kerr-fold-rates}
\end{equation}
The corresponding null eigenvector is
\begin{equation}
 \boxed{
 e_0\propto
 \begin{pmatrix}1\\ \kappa(a)\end{pmatrix},
 \qquad
 \kappa(a)
 =\frac{4\sqrt{2}\,a\sqrt{1+a^2}}{a^2+3}.
 }
 \label{eq:Kerr-soft-vector}
\end{equation}
Thus the soft fluctuation obeys
$\delta a=\kappa(a)\,\delta\rp$.  The same vector annihilates the Jacobian
of the source map $(\rp,a)\mapsto(\beta,\Omega)$ on the fold, confirming
that it is precisely the state-space direction along which the sources do
not change to first order.  At $a=0$, $\kappa=0$ and the mode is purely
radial; rotation continuously mixes the radial and angular fluctuations.
The rates are even under $a\to-a$, while the mixing coefficient is odd.
Representative values along the rotating fold are given in
Table~\ref{tab:Kerr-fold-rates}.

\begin{table}[t]
 \centering
 \renewcommand{\arraystretch}{1.18}
 \begin{tabular}{cccccc}
 \toprule
 $a$ & $\rp^{\mathrm{sp}}$ & $\Omega_{\mathrm{sp}}$
 & $\beta_{\mathrm{sp}}$ & $\widehat\lambda_{\max}^{\mathrm{sp}}$
 & $\kappa$\\
 \midrule
 $0.0$ & $0.707107$ & $0.000000$ & $2.221441$ & $0.795495$ & $0.000000$\\
 $0.3$ & $0.738241$ & $0.729921$ & $2.479127$ & $1.082887$ & $0.573391$\\
 $0.6$ & $0.824621$ & $0.969231$ & $2.913331$ & $2.347335$ & $1.178030$\\
 $0.9$ & $0.951315$ & $0.999708$ & $3.129040$ & $11.735815$ & $1.797760$\\
 \bottomrule
 \end{tabular}
 \caption{Representative points on the full rotating fold.
 The dimensionless hard rate is
 $\widehat\lambda=(2G/\pi^2)\lambda_{\mathrm{PL}}$, and
 $\kappa=\delta a/\delta\rp$ is the soft-mode mixing.  The increase near
 $a=1$ reflects the $\Xi^{-1}$ ultraspinning singularity; the other rate
 remains exactly zero on the fold.}
 \label{tab:Kerr-fold-rates}
\end{table}

\paragraph{Complex continuation.}

For $(\rp,a)\in\C^2$, Eqs.~\eqref{eq:H11}--\eqref{eq:H12} still give the
complete Hessian by holomorphic continuation, but
Eq.~\eqref{eq:full-Kerr-eigenvalues} is no longer the Takagi spectrum.
For a nondegenerate complex saddle, substitution into
Eq.~\eqref{eq:Takagi-2by2} gives two strictly positive flow rates; at a
complex degeneracy one or both Takagi values can vanish.  For real
$(\beta,\Omega,G)$, a complex-conjugate saddle pair has
$H_{\bar\sigma}=\overline{H_\sigma}$ and hence identical Takagi values.
These Takagi data are local; whether either member contributes still requires
its global intersection number.

\subsubsection{Reduced Gaussian phase across the fold}

Takagi values determine the local growth and decay rates but not the phase of
the reduced Gaussian prefactor.  For real Kerr saddles, this phase follows
directly from the factorized Hessian determinant.
For real saddles in the physical domain,
Eq.~\eqref{eq:detH-factorized} gives
\begin{equation}
 \arg\det H=
 \begin{cases}
  0, & 2\rp^2-a^2-1>0,\\
  \pi, & 2\rp^2-a^2-1<0.
 \end{cases}
\end{equation}
Therefore
\begin{equation}
 \boxed{
 \phiM=
 \begin{cases}
  0, & 2\rp^2-a^2-1>0,\\
  \pi/2, & 2\rp^2-a^2-1<0,
 \end{cases}
 \pmod\pi.
 }
 \label{eq:Maslov-real}
\end{equation}
The phase is undefined at the fold itself, where the Gaussian determinant
vanishes.
The resulting sign and phase classification is summarized in
Table~\ref{tab:Maslov-Kerr}.

\begin{table}[htbp]
 \centering
 \small
 \renewcommand{\arraystretch}{1.2}
 \begin{tabular}{lccccc}
 \toprule
 Region
 & $F=2\rp^2-a^2-1$ & $\sgn(\det H)$
 & $\widehat\lambda_{\min}$ & $\arg\det H$ & $\phiM$\\
 \midrule
 Thermal saddle & --- & $+$ & $>0$ & $0$ & $0$\\
 Large/HP side & $F>0$ & $+$ & $>0$ & $0$ & $0$\\
 Fold/spinodal & $F=0$ & $0$ & $0$ & undefined & undefined\\
 Small/unstable side & $F<0$ & $-$ & $>0$ & $\pi$ & $\pi/2\pmod\pi$\\
 \bottomrule
 \end{tabular}
 \caption{Phase of the reduced Gaussian determinant for the real singly
 rotating Kerr family.  The fold is a local Hessian degeneracy; calling it
 a Fisher edge additionally requires contributing thimbles and saddle
 interference.  No functional determinant of the full gravitational
 fluctuation operator is included.}
 \label{tab:Maslov-Kerr}
\end{table}

\begin{figure}[t]
 \centering
 \includegraphics[width=\textwidth]{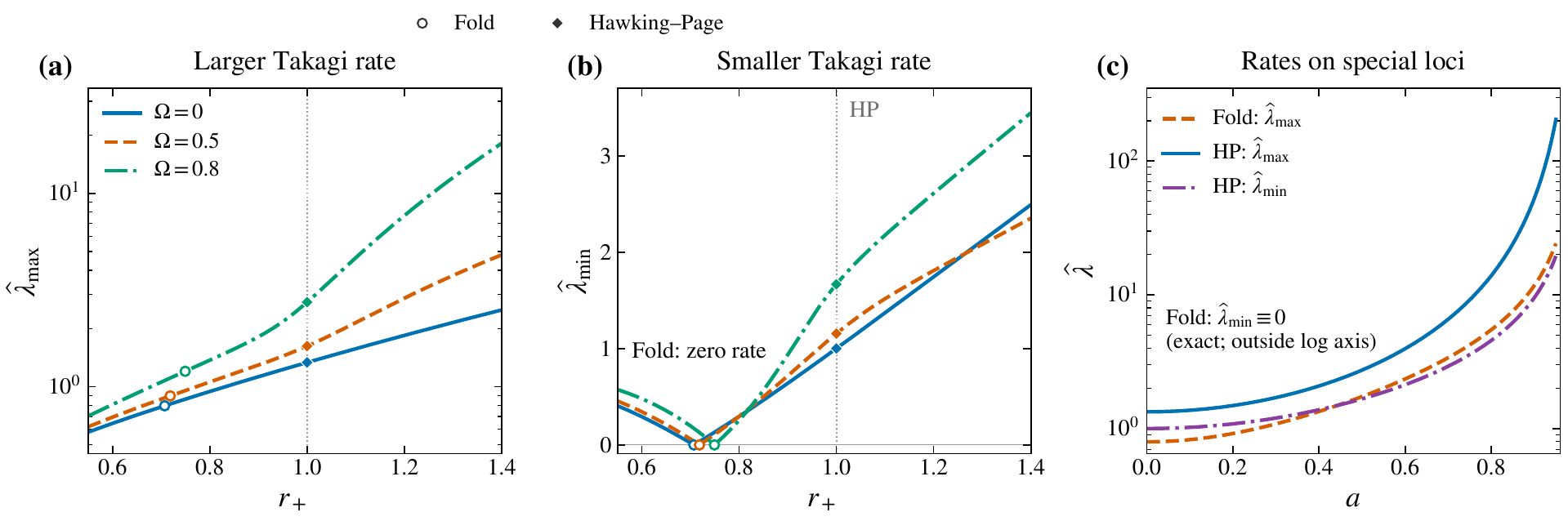}
 \caption{Dimensionless local Picard--Lefschetz rates of the reduced
 exponent, $\widehat\lambda=(2G/\pi^2)\lambda_{\mathrm{PL}}$.
 (a,b) Larger and smaller Takagi values along fixed-$\Omega$ slices
 $\Omega=0,0.5,0.8$ (solid blue, dashed orange, and dash-dotted green).
 Open circles mark the folds; diamonds and the dotted vertical guide
 mark Hawking--Page at $r_+=1$. One rate vanishes at the fold, whereas
 both are nonzero at Hawking--Page. Panel (b) has a linear vertical axis
 to display the exact zeros. (c) Exact rates along the rotating fold and
 Hawking--Page loci, compared at equal $a$ rather than equal $\Omega$.
 The fold rate $\widehat\lambda_{\min}\equiv0$ is stated separately.
 Panels (a,c) use logarithmic vertical axes. These local rates do not
 determine global thimble coefficients.}
 \label{fig:stability-exponents}
\end{figure}

Once an analytic-continuation and thimble-orientation convention is fixed,
the two real chambers differ by $\pi/2$ modulo $\pi$ in the square-root
phase of the reduced Gaussian determinant.  This phase difference neither
removes the unstable saddle nor fixes its contour coefficient.  Whether
that saddle contributes is determined by the global intersection number
$n_\sigma$ in Eq.~\eqref{eq:thimble-decomposition}.
Figure~\ref{fig:stability-exponents}(a,b) displays the two ordered rates
along fixed angular potential slices $\Omega=0,0.5,0.8$, with $a$ varying according to
Eq.~\eqref{eq:a-fixed-Omega}. These slices follow the physical ensemble:
each passes through the small/large fold and then through the Hawking--Page
point on the large branch. The open circles isolate the local signature
of the fold: $\widehat\lambda_{\min}=0$, while
$\widehat\lambda_{\max}$ remains finite. The smaller rate is shown on a
linear scale so that the exact zero is visible. Since Takagi values are
nonnegative magnitudes, this curve touches zero rather than changing sign;
it is the signed Hessian eigenvalue, or equivalently the fold factor $F$,
that distinguishes the two sides. Positivity of a Takagi rate by itself is
therefore not a thermodynamic stability criterion. Near a generic fold,
$\widehat\lambda_{\min}$ is linear in $|\rp-\rp^{\mathrm{sp}}|$ along
a fixed-$\Omega$ slice, whereas $\beta_{\mathrm{sp}}-\beta$ is quadratic
in that displacement. The same zero thus gives the square-root dependence
on the source displacement described in Eq.~\eqref{eq:soft-rate-scaling}.

The diamonds at $r_+=1$ emphasize the different mechanism at Hawking--Page:
both local rates remain nonzero when the black hole and thermal AdS actions
cross. Figure~\ref{fig:stability-exponents}(c) extends this comparison along
the full rotating special loci using the exact expressions
\eqref{eq:Kerr-HP-rates} and \eqref{eq:Kerr-fold-rates}. Equal values of $a$
on the two loci correspond to different angular potentials, so this
panel compares the loci themselves rather than a single fixed-$\Omega$
ensemble. The hard fold rate is finite at every generic point $a<1$;
its increase toward $a=1$ reflects the excluded ultraspinning singularity.
The identically zero fold rate is stated separately because it cannot be
plotted on a logarithmic axis. These observations identify the local
saddle merger without assigning its contour weight or implying a
zero of the partition function.

\FloatBarrier

\subsection{Fold uniformization, contour selection, and Fisher zeros}
\label{sec:airy}

\subsubsection{Local Airy normal form}

At a generic point on the fold, one Hessian direction is nondegenerate and the
other is soft.  After a smooth local change of variables and rotation onto the
steepest-descent directions, the exponent can be brought to the normal form
\begin{equation}
 \Phi_{\mathrm{loc}}(p,q)
 =\Phi_{\mathrm{reg}}(\delta\lambda)
 -\frac{C}{2G}p^2
 +\frac{1}{G}\left(A\,\delta\lambda\,q+\frac{B}{3}q^3\right)
 +\cdots,
 \qquad ABC\neq0,
 \label{eq:fold-normal-form}
\end{equation}
where $\delta\lambda$ is a transverse displacement in source space and
$\Phi_{\mathrm{reg}}(0)=\Phi_{\mathrm{sp}}$; on the fixed-$\Omega$ slices
used below one may take
$\delta\lambda=\beta-\beta_{\mathrm{sp}}(\Omega)$.  The
Gaussian integral over $p$ supplies a factor $G^{1/2}$, while the Airy
integral over $q$ supplies $G^{1/3}$.  The two-dimensional uniform
approximation is therefore
\begin{equation}
 \boxed{
 Z_{\mathrm{fold}}
 \sim
 \mathcal{C}\,e^{\Phi_{\mathrm{reg}}(\delta\lambda)}
 G^{5/6}\,
 \operatorname{Ai}\!\left(
 c\,\frac{\delta\lambda}{G^{2/3}}
 \right),
 }
 \label{eq:Airy}
\end{equation}
where $\mathcal{C}$ and $c$ are nonzero constants fixed by the local
derivatives, measure, and contour orientation.  Equation~\eqref{eq:Airy}
is the contribution of an elementary Airy thimble, with the cube-root
branch selecting the appropriate rotated $\operatorname{Ai}$ function.
The displayed
$G^{5/6}$ is the power supplied by the two reduced integrations, assuming
that the smooth reduced measure carries no additional power of $G$.  The
canonical contour gives the unrotated form shown above. More generally,
for bounded $\zeta=\delta\lambda/G^{2/3}$ a fixed local contour gives
\begin{align}
 Z_{\mathrm{fold},\Gamma}
 &=e^{\Phi_{\mathrm{reg}}(\delta\lambda)}G^{5/6}
 \left[\mathcal F_\Gamma(\zeta)+O(G^{1/3})\right],
 \label{eq:fold-general-uniform}\\
 \mathcal F_\Gamma(\zeta)
 &=d_0\operatorname{Ai}(c\zeta)
   +d_1\operatorname{Ai}(e^{2\pi i/3}c\zeta).
 \nonumber
\end{align}
The coefficients $d_0,d_1$ include the leading local amplitude and must be
fixed by the complete contour. The error is additive and uniform on compact
$\zeta$ sets, including zeros of $\mathcal F_\Gamma$; amplitude corrections
can involve derivatives of the Airy functions. Thus the isolated-saddle
Gaussian divergence is replaced by a finite local approximation. Cancellation
of one isolated Gaussian term outside the coalescence window does not fix
$d_0,d_1$ within it.

Because the soft Hessian eigenvalue varies as the square root of a generic
transverse control parameter,
\begin{equation}
 \lPL{\min}\sim G^{-1}|\delta\lambda|^{1/2},
 \label{eq:soft-rate-scaling}
\end{equation}
the naive one-loop factor behaves as
\begin{equation}
 |\det H|^{-1/2}\sim G|\delta\lambda|^{-1/4}.
 \label{eq:one-loop-scaling}
\end{equation}
This divergence is the signal that Eq.~\eqref{eq:Airy}, not the isolated
Gaussian saddle approximation, must be used.  Appendix~\ref{app:fold-Airy}
derives Eqs.~\eqref{eq:fold-normal-form}--\eqref{eq:one-loop-scaling},
including their coefficients and the Kerr soft and hard directions.
Table~\ref{tab:one-loop-summary} compares the corresponding Gaussian data
and records where the Airy replacement is required.

The cubic stationary-point separation scales as $|\delta\lambda|^{1/2}$,
while the exponent splitting scales as $|\delta\lambda|^{3/2}/G$.
Requiring the latter to remain of order unity gives the Airy source window
$\delta\lambda=O(G^{2/3})$. Section~\ref{sec:extended} recovers the same
cubic scaling at each isolated fixed-$J$ spinodal and tests the associated
three-halves Gibbs gap in Figure~\ref{fig:canonical-three-halves}.
When the two canonical folds merge, the cubic coefficient vanishes and a
quartic Pearcey approximation replaces the separate Airy descriptions.
Subsection~\ref{sec:Airy-Pearcey-matching} gives the explicit local matching
and connects its scaling variables to the thermodynamic paths. The common
normal form does not equate the grand canonical and fixed-$J$ contours.
\begin{table}[htbp]
 \centering
 \small
 \renewcommand{\arraystretch}{1.35}
 \begin{tabular}{@{}
 >{\raggedright\arraybackslash}p{0.16\textwidth}
 >{\raggedright\arraybackslash}p{0.43\textwidth}
 >{\raggedright\arraybackslash}p{0.31\textwidth}@{}}
 \toprule
 Saddle or locus & Takagi-rate set $\{\lambda_{\mathrm{PL},1},
 \lambda_{\mathrm{PL},2}\}$ & Gaussian factor
 $\mathcal A_G=|\det H|^{-1/2}$\\
 \midrule
 Thermal
 & $\left\{\dfrac{3\pi\beta}{4G},\dfrac{3\pi\beta}{4G}\right\}$
 & $\dfrac{4G}{3\pi\beta}$\\[5pt]
 Generic Kerr, $F\neq0$
 & $\dfrac{\pi^2}{2G}
 \left\{|\widehat e_+|,|\widehat e_-|\right\}$
 & $\dfrac{2G\rp\XiA^2s}
 {\pi^2uv\sqrt{(a^2+3)|F|}}$\\[7pt]
 Kerr Hawking--Page
 & $\dfrac{\pi^2}{2G}
 \left\{
 \dfrac{4(1+a^2)}{(1-a^2)^2(a^2+3)},
 \dfrac{1+a^2}{1-a^2}
 \right\}$
 & $\dfrac{G(1-a^2)^{3/2}\sqrt{a^2+3}}
 {\pi^2(1+a^2)}$\\[9pt]
 Rotating fold
 & $\dfrac{\pi^2}{2G}
 \left\{\widehat\lambda_{\max}^{\mathrm{sp}},0\right\}$
 & Divergent Gaussian; replace by
 Eq.~\eqref{eq:Airy}\\
 \bottomrule
 \end{tabular}
 \caption{Local reduced Gaussian data.  Here
 $F=2\rp^2-a^2-1$, the real Kerr eigenvalues $\widehat e_\pm$ are given in
 Eq.~\eqref{eq:full-Kerr-eigenvalues}, and
 $\widehat\lambda_{\max}^{\mathrm{sp}}$ is
 Eq.~\eqref{eq:Kerr-fold-rates}.  The table records local Gaussian data
 only; it does not determine global intersection numbers.}
 \label{tab:one-loop-summary}
\end{table}

\subsubsection{Symmetric contour prescription and Hawking--Page Fisher zeros}

The Airy approximation is a local statement about the coalescing small and
large black hole critical points.  It becomes a zero of the partition function
prediction only after the image of the original cycle in the local relative
homology fixes the corresponding Airy solution.  Ref.~\cite{Singhi:2025}
regulates the real Stokes surface by taking $\Im G\neq0$.  For the two signs
of the regulator, its contour decompositions are
\begin{equation}
 \begin{aligned}
  \Gamma^-&=\frac12\mathcal J_0^-
  +\mathcal J_-^-+\mathcal J_+^- &&(\Im G<0),\\
  \Gamma^+&=\frac12\mathcal J_0^+
  -\mathcal J_-^++\mathcal J_+^+ &&(\Im G>0).
 \end{aligned}
 \label{eq:Kerr-selected-cycle}
\end{equation}
Here $0,+,-$ denote thermal AdS, the large black hole, and the unstable
small black hole.  The two regulated integrals are conjugate.  The symmetric
limit is taken for the complete lateral sums, not for separately identified
terms on the Stokes surface.  With the orientations in
Eq.~\eqref{eq:Kerr-selected-cycle}, the coefficient-weighted unstable-saddle
Gaussian is odd under lateral conjugation; its real-$G$ determinant phase is
$\pi/2$.  The leading saddle expansion of the conjugation-symmetric limit is
therefore
\begin{equation}
 \boxed{
 Z_{\mathrm{phys}}^{\mathrm{(sym)}}
 \sim\frac12 Z_0+Z_+.
 }
 \label{eq:Kerr-symmetric-sum}
\end{equation}
Thus neither regulated contour has $n_-=0$; the unstable sector is absent
only after the two complete lateral expansions are combined.  Equation
\eqref{eq:Kerr-symmetric-sum} is a leading saddle statement, not an exact
termwise identity between thimbles at real $G$. Beyond the black hole fold,
the corresponding leading prescription contains only $\frac12 Z_0$.
Neither statement is a uniform approximation within
$|\beta-\beta_{\mathrm{sp}}|=O(G^{2/3})$, and neither proves cancellation
of the Airy combination in Eq.~\eqref{eq:fold-general-uniform}.
For the direct integral with a flat measure defined below, we establish absence
of zeros near the real fold by a separate bound on the complete original
cycle in Subsection~\ref{sec:fold-zero-free}.

\begin{figure}[htbp]
 \centering
 \IfFileExists{kerr_ads_fisher_zeros.pdf}{%
   \includegraphics[width=\textwidth]{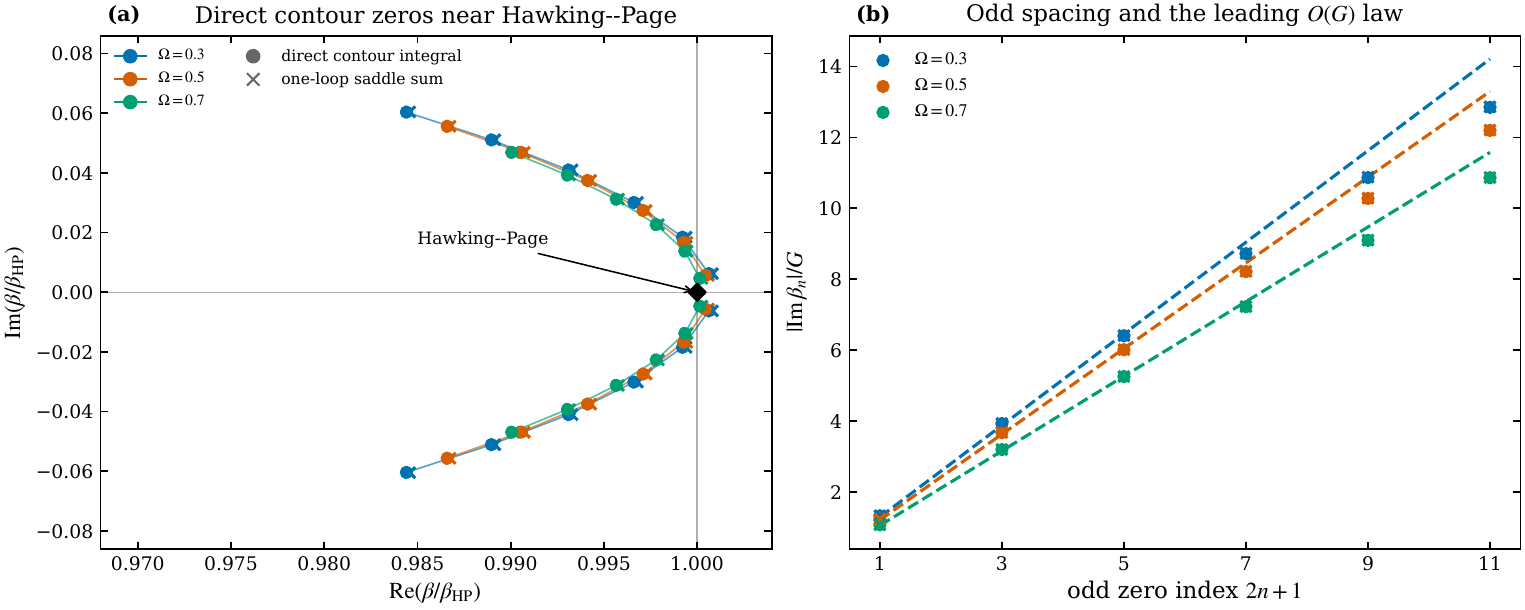}%
 }{%
   \fbox{\parbox[c][5.0cm][c]{0.93\textwidth}{\centering
   Missing figure \texttt{kerr\_ads\_fisher\_zeros.pdf}.}}%
 }
 \caption{Fisher zeros of the selected flat-measure Kerr--AdS$_5$
 mini-superspace contour at fixed real angular potential and $G=10^{-2}$.
 (a) Circles are direct zeros of the two-dimensional contour integral
 \eqref{eq:Kerr-Fisher-direct-integral}; crosses solve the independent
 reduced Gaussian saddle equation \eqref{eq:Kerr-Fisher-zero-condition}.  Their
 agreement tests the saddle calculation.  The conjugate sequences converge
 to the rotating Hawking--Page point, not the black hole fold.  (b) Circles
 and crosses show the corresponding imaginary parts; dashed lines are the
 leading odd-spacing law \eqref{eq:Kerr-Fisher-flat-measure}. The Gaussian
 comparison uses thermal AdS and the large Kerr saddle. Independently,
 Eq.~\eqref{eq:fold-zero-free-disk} excludes zeros near each real fold
 for sufficiently small $G$ at fixed $|\Omega|<1$; it does not assert
 cancellation of the local Airy contribution.}
 \label{fig:Kerr-Fisher-zeros}
\end{figure}

For the symmetric prescription in Eq.~\eqref{eq:Kerr-symmetric-sum}, the
contributing two-saddle competition occurs instead at the Hawking--Page
anti-Stokes line.  Analytically continuing
$\beta$ at fixed real $\Omega$, without crossing a new Stokes wall, gives
the reduced Gaussian sum
\begin{equation}
 Z_{\mathrm{phys}}^{(1)}
 \sim
 \frac12\mathcal A_0
 +\mathcal A_+e^{\Phi_+},
 \qquad
 \mathcal A_\sigma
 =
 \mu_\sigma\frac{2\pi}{\sqrt{\det(-H_\sigma)}}.
 \label{eq:Kerr-Fisher-sum}
\end{equation}
Here $\Phi_0=0$, and $+$ is the analytic continuation of the stable large
Kerr saddle.  These $\mathcal A_\sigma$ are reduced Gaussian factors, not
functional determinants of the full gravitational theory.  The quantity
$\mathcal A_0$ is the Gaussian on the full
radially extended $\mathcal J_0$ and contains no factor $1/2$; the sole
thermal half-weight is displayed explicitly.  The Fisher zeros
\cite{YangLee:1952,LeeYang:1952,Fisher:1965} therefore obey
\begin{equation}
 \boxed{
 \Phi_+(\beta_n,\Omega)
 +\log\!\left(\frac{2\mathcal A_+(\beta_n,\Omega)}
 {\mathcal A_0(\beta_n,\Omega)}\right)
 =i(2n+1)\pi,
 \qquad n\in\mathbb Z.
 }
 \label{eq:Kerr-Fisher-zero-condition}
\end{equation}
This single complex equation imposes both equal magnitude and opposite phase.
For a flat measure, the original real cycle can also be integrated directly:
\begin{equation}
 \boxed{
 Z_{\Gamma_{\mathbb R}}^{\mathrm{flat}}(\beta,\Omega)
 =
 \int_0^\infty\dd\rp
 \int_{-1}^{1}\dd a\;
 e^{\Phi(\rp,a;\beta,\Omega)}.
 }
 \label{eq:Kerr-Fisher-direct-integral}
\end{equation}
For fixed real $|\Omega|<1$, the integral converges when
$\Re\beta>0$.  We locate its complex zeros by
two-dimensional Gauss--Legendre quadrature and a complex root solve, and use
Eq.~\eqref{eq:Kerr-Fisher-zero-condition} as the independently computed
reduced Gaussian comparison.  For Fig.~\ref{fig:Kerr-Fisher-zeros} we use a
$300\times300$ product rule on
$(\rp,a)\in[0,4]\times[-1,1]$.  Repeating the calculation with a
$220\times220$ rule on $[0,3.5]\times[-1,1]$ changes the first displayed zero
by less than $3\times10^{-14}$ and the farthest displayed zero by less than
$6\times10^{-6}$.

The thermal endpoint of the direct half-domain integral is not, beyond
quadratic order, one half of the expansion on the radially extended
$\Jth_0$.  For the flat measure,
\begin{equation}
 Z_{0,\Gamma_{\mathbb R}}^{\mathrm{flat}}
 =\frac{4G}{3\beta}
 \left[1+\frac{4\sqrt{2/3}}{\beta^{3/2}}\sqrt G+O(G)\right].
 \label{eq:thermal-boundary-expansion}
\end{equation}
The leading term is $\frac12\mathcal A_0$; the $O(\sqrt G)$ correction comes
from the radial cubic term in $\Phi$.

\subsubsection{A uniform zero-free neighborhood of the real fold}
\label{sec:fold-zero-free}

The integral with a flat measure \eqref{eq:Kerr-Fisher-direct-integral} admits a
uniform estimate through the fold without first resolving the individual
coalescing thimbles. Fix real $|\Omega|<1$ and take $G\to0^+$.
Introduce the $G$-independent entropy and grand canonical energy
\begin{equation}
 \mathcal S=GS=\frac{\pi^2\rp u}{2\XiA},
 \qquad
 \mathcal E=G(E-\Omega J)=\frac{\pi u v w}{8\XiA^2},
 \qquad G\Phi=\mathcal S-\beta\mathcal E.
 \label{eq:fold-global-scaled-action}
\end{equation}
Here $u,v,w$ are defined in Eq.~\eqref{eq:shorthand}. Let
$d=\sqrt{1-\Omega^2}>0$ and $a_{\mathrm{HP}}=\Omega/(1+d)$.
The exact identity
\begin{equation}
 w=(2+d)\XiA+(1+d)(a-a_{\mathrm{HP}})^2
 \label{eq:fold-global-square}
\end{equation}
implies, throughout the original real cycle and away from its origin,
\begin{equation}
 \frac{\mathcal S}{\mathcal E}
 =\frac{4\pi\rp\XiA}{(1+\rp^2)w}
 \leq\frac{2\pi}{2+d}=\beta_{\mathrm{HP}}(\Omega),
 \qquad
 \mathcal E\geq\frac{\pi(2+d)}{8}(\rp^2+a^2).
 \label{eq:fold-global-bound}
\end{equation}
We used $2\rp\leq1+\rp^2$, $v\geq1$, and $0<\XiA\leq1$.
The first inequality extends continuously to the origin in the form
$\mathcal S\leq\beta_{\mathrm{HP}}\mathcal E$. Equality away from the
origin occurs only at the Hawking--Page saddle.

Let $K$ be any compact subset of
$\{\beta\in\C:\Re\beta>\beta_{\mathrm{HP}}(\Omega)\}$ and put
$\delta_K=\min_{\beta\in K}(\Re\beta-\beta_{\mathrm{HP}})>0$.
Equation~\eqref{eq:fold-global-bound} yields the global estimate
\begin{equation}
 \left|e^{\Phi}\right|
 \leq \exp\!\left[-\frac{\delta_K\mathcal E}{G}\right]
 \leq \exp\!\left[-\frac{\pi(2+d)\delta_K}{8G}
 (\rp^2+a^2)\right],\qquad \beta\in K.
 \label{eq:fold-uniform-majorant}
\end{equation}
Thus every region a fixed distance from the thermal endpoint is
exponentially suppressed, uniformly for complex $\beta\in K$. This bound
also controls the noncompact radial tail and the limits $a\to\pm1$.

Rescaling $(\rp,a)=\sqrt G(x,y)$ gives
\begin{equation}
 \Phi=-\frac{3\pi\beta}{8}(x^2+y^2)
 +\sqrt G\left[\frac{\pi^2}{2}x(x^2+y^2)
 +\frac{\pi\beta\Omega}{4}y(x^2+y^2)\right]
 +O\!\left(G(x^2+y^2)^2\right)
 \label{eq:fold-endpoint-rescaling}
\end{equation}
on fixed scaled neighborhoods. The Gaussian bound
\eqref{eq:fold-uniform-majorant} permits uniform dominated convergence
on the expanding half-plane. Expanding in a fixed small neighborhood of
the original endpoint, and bounding its complement exponentially, also
makes the endpoint expansion \eqref{eq:thermal-boundary-expansion}
uniform.\footnote{For an explicit remainder bound, put $R^2=x^2+y^2$,
$\Psi_2=-3\pi\beta R^2/8$ and denote the cubic bracket in
Eq.~\eqref{eq:fold-endpoint-rescaling} by $P_3$. In a fixed small
original-coordinate neighborhood, $D=\Phi-\Psi_2
=\sqrt G P_3+O_K(GR^4)$ and $|D|\leq C_K\sqrt G R^3$.
Both $\Re\Phi$ and $\Re\Psi_2$ are bounded above by $-c_KR^2$.
Using $e^D-1-D=D^2\int_0^1(1-t)e^{tD}\dd t$ therefore gives
$|e^\Phi-e^{\Psi_2}-\sqrt G P_3e^{\Psi_2}|
\leq C_KG(R^4+R^6)e^{-c_KR^2}$.
Its integral is $O_K(G)$ in scaled coordinates; the exterior is
exponentially suppressed by Eq.~\eqref{eq:fold-uniform-majorant}.
The positive constants may depend on $K$ and fixed $\Omega$, but not on $G$.}
We obtain
\begin{equation}
 \boxed{
 Z_{\Gamma_{\mathbb R}}^{\mathrm{flat}}(\beta,\Omega)
 =\frac{4G}{3\beta}
 \left[1+\frac{4\sqrt{2/3}}{\beta^{3/2}}\sqrt G+O_K(G)\right].
 }
 \label{eq:fold-uniform-endpoint}
\end{equation}
The branch of $\beta^{3/2}$ is analytic in the right half-plane and positive
on the positive real axis. The $y$-odd cubic integrates to zero; the radial
cubic supplies the displayed correction. In particular,
$3\beta Z_{\Gamma_{\mathbb R}}^{\mathrm{flat}}/(4G)\to1$ uniformly on $K$.
Its distance from $1$ is therefore less than $1/2$ for sufficiently small
$G$, proving that the full integral has no zeros on $K$.

Since $\beta_{\mathrm{sp}}(\Omega)>\beta_{\mathrm{HP}}(\Omega)$, choose
any fixed $0<\rho<\beta_{\mathrm{sp}}-\beta_{\mathrm{HP}}$. The preceding
argument proves
\begin{equation}
 \boxed{
 Z_{\Gamma_{\mathbb R}}^{\mathrm{flat}}(\beta,\Omega)\neq0
 \quad\text{for}\quad |\beta-\beta_{\mathrm{sp}}|\leq\rho,
 \qquad 0<G<G_0(\Omega,\rho).
 }
 \label{eq:fold-zero-free-disk}
\end{equation}
Every bounded Airy scaling window
$\beta=\beta_{\mathrm{sp}}+G^{2/3}\zeta$ lies in this disk when $G$ is
sufficiently small. Hence no sequence of zeros of this full reduced
integral can approach the real fold on that scale.

This conclusion concerns thermal dominance, not disappearance of the
local Airy sector. Indeed, writing
$\Delta_\Omega=G I_E^{\mathrm{sp}}
=\pi\beta_{\mathrm{sp}}(1+3a_{\mathrm{sp}}^2)/32>0$, analyticity of the
regular action in Eq.~\eqref{eq:fold-general-uniform} gives, for bounded
$\zeta$ and finite local contour coefficients,
\begin{equation}
 \frac{|Z_{\mathrm{fold},\Gamma}|}{|4G/(3\beta)|}
 =O\!\left(G^{-1/6}
 e^{-\Delta_\Omega/G+O(G^{-1/3})}\right)\longrightarrow0.
 \label{eq:fold-uniform-suppression}
\end{equation}
Zeros of the local Airy combination therefore need not be zeros of the
complete partition function. The proof of
Eq.~\eqref{eq:fold-zero-free-disk} uses the original flat-measure cycle
itself; it neither assigns the Airy coefficients of the lateral sums nor
identifies the direct endpoint integral with half an extended thermal
thimble beyond Gaussian order. Its conclusion is asymptotic at fixed
$|\Omega|<1$. It does not establish zero exclusion for arbitrary finite
$G$, for other integration cycles, or in a simultaneous
$G\to0$, $|\Omega|\to1$ limit, where the separation from Hawking--Page
shrinks.

\subsubsection{Hawking--Page asymptotics and numerical zeros}

At the transition, write
\begin{equation}
 a_{\mathrm{HP}}
 =\frac{\Omega}{1+\sqrt{1-\Omega^2}},
 \qquad
 \beta_{\mathrm{HP}}
 =\frac{2\pi(1+a_{\mathrm{HP}}^2)}
 {3+a_{\mathrm{HP}}^2}.
 \label{eq:Kerr-HP-source}
\end{equation}
The envelope theorem and the exact thermodynamic charges give
\begin{equation}
 \left.
 -\frac{\dd\Phi_+}{\dd\beta}
 \right|_{\mathrm{HP},\Omega}
 =
 (E-\Omega J)_{\mathrm{HP}}
 \equiv L_\Omega
 =
 \frac{\pi(3+a_{\mathrm{HP}}^2)}
 {4G(1-a_{\mathrm{HP}}^2)}.
 \label{eq:Kerr-HP-latent-charge}
\end{equation}
Consequently, if
$R_\Omega=\mathcal A_0/(2\mathcal A_+)|_{\mathrm{HP}}$, the zeros nearest
the real axis are
\begin{equation}
 \boxed{
 \beta_n
 =
 \beta_{\mathrm{HP}}
 -\frac{\log R_\Omega+i(2n+1)\pi}{L_\Omega}
 +O(G^{3/2}).
 }
 \label{eq:Kerr-Fisher-zero-asymptotic}
\end{equation}
For the flat mini-superspace measure $\mu_0=\mu_+=1$, the Hessians derived
above give
\begin{equation}
 R_\Omega
 =
 \frac{\sqrt{3+a_{\mathrm{HP}}^2}}
 {3(1-a_{\mathrm{HP}}^2)^{3/2}},
 \qquad
 \Im\beta_n
 =
 -\frac{4G(1-a_{\mathrm{HP}}^2)}
 {3+a_{\mathrm{HP}}^2}(2n+1)+O(G^2).
 \label{eq:Kerr-Fisher-flat-measure}
\end{equation}
Thus, for this flat-measure reduced integral, the zeros form conjugate pairs,
approach the Hawking--Page point
linearly in $G$, and have odd spacing controlled by the rotating latent
charge.  This is the first-order Fisher zero scaling; it is different from
the $G^{2/3}$ scaling of a contour-supported Airy edge.
The first direct zero for each angular potential is listed in
Table~\ref{tab:Kerr-Fisher-first-zero}.

\begin{table}[htbp]
 \centering
 \small
 \renewcommand{\arraystretch}{1.15}
 \begin{tabular}{ccc}
 \toprule
 $\Omega$ & $\beta_{\mathrm{HP}}$
 & first lower-half-plane zero $\beta_0$\\
 \midrule
 $0.3$ & $2.127053$ & $2.128395-0.013251\,i$\\
 $0.5$ & $2.192299$ & $2.193282-0.012366\,i$\\
 $0.7$ & $2.314980$ & $2.315311-0.010720\,i$\\
 \bottomrule
 \end{tabular}
 \caption{First Fisher zero of the selected flat-measure reduced contour,
 obtained by direct quadrature of
 Eq.~\eqref{eq:Kerr-Fisher-direct-integral} with $G=10^{-2}$.  The conjugate
 zero lies in the upper half-plane.  A full gravitational fluctuation
 determinant is not included; it can change the $O(G)$ prefactor shift
 through $R_\Omega$, but not the classical action difference or the selected
 saddle pair.}
 \label{tab:Kerr-Fisher-first-zero}
\end{table}

\FloatBarrier
\subsection{Complex saddles and Stokes geometry}
\label{sec:complex}

We next separate the phase geometry of the analytically continued saddles
from their local Hessian data.  The Schwarzschild slice provides an analytic
model for the Stokes network, which then extends to fixed-$\Omega$ rotating
slices.
For source values beyond the fold, the two real black hole saddles continue as
a complex-conjugate pair.  The Hessian formulas
Eqs.~\eqref{eq:H11}--\eqref{eq:H12} remain valid by holomorphic continuation.
Their local flow rates must be computed from the Takagi values.

\subsubsection{Schwarzschild continuation and Stokes network}
\label{sec:Schwarzschild-Stokes}

In the Schwarzschild sector, the saddle equation gives, on the positive real
$\beta$ axis,
\begin{equation}
 \rp^{\star}
 =\frac{\pi\pm i\sqrt{2\beta^2-\pi^2}}{2\beta},
 \qquad
 \beta>\frac{\pi}{\sqrt{2}}.
 \label{eq:complex-Schwarzschild}
\end{equation}
The two saddles have conjugate actions and conjugate Hessians.  Their local
one-loop data are well defined, but this does not establish that they
contribute to the original gravitational contour.

The Stokes geometry becomes explicit after complexifying the inverse
temperature.  Define
\begin{equation}
 \widehat{\beta}=\frac{\sqrt{2}\beta}{\pi},
 \qquad
 r_\pm
 =\frac{1\pm\sqrt{1-\widehat{\beta}^{\,2}}}
 {\sqrt{2}\widehat{\beta}}.
\end{equation}
Using the on-shell Schwarzschild exponent
$\Phi(r)=\pi\beta r^2(r^2-1)/(8G)$, the difference between the two
black hole saddles is
\begin{equation}
 \boxed{
 \Delta\widehat{\Phi}
 \equiv
 \frac{4G}{\sqrt{2}\pi^2}\left(\Phi_+-\Phi_-\right)
 =\frac{\left(1-\widehat{\beta}^{\,2}\right)^{3/2}}
 {\widehat{\beta}^{\,3}}.
 }
 \label{eq:Schwarzschild-action-gap}
\end{equation}
The sign of the square root only exchanges the labels $+$ and $-$ and
therefore does not change either phase locus.  A branch-independent way to
plot them is to introduce
\begin{equation}
 Q(\widehat{\beta})
 =\left(\Delta\widehat{\Phi}\right)^2
 =\frac{\left(1-\widehat{\beta}^{\,2}\right)^3}
 {\widehat{\beta}^{\,6}}.
 \label{eq:Schwarzschild-Q}
\end{equation}
The Stokes phase-alignment curves obey
$\Im Q=0$ with $\Re Q>0$; they are candidate Stokes walls until a connecting
upward flow is verified.  The anti-Stokes curves obey
$\Im Q=0$ with $\Re Q<0$.  On the physical real axis,
$0<\widehat{\beta}<1$ is Stokes phase-aligned and contains two real
black hole saddles, whereas $\widehat{\beta}>1$ is anti-Stokes and contains
the complex-conjugate pair.  The two networks meet at the fold
$\widehat{\beta}=1$.

\begin{figure}[htbp]
 \centering
 \includegraphics[width=\textwidth]{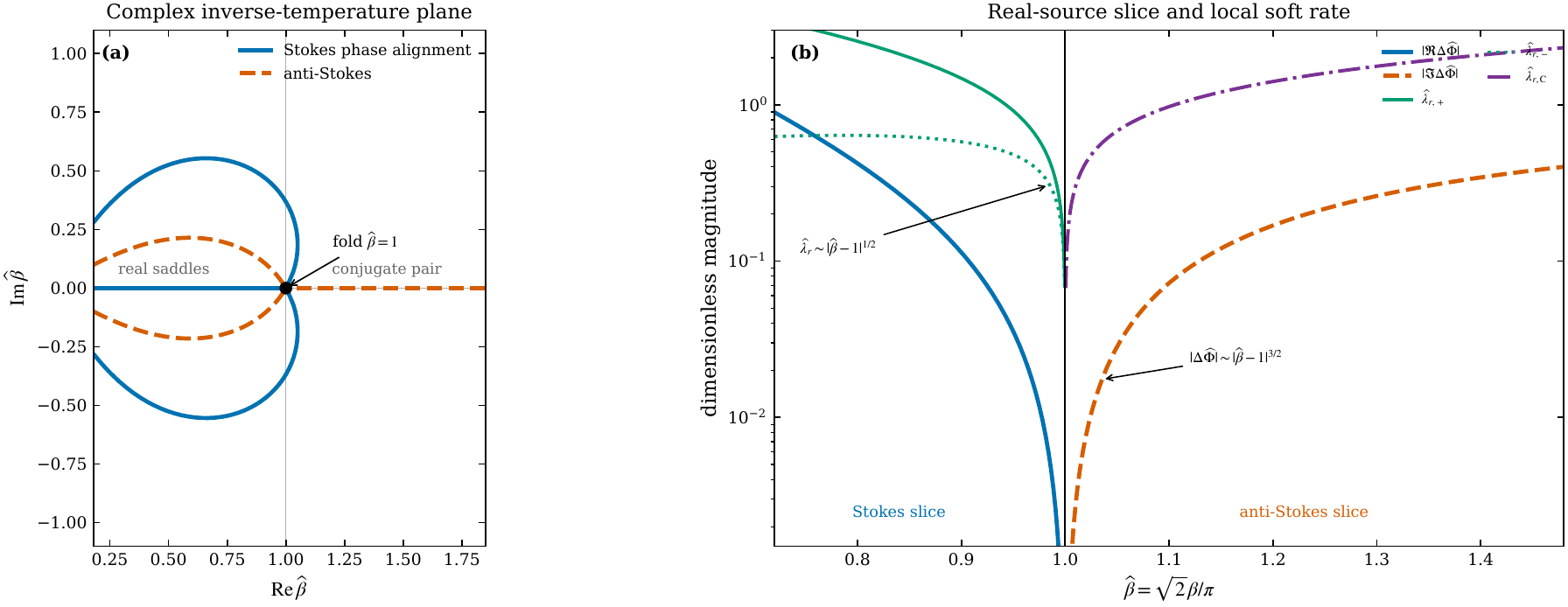}
 \caption{Stokes and anti-Stokes structure of the two Schwarzschild
 black hole saddles in the complex normalized inverse-temperature plane.
 (a) Solid blue curves satisfy the Stokes phase condition
 $\Im\Delta\widehat{\Phi}=0$; a connecting upward flow is still required for
 an actual thimble jump.  Dashed orange curves satisfy the anti-Stokes
 condition $\Re\Delta\widehat{\Phi}=0$.  All six local rays meet at the
 black hole fold $\widehat{\beta}=1$.  (b) Along the physical real-source
 slice, the action gap is real below the fold and imaginary above it.  Its
 magnitude vanishes with exponent $3/2$, whereas the radial Takagi rate
 $\widehat{\lambda}_r$ vanishes with exponent $1/2$.  The two real branches
 are shown separately below the fold; the conjugate saddles have a common
 rate above it.}
 \label{fig:Schwarzschild-Stokes}
\end{figure}

This also makes the relation to the local gradient-flow exponent precise.
Near the fold,
\begin{equation}
 \abs{\Delta\widehat{\Phi}}
 \sim 2^{3/2}\abs{\widehat{\beta}-1}^{3/2},
 \qquad
 \widehat{\lambda}_{r}
 \equiv\frac{2G}{\pi^2}\abs{H_{11}}
 \sim3\abs{\widehat{\beta}-1}^{1/2}.
 \label{eq:Stokes-fold-scalings}
\end{equation}
Thus the Stokes/anti-Stokes network is global action-difference data, while
the vanishing Takagi rate is local Hessian data; their common endpoint and
different critical powers are the characteristic Airy-fold structure.
Figure~\ref{fig:Schwarzschild-Stokes} displays both networks and their
distinct near-fold scalings.

On the physical real-source slice, the flow analysis of
Ref.~\cite{Singhi:2025} determines the relevant contour weights for the
chosen mini-superspace cycle.  Our results provide the local stability,
determinant phase, and uniform fold behavior.  A claim of dominance by a
complex saddle requires both ingredients:
\begin{equation}
 \boxed{
 \text{local saddle data}
 \quad+\quad
 c_\sigma\neq0.
 }
 \label{eq:local-global}
\end{equation}
The real part of a complex on-shell action, considered in isolation, is not a
sufficient criterion.

\subsubsection{Rotating Kerr--AdS\texorpdfstring{$_5$}{5} Stokes slices}
\label{sec:Kerr-Stokes}

The same pairwise construction extends away from the Schwarzschild axis.
Fix a real $0<\Omega<1$ and analytically continue
$\beta$.  Eliminating $a$ from the two saddle equations gives the cubic
\begin{equation}
 \boxed{
 2\beta^2\rp^3-6\pi\beta\rp^2
 + \left[\beta^2(1+\Omega^2)+4\pi^2\right]\rp
 -2\pi\beta=0.
 }
 \label{eq:Kerr-fixed-Omega-cubic}
\end{equation}
For every root, the corresponding rotation parameter is
\begin{equation}
 a(\rp;\beta,\Omega)
 =
 \frac{\beta\rp(1+\Omega^2+2\rp^2)-2\pi\rp^2}
 {2\pi\Omega}.
 \label{eq:Kerr-a-from-root}
\end{equation}
Two roots are the analytic continuations of the physical small and large Kerr
black holes and meet at $\beta=\beta_{\mathrm{sp}}(\Omega)$.  On the real
physical slice, the remaining algebraic root lies outside the
$|a|<1$ Kerr family.  After $\beta$ is complexified it remains a legitimate
algebraic saddle and can enter other pairwise degeneracies.  The diagnostic
below deliberately follows only the analytic continuation of the physical
small/large pair.

\begin{figure}[htbp]
 \centering
 \IfFileExists{kerr_ads_stokes_network.pdf}{%
   \includegraphics[width=\textwidth,height=5cm,keepaspectratio]{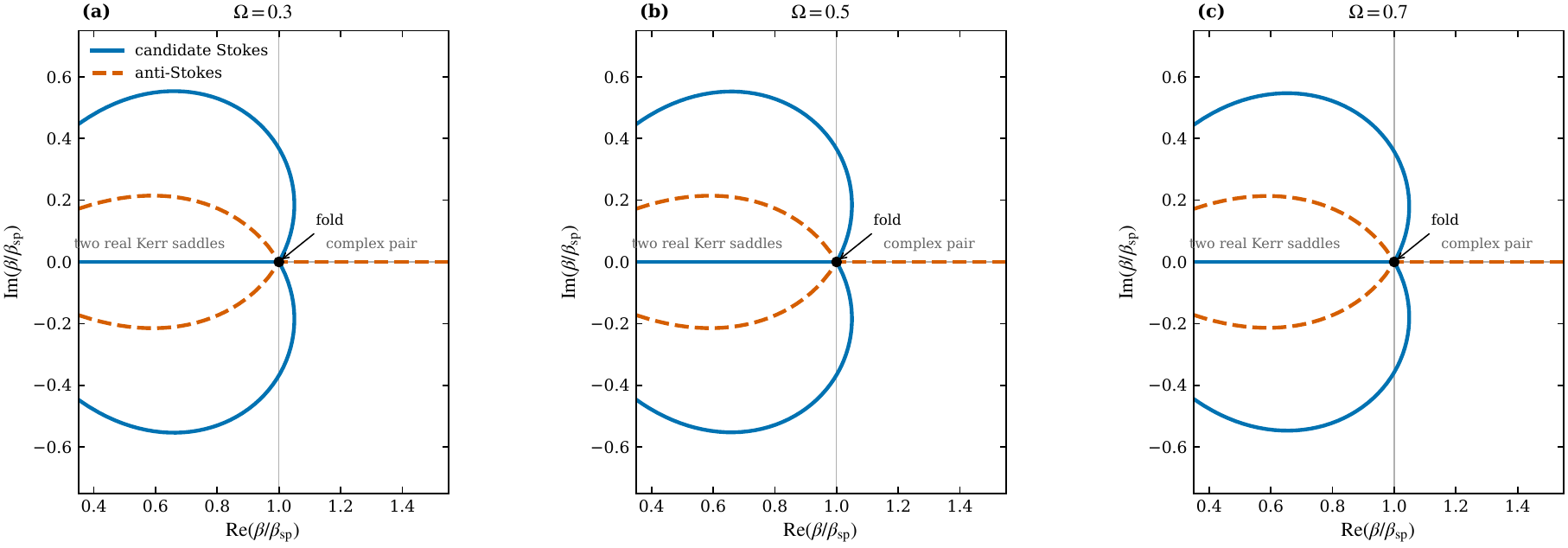}%
 }{%
   \fbox{\parbox[c][5.0cm][c]{0.93\textwidth}{\centering
   Missing figure \texttt{kerr\_ads\_stokes\_network.pdf}.}}%
 }
 \caption{Rotating Kerr--AdS$_5$ phase networks in the complex inverse-
 temperature plane at fixed real angular potential.  Solid blue curves obey
 the phase-alignment condition for the analytically continued small/large
 Kerr pair and are candidate Stokes curves; dashed orange curves are
 pairwise anti-Stokes equal-magnitude loci.  The three panels show
 $\Omega=0.3,0.5,0.7$.  In each panel
 $\widehat\beta_\Omega=\beta/\beta_{\mathrm{sp}}(\Omega)$, so the black point
 at $(1,0)$ is the rotating fold.  These plots do not include the third
 algebraic saddle and are therefore not a complete Stokes diagram.  Actual
 thimble jumps require connecting flows and intersection numbers; the
 real-$\beta$ contour data are taken from Ref.~\cite{Singhi:2025}.}
 \label{fig:Kerr-Stokes}
\end{figure}

Let $\Phi_\pm(\beta,\Omega)$ denote the on-shell exponents of the physical
pair and define the label-independent diagnostic
\begin{equation}
 Q_\Omega(\beta)
 =
 \left[\Phi_+(\beta,\Omega)-\Phi_-(\beta,\Omega)\right]^2.
 \label{eq:Kerr-Q}
\end{equation}
Then $\Im Q_\Omega=0$ with $\Re Q_\Omega>0$ gives phase-aligned candidate
Stokes curves, while $\Im Q_\Omega=0$ with $\Re Q_\Omega<0$ gives pairwise
anti-Stokes loci.  Figure~\ref{fig:Kerr-Stokes} shows three fixed-$\Omega$
slices of this pairwise diagnostic,
normalized so the rotating fold is always at
$\beta/\beta_{\mathrm{sp}}=1$.  A candidate Stokes wall becomes an actual
thimble jump only when it supports the connecting flow of the global
mini-superspace contour.  Ref.~\cite{Singhi:2025} fixes the relevant
decomposition on the physical real-$\beta$ Stokes slice, but does not by
itself determine the intersection numbers on every curve in the displayed
complex-$\beta$ plane.

\FloatBarrier
\subsection{KSW allowability of Kerr--AdS\texorpdfstring{$_5$}{5} metrics}
\label{sec:KSW}

The thermodynamic Hessian and the spacetime metric answer different
questions.  The Hessian in Subsection~\ref{sec:hessian} and the fold analysis
in Subsection~\ref{sec:airy} govern local flow in the two-dimensional space
of collective variables.
The Kontsevich--Segal condition instead selects a domain of complex
spacetime metrics on which the Gaussian actions of real differential forms
have positive real part~\cite{KontsevichSegal}.  Witten proposed using this
allowable domain as a restriction on complex saddles in the gravitational
path integral~\cite{WittenComplexMetrics}.  The diagonal characterization
below is a mathematical statement about allowable metrics; its use as a
complete selection rule for quantum gravity remains a proposal.

\paragraph{Pointwise definition.}
For a real $p$-form $F$, define the pointwise quadratic density
\begin{equation}
 \mathcal Q_p[F]
 =
 \frac{\sqrt{\det g}}{p!}\,
 g^{i_1j_1}\cdots g^{i_pj_p}
 F_{i_1\cdots i_p}F_{j_1\cdots j_p}.
 \label{eq:KSW-pform-density}
\end{equation}
The square-root branch is the one obtained continuously from the Euclidean
component.  A complex metric is allowable when
\begin{equation}
 \Re\mathcal Q_p[F]>0
 \quad\text{for every nonzero real $F$ and every $p=0,\ldots,5$.}
 \label{eq:KSW-pform-positivity}
\end{equation}
At a spacetime point this is equivalent to the existence of a
\emph{real} cotangent-space basis in which
\begin{equation}
 g=\sum_{A=1}^{5}\lambda_A(e^A)^2,
 \qquad
 \lambda_A\in\C\setminus\{0\}.
 \label{eq:KSW-real-congruence}
\end{equation}
If $I\subset\{1,\ldots,5\}$ labels one component of a real form, its
coefficient in Eq.~\eqref{eq:KSW-pform-density} is
\begin{equation}
 q_I
 =
 \left(\prod_{A=1}^{5}\lambda_A\right)^{1/2}
 \prod_{A\in I}\lambda_A^{-1}.
 \label{eq:KSW-component-coefficient}
\end{equation}
Writing $\theta_A=\arg\lambda_A\in(-\pi,\pi)$, positivity of
$\Re q_I$ for every subset $I$ is equivalent to
\begin{equation}
 \boxed{
 \Theta_{\mathrm{KSW}}
 \equiv\sum_{A=1}^{5}\abs{\theta_A}<\pi .
 }
 \label{eq:KSW-phase}
\end{equation}
This is a real-congruence statement, not an ordinary complex-eigenvalue
test on the coordinate matrix $g_{ij}$.

\paragraph{Quasi-Euclidean rotating block.}
Away from the rotational axis and the horizon cap, the only non-real block
of a one-spin quasi-Euclidean metric can be written
\begin{equation}
 g_{(2)}
 =
 \begin{pmatrix}
  \mathsf A&i\mathsf C\\
  i\mathsf C&\mathsf B
 \end{pmatrix},
 \qquad
 \mathsf A,\mathsf B,\mathsf C\in\R .
 \label{eq:KSW-block}
\end{equation}
Its determinant is
$D=\mathsf A\mathsf B+\mathsf C^2$.  Since $D$ is real, the $p=0$
condition first requires $D>0$: $D=0$ is degenerate, while $D<0$ makes
$\sqrt D$ purely imaginary.  With $D>0$, the one-form
quadratic form appearing in Eq.~\eqref{eq:KSW-pform-density} obeys
\begin{equation}
 \Re\!\left(\sqrt D\,g_{(2)}^{-1}\right)
 =
 \frac{1}{\sqrt D}
 \begin{pmatrix}
  \mathsf B&0\\
  0&\mathsf A
 \end{pmatrix}.
 \label{eq:KSW-block-necessity}
\end{equation}
Thus allowability requires $\mathsf A>0$ and $\mathsf B>0$.  Conversely,
when both inequalities hold, use the real rescaling
$x=\sqrt{\mathsf A}\,\dd\tau$,
$y=\sqrt{\mathsf B}\,\dd\widetilde\varphi$ and the real rotation
$e^\pm=(x\pm y)/\sqrt2$.  Then
\begin{equation}
 \dd s^2_{(2)}
 =(1+i\chi)(e^+)^2+(1-i\chi)(e^-)^2,
 \qquad
 \chi=\frac{\mathsf C}{\sqrt{\mathsf A\mathsf B}}.
 \label{eq:KSW-block-diagonal}
\end{equation}
The remaining three directions have positive real coefficients and hence
zero phase.  The complete phase budget is therefore
\begin{equation}
 \Theta_{\mathrm{KSW}}
 =2\arctan|\chi|<\pi .
 \label{eq:KSW-block-phase}
\end{equation}
Equations~\eqref{eq:KSW-block-necessity}--\eqref{eq:KSW-block-phase}
prove, rather than assume, that the block is allowable if and only if
$\mathsf A>0$ and $\mathsf B>0$.

To relate these coefficients to Lorentzian geometry, write the relevant
stationary block as
\begin{equation}
 \dd s_L^2
 =-N^2\dd t^2
 {}+\mathsf B(\dd\varphi-\omega\dd t)^2
 {}+\dd s_\perp^2,
 \qquad \mathsf B>0.
 \label{eq:KSW-Lorentzian-ADM}
\end{equation}
Let $V=\partial_t+\Omega\partial_\varphi$ be the horizon generator and set
$\widetilde\varphi=\varphi-\Omega t$.  After $t=-i\tau$,
\begin{equation}
 \dd s_{\mathrm{QE}}^2
 =N^2\dd\tau^2
 {}+\mathsf B\left[
 \dd\widetilde\varphi-i(\Omega-\omega)\dd\tau
 \right]^2+\dd s_\perp^2 .
 \label{eq:KSW-quasi-Euclidean}
\end{equation}
Comparison with Eq.~\eqref{eq:KSW-block} gives
\begin{equation}
 \mathsf A=N^2-\mathsf B(\Omega-\omega)^2=-g_L(V,V),
 \qquad
 \mathsf C=-\mathsf B(\Omega-\omega),
 \qquad
 D=N^2\mathsf B .
 \label{eq:KSW-ADM-coefficients}
\end{equation}
Consequently,
\begin{equation}
 \boxed{
 \begin{gathered}
 \text{pointwise quasi-Euclidean KSW allowability on the open exterior}
 \\
 \Longleftrightarrow\quad
 g_L(V,V)<0
 \quad\text{at every finite exterior point}
 \end{gathered}
 }
 \label{eq:KSW-timelike-equivalence}
\end{equation}
More explicitly, the local phase consumption away from the cap is
\begin{equation}
 \Theta_{\mathrm{KSW}}(x)
 =
 2\arctan\!\left[
 \frac{\sqrt{\mathsf B}\,|\Omega-\omega|}
 {\sqrt{N^2-\mathsf B(\Omega-\omega)^2}}
 \right].
 \label{eq:KSW-local-budget}
\end{equation}
The rotation axis and horizon are coordinate degeneracies of
Eqs.~\eqref{eq:KSW-block} and \eqref{eq:KSW-Lorentzian-ADM}, not
degeneracies of the smooth spacetime metric; they are treated in a regular
Cartesian frame and in the regular Euclidean cigar coordinates,
respectively.

\paragraph{Kerr--AdS$_5$ allowability.}
The azimuth $\phi$ in Eq.~\eqref{eq:metric} is the Boyer--Lindquist azimuth.
The nonrotating boundary azimuth is
$\varphi_\infty=\phi+at$.  Hence, in the coordinates of
Eq.~\eqref{eq:metric}, the physical angular potential corresponds to
\begin{equation}
 V
 =\partial_t+\widetilde\Omega\,\partial_\phi,
 \qquad
\widetilde\Omega=\Omega-a
 =\frac{a(1-a^2)}{\rp^2+a^2}.
 \label{eq:KSW-coordinate-angular-velocity}
\end{equation}
Introduce
\begin{equation}
 x=r^2,\qquad R=\rp^2,\qquad y=\cos^2\theta,
 \qquad
 \rho^2=x+a^2y.
 \label{eq:KSW-xyR}
\end{equation}
Since
$\Delta_r=(x-R)(x+R+a^2+1)$, direct contraction of
Eq.~\eqref{eq:metric} with $V$ gives
\begin{align}
 \rho^2 g_L(V,V)
 ={}&
 -\Delta_r
 \left(1-\frac{a\widetilde\Omega(1-y)}{1-a^2}\right)^2
 \nonumber\\
 &+(1-a^2y)(1-y)
 \left(a-\frac{(x+a^2)\widetilde\Omega}{1-a^2}\right)^2 .
 \label{eq:KSW-Kerr-norm-unsimplified}
\end{align}
The two squared factors reduce to
\begin{equation}
 1-\frac{a\widetilde\Omega(1-y)}{1-a^2}
 =\frac{R+a^2y}{R+a^2},
 \qquad
 a-\frac{(x+a^2)\widetilde\Omega}{1-a^2}
 =-\frac{a(x-R)}{R+a^2}.
 \label{eq:KSW-Kerr-square-identities}
\end{equation}
Therefore the norm factorizes as
\begin{equation}
 \boxed{
 -g_L(V,V)
 =
 \frac{x-R}{\rho^2(R+a^2)^2}\,
 \mathcal P(x,y),
 }
 \label{eq:KSW-Kerr-norm-factor}
\end{equation}
where
\begin{align}
 \mathcal P(x,y)
 ={}&
 (2R+a^2+1)(R+a^2y)^2
 \nonumber\\
 &+(x-R)
 \left[
 R^2-a^2+a^2y(2R+a^2+1)
 \right].
 \label{eq:KSW-Kerr-positive-polynomial}
\end{align}
For $x>R$ and $0\leq y\leq1$, every factor in
Eqs.~\eqref{eq:KSW-Kerr-norm-factor} and
\eqref{eq:KSW-Kerr-positive-polynomial} is positive when
$R^2>a^2$, or equivalently $\rp^2>|a|$.  Thus $V$ is timelike everywhere
outside the horizon.  If $R^2<a^2$, choosing $y=0$ makes the coefficient
of $x-R$ in $\mathcal P$ negative, so $V$ becomes spacelike sufficiently
far from the black hole.  At equality, $V$ remains timelike at every finite
exterior point, but its norm in the conformally rescaled boundary metric
becomes null and $\Theta_{\mathrm{KSW}}$ approaches $\pi$ at infinity.
Thus pointwise open-exterior allowability survives at equality, while a
uniform strict asymptotic phase margin is lost.  This reproduces the strict
Hawking--Reall global condition
directly~\cite{HawkingReall:2000}.

The thermodynamic map gives the equivalent boundary form
\begin{align}
 1-\Omega^2
 &=
 \frac{(\rp^2+a^2)^2-a^2(1+\rp^2)^2}
 {(\rp^2+a^2)^2}
 \nonumber\\
 &=
 \boxed{
 \frac{(1-a^2)(\rp^4-a^2)}
 {(\rp^2+a^2)^2}
 }.
 \label{eq:KSW-factor}
\end{align}
Since $|a|<1$, this proves
\begin{equation}
 \boxed{
 \text{KSW allowable with a strict asymptotic margin}
 \quad\Longleftrightarrow\quad
 |\Omega|<1
 \quad\Longleftrightarrow\quad
 \rp^2>|a| .
 }
 \label{eq:KSW-bound}
\end{equation}
If only pointwise allowability at finite points of the open exterior is
required, the limiting equality $|\Omega|=1$, or $\rp^2=|a|$, is also
included.  With a strict margin imposed up to conformal infinity, the
physical grand canonical domain is exactly the KSW-allowable domain of the
real quasi-Euclidean Kerr saddles.

\paragraph{Relation to the thermodynamic fold.}
Two useful checks show that KSW allowability and Hessian degeneracy are
independent.  On the Hawking--Page and fold loci,
\begin{equation}
 \begin{aligned}
 \text{Hawking--Page:}\qquad
 &\rp^2=1,
 &1-\Omega_{\mathrm{HP}}^2
 &=\frac{(1-a^2)^2}{(1+a^2)^2},\\[4pt]
 \text{fold/spinodal:}\qquad
 &\rp^2=\frac{1+a^2}{2},
 &1-\Omega_{\mathrm{sp}}^2
 &=\frac{(1-a^2)^3}{(1+3a^2)^2}.
 \end{aligned}
 \label{eq:KSW-checks}
\end{equation}
Both expressions are strictly positive for every finite $|a|<1$.  In
particular,
\begin{equation}
 \left.\left(\rp^2-|a|\right)\right|_{\mathrm{fold}}
 =\frac{(1-|a|)^2}{2}>0,
 \label{eq:KSW-fold-gap}
\end{equation}
so the vanishing Takagi value at the fold does not signal a failure of KSW
allowability.  The two boundaries meet only in the singular ultraspinning
limit $|a|\to1$.

\paragraph{Complex saddles.}
For genuinely complex saddles, Eq.~\eqref{eq:KSW-phase} remains the
pointwise definition, but the shortcut
Eq.~\eqref{eq:KSW-bound} does not extend holomorphically.  One must first
specify a real middle-dimensional spacetime cycle, including its radial
contour.  At each point one must pull back the complex metric to that real
five-cycle, find its real-congruence coefficients $\lambda_A$, and verify
both nondegeneracy and
$\sum_A|\arg\lambda_A|<\pi$.  Therefore the complex solutions in
Eq.~\eqref{eq:complex-Schwarzschild} cannot be declared allowable from
their thermodynamic parameters alone.
Table~\ref{tab:KSW-logical-tests} summarizes the independent roles of the
field equations, KSW condition, thimble coefficients, and thermodynamic
Hessian.

\begin{table}[htbp]
 \centering
 \small
 \renewcommand{\arraystretch}{1.18}
 \begin{tabular}{@{}lll@{}}
 \toprule
 Test & Object being tested & Required condition\\
 \midrule
 Classical saddle
 & Complex spacetime fields
 & Einstein equations and boundary data\\
 KSW allowability
 & Chosen real spacetime five-cycle
 & $\Theta_{\mathrm{KSW}}(x)<\pi$ at every point\\
 Thimble contribution
 & Cycle in complexified field space
 & $c_\sigma\neq0$\\
 Local one-loop data
 & Thermodynamic mini-superspace
 & Takagi spectrum of $H_\sigma$\\
 \bottomrule
 \end{tabular}
 \caption{Four distinct tests.  KSW allowability is a
 pointwise condition on the spacetime metric and is neither implied by nor
 sufficient for a nonzero Lefschetz-thimble coefficient.}
 \label{tab:KSW-logical-tests}
\end{table}

\FloatBarrier

\section{Extended thermodynamics and the
canonical cusp}
\label{sec:extended}

The grand canonical analysis above was performed at fixed AdS radius.  We now
pass to extended thermodynamics, where
$P=-\Lambda/(8\pi)$ is varied and the angular momentum is held fixed.  At
fixed $(J,P)$, a canonical spinodal is a fold of the entropy integral, and
the merger of the two folds at the small/large black hole critical point is
a cusp.  We first derive this local saddle geometry and its Pearcey
uniformization.  We then insert the Kerr--AdS$_5$ equation of state, construct
the fixed-$J$ Gibbs branches, and continue the spinodals into the complex
domain.  The canonical cusp is ensemble-dependent and is not an endpoint of
the grand canonical Hawking--Page line.

In extended thermodynamics the mass $M(S,J,P)$ is the enthalpy and
$T=(\partial M/\partial S)_{J,P}$.  The local construction below requires
only these thermodynamic relations; the explicit Kerr--AdS$_5$ functions are
given in Subsection~\ref{sec:extended-thermodynamics}.  In this section only, we
restore the AdS radius $\ell$ and set $G=1$ to match the standard
extended-thermodynamics conventions.  Factors of Newton's constant can be
restored using the geometrized variables $J_{\rm geo}=GJ$,
$S_{\rm geo}=GS$, and $P_{\rm geo}=GP$.

\subsection{Canonical folds and cusp uniformization}
\label{sec:canonical-cusp}

At fixed $(J,P)$, the canonical entropy integral has exponent
\begin{equation}
 \Phi_{J,P}(S;\beta)=S-\beta M(S,J,P).
 \label{eq:canonical-exponent}
\end{equation}
Its stationary equation is $T=\beta^{-1}$, and its local Hessian at a
stationary point is
\begin{equation}
 H_{\rm can}
 =
 \frac{\partial^2\Phi_{J,P}}{\partial S^2}
 =
 -\beta
 \left(\frac{\partial T}{\partial S}\right)_{J,P}
 =
 -\frac{1}{C_{J,P}},
 \qquad
 C_{J,P}
 =
 T\left(\frac{\partial S}{\partial T}\right)_{J,P}.
 \label{eq:canonical-Hessian}
\end{equation}
A canonical fold occurs when the map $S\mapsto T$ at fixed $(J,P)$ ceases
to be locally invertible,
\begin{equation}
 \left(\frac{\partial T}{\partial S}\right)_{J,P}=0.
 \label{eq:canonical-spinodal}
\end{equation}
At this point the one-dimensional Takagi rate $|H_{\rm can}|$ vanishes and
$C_{J,P}$ diverges.  Two folds merge at a canonical critical point, where
\begin{equation}
 \left(\frac{\partial T}{\partial S}\right)_{J,P}=0,
 \qquad
 \left(\frac{\partial^2T}{\partial S^2}\right)_{J,P}=0.
 \label{eq:canonical-criticality}
\end{equation}
Consequently, a stationary saddle at the merger satisfies
\begin{equation}
 \Phi_{J,P}'=\Phi_{J,P}''=\Phi_{J,P}'''=0,
 \label{eq:canonical-cusp}
\end{equation}
with a generically nonzero quartic term.  The merger is therefore a canonical
cusp, distinct from the codimension-one grand canonical fold analyzed in
Section~\ref{sec:phase} and Subsection~\ref{sec:airy}.

Let $(S_c,P_c,T_c)$ denote such a critical point.  Its location for singly
rotating Kerr--AdS$_5$ will be obtained in
Subsection~\ref{sec:fixedJ-branches}.  With $\beta_c=T_c^{-1}$, set
$x=S-S_c$, $\delta\beta=\beta-\beta_c$, and
$\delta P=P-P_c$, with $J$ fixed.  Since
$\partial_S\Phi_{J,P}=1-\beta T(S,J,P)$, expansion about the critical point
gives
\begin{align}
 \partial_S\Phi_{J,P}
 ={}&
 -T_c\,\delta\beta
 -\beta_c T_P\,\delta P
 -\beta_c T_{SP}\,\delta P\,x
 \nonumber\\
 &-\frac{\beta_c}{6}T_{SSS}\,x^3
 +O\!\left(
 \delta\beta^2,\delta P^2,\delta\beta\,\delta P,
 \delta P\,x^2,\delta\beta\,x^3,x^4
 \right),
 \label{eq:canonical-cusp-expansion}
\end{align}
where all derivatives of $T(S,J,P)$ are evaluated at the critical point.
For a generic cusp, $T_{SP}\neq0$ and $T_{SSS}\neq0$.  In particular,
\begin{equation}
 \left.\partial_S^4\Phi_{J,P}\right|_c
 =-\beta_c T_{SSS}\neq0 .
 \label{eq:canonical-quartic}
\end{equation}
After analytic shifts and rescalings of $x$ and of the two source
displacements, Eq.~\eqref{eq:canonical-cusp-expansion} takes the universal
normal form
\begin{equation}
 \Phi_{\mathrm{loc}}
 =
 \Phi_{\mathrm{reg}}
 {}+\frac{1}{\varepsilon}
 \left(
 -\frac{D}{4}q^4+\frac{X}{2}q^2+Yq
 \right)+\cdots,
 \qquad D\neq0.
 \label{eq:canonical-cusp-normal-form}
\end{equation}
Here $\varepsilon$ extracts the semiclassical scale
($\varepsilon\propto G$ after restoring Newton's constant), while $(X,Y)$
are analytic linear combinations of $(\delta P,\delta\beta)$ with
nonzero Jacobian at the critical point.  The phase of $D$ and the fourth
root used below are fixed by the steepest-descent contour.

The three local entropy saddles obey
\begin{equation}
 -Dq^3+Xq+Y=0,
 \qquad
 \Delta_{\mathrm{cusp}}
 =4X^3-27DY^2.
 \label{eq:canonical-cusp-discriminant}
\end{equation}
Thus $\Delta_{\mathrm{cusp}}=0$ is the semicubical caustic formed by the two
canonical folds.  On a real symmetric slice with $Y=0$ and real
$X/D>0$, the two outer
saddles have equal exponent and the middle saddle lies between them.  For
the Kerr branches constructed in Subsection~\ref{sec:fixedJ-branches}, the
Gibbs swallowtail is the thermodynamic projection of this cusp family; it
should not be confused with the codimension-three swallowtail catastrophe,
whose normal form is quintic.

The separate Gaussian approximations fail when the three saddles approach
the cusp.  Their uniform replacement is the contour-dependent, Laplace-type
Pearcey integral~\cite{Pearcey:1946}
\begin{equation}
 \mathfrak P_\Gamma(\xi,\eta)
 =
 \int_\Gamma\dd t\,
 \exp\!\left(
 -\frac{t^4}{4}+\frac{\xi t^2}{2}+\eta t
 \right).
 \label{eq:canonical-Pearcey}
\end{equation}
Indeed, setting $q=(\varepsilon/D)^{1/4}t$ in
Eq.~\eqref{eq:canonical-cusp-normal-form} gives
\begin{equation}
 \boxed{
 Z_{\mathrm{cusp}}^{\mathrm{loc}}
 \sim
 \mathcal N e^{\Phi_{\mathrm{reg}}}
 \varepsilon^{1/4}D^{-1/4}
 \mathfrak P_{\Gamma_{\mathrm{loc}}}\!\left(
 \frac{X}{D^{1/2}\varepsilon^{1/2}},
 \frac{Y}{D^{1/4}\varepsilon^{3/4}}
 \right).
 }
 \label{eq:canonical-Pearcey-uniform}
\end{equation}
Here $\mathcal N$ is a smooth, nonzero prefactor containing the regular
measure and the noncritical fluctuations.  The universal cusp window is
$X=O(\varepsilon^{1/2})$, $Y=O(\varepsilon^{3/4})$, and candidate zeros of
the local uniform approximation satisfy
\begin{equation}
 \mathfrak P_{\Gamma_{\mathrm{loc}}}(\xi,\eta)=0.
 \label{eq:canonical-Pearcey-zeros}
\end{equation}

The choice of $\Gamma_{\mathrm{loc}}$ remains global.  Away from the cusp,
the analytically continued canonical integral has the thimble expansion
\begin{equation}
 Z_{J,P}(\beta)
 \sim
 \sum_\sigma n_\sigma\,
 \mathcal A_\sigma e^{\Phi_\sigma},
 \qquad
 \mathcal A_\sigma
 =
 \mu_\sigma
 \left(\frac{2\pi}{-H_{{\rm can},\sigma}}\right)^{1/2},
 \label{eq:canonical-thimble-sum}
\end{equation}
where the square-root phase is fixed by the thimble orientation.  If exactly
two saddles $i,j$ contribute, a zero requires both
\begin{align}
 \Re(\Phi_i-\Phi_j)
 &=
 -\log\left|
 \frac{n_i\mathcal A_i}{n_j\mathcal A_j}
 \right|,
 \nonumber\\
 \Im(\Phi_i-\Phi_j)
 &=
 (2k+1)\pi
 -\arg\left(
 \frac{n_i\mathcal A_i}{n_j\mathcal A_j}
 \right),
 \qquad k\in\mathbb Z.
 \label{eq:canonical-two-saddle-zero}
\end{align}
On shell $\Phi_\sigma=-\beta\mathcal G_\sigma$, with
$\mathcal G=M-TS$.  The self-intersection of the real Gibbs swallowtail
gives the leading equal-magnitude, or anti-Stokes, condition, but does not
impose phase cancellation.  The discriminant
\eqref{eq:canonical-cusp-discriminant}, the Pearcey approximation
\eqref{eq:canonical-Pearcey-uniform}, and the local zero condition
\eqref{eq:canonical-Pearcey-zeros} determine the local cusp structure.
A physical complex zero diagram additionally requires the canonical contour,
its intersection numbers $n_\sigma$, and the one-loop phases.

The local normal form and its scaling window are independent of the detailed
equation of state.  The critical scales and real branches follow from the
Kerr--AdS$_5$ thermodynamic relations.

\paragraph{Airy limits of the Pearcey cusp and the scaling limit}
\label{sec:Airy-Pearcey-matching}

Sections~\ref{sec:exponents} and~\ref{sec:extended} describe related local
saddle geometries in different ensembles. The precise connection is that
each separated fold of the canonical Pearcey family has an Airy
uniformization, of the same cubic type as the grand canonical fold in
Eq.~\eqref{eq:fold-normal-form}. This is a correspondence of normal forms
and scaling powers; it does not identify the saddle locations, integration
cycles, or full partition functions of the two ensembles.

Write the polynomial in Eq.~\eqref{eq:canonical-cusp-normal-form} as
$f(q;X,Y)=-Dq^4/4+Xq^2/2+Yq$. A generic point on its caustic is
\begin{equation}
 X_f=3Dq_f^2,\qquad Y_f=-2Dq_f^3,\qquad q_f\ne0.
 \label{eq:Pearcey-fold-location}
\end{equation}
There $q_f$ is a double stationary point; the third stationary point is
$-2q_f$ and remains separated. At fixed $X=X_f$, set $Y=Y_f+h$ and
$q=q_f+u$. The exact polynomial expansion is
\begin{equation}
 f(q_f+u;X_f,Y_f+h)
 =f_{\mathrm{reg}}(h)+hu+\frac{B_f}{3}u^3-\frac D4u^4,
 \qquad B_f=-3Dq_f\ne0.
 \label{eq:Pearcey-to-Airy-polynomial}
\end{equation}
For a general transverse displacement, an analytic shift removes the
quadratic term and replaces $h$ by
$\delta Y+q_f\delta X+O(\delta X^2,\delta X\delta Y,\delta Y^2)$.
Thus the cubic term has precisely the structure used in
Section~\ref{sec:airy}. With cube-root branches fixed by the local cycle,
\begin{align}
 u&=\left(\frac{\varepsilon}{B_f}\right)^{1/3}t,
 &z&=-\frac{h}{B_f^{1/3}\varepsilon^{2/3}},\nonumber\\
 \frac{f-f_{\mathrm{reg}}}{\varepsilon}
 &=\frac{t^3}{3}-zt
 -\frac{D\varepsilon^{1/3}}{4B_f^{4/3}}t^4.
 \label{eq:Pearcey-to-Airy-rescaling}
\end{align}
Consequently, the coalescing pair reduces to a contour-selected Airy
combination when $z$ is bounded and
\begin{equation}
 \frac{|D|\varepsilon^{1/3}}{|B_f|^{4/3}}\ll1,
 \qquad\hbox{equivalently}\qquad
 |q_f|\gg\left(\frac{\varepsilon}{|D|}\right)^{1/4},
 \label{eq:Airy-Pearcey-overlap}
\end{equation}
up to fixed numerical factors. The separated third saddle must still be
included if the chosen contour contains its thimble. As $q_f\to0$, the
cubic coefficient vanishes and this overlap condition fails: the quartic
Pearcey form is then required for the three-saddle cluster.  

The powers of the uniform approximation must also be compared at the same
level. One soft cubic integration supplies $\varepsilon^{1/3}$; the
additional nondegenerate integration in Section~\ref{sec:airy} supplies
$G^{1/2}$, giving the displayed $G^{5/6}$. One soft quartic entropy
integration supplies $\varepsilon^{1/4}$ in
Eq.~\eqref{eq:canonical-Pearcey-uniform}. These different total prefactors
are consistent; their comparison requires the measure and any regular
modes to be specified.

\paragraph{Matching to the thermodynamic plots.}
The stationary values of the cubic polynomial in
Eq.~\eqref{eq:Pearcey-to-Airy-polynomial} split as $|h|^{3/2}$.
Because $\Phi_\sigma=-\beta\mathcal G_\sigma$, a generic transverse
thermal displacement has
$\Delta\mathcal G_{\mathrm{fold}}\propto|T-T_{\mathrm{sp}}|^{3/2}$.
Figure~\ref{fig:canonical-three-halves} tests this Airy-fold prediction
using the full thermodynamics, including its coefficient. An exponent
splitting of order unity requires $|h|^{3/2}/\varepsilon=O(1)$, which
explains the Airy scaling $h=O(\varepsilon^{2/3})$; the $3/2$ and
$2/3$ powers refer to different quantities.

The cusp gives two complementary thermodynamic paths. In real scaling
coordinates with $D>0$, the leading coexistence slice is $Y=0$, $X>0$.
The outer stationary points are $q_\pm=\pm\sqrt{X/D}$ and their exponent
relative to the middle point is $X^2/(4D\varepsilon)$. Thus the entropy
jump, and hence the latent heat, vanish as $X^{1/2}$, while the uniform
cusp window has $X=O(\varepsilon^{1/2})$. Analytic mixing of the physical
sources places coexistence on this slice to leading order, with
$X\propto1-P/P_c$; this is the square-root law tested in
Figure~\ref{fig:canonical-latent-heat}. On the leading field axis $X=0$,
$q^3=Y/D$ and $f(q)-f(0)=3Dq^4/4$. Its $|Y|^{4/3}$ stationary action
explains both $Y=O(\varepsilon^{3/4})$ and the critical-isobar response
$C_{J,P_c}\propto|T-T_c|^{-2/3}$. In the physical variables of
Eq.~\eqref{eq:canonical-critical-response-scaling}, this also implies
\begin{equation}
 \mathcal G_{\mathrm{eq}}(T,P_c)
 =\mathcal G_c-S_c\tau
 -\frac{3}{4B_c^{1/3}}|\tau|^{4/3}
 +o(|\tau|^{4/3}),\qquad \tau=T-T_c.
 \label{eq:cusp-critical-free-energy}
\end{equation}
The $3/2$ fold law and the $4/3$ cusp-field law therefore apply on distinct
paths. The plots verify stationary-point scaling and thermodynamic
amplitudes; they are not numerical evaluations of a contour-selected
Pearcey partition function. In particular, none of these local matches
fixes the contour coefficients needed for Fisher zeros, or alters the
grand canonical zero-free result of
Subsection~\ref{sec:fold-zero-free}.

\subsection{Fixed-\texorpdfstring{$J$}{J} branches, critical point, and
complex continuation}
\label{sec:fixedJ-branches}
We now apply the preceding thermodynamic relations to singly rotating Kerr--AdS$_5$ at fixed angular momentum. For each pressure, the fixed-\(J\) constraint determines the black hole branches: their Gibbs potentials identify coexistence, while the turning points of \(T(S)\) locate the spinodals. We first establish the real phase structure and test the latent-heat, fold-gap, and critical-response scalings connected to the Airy and Pearcey normal forms in Section~\ref{sec:Airy-Pearcey-matching}. We then present the pressure-volume description and follow the spinodal pair into the complex-temperature plane above the critical pressure, keeping its continuation distinct from real equilibrium phase coexistence.
\subsubsection{Extended Kerr--AdS\texorpdfstring{$_5$}{5} thermodynamics}
\label{sec:extended-thermodynamics}

For singly rotating Kerr--AdS$_5$, define
\begin{equation}
 \Xi_\ell=1-\frac{a^2}{\ell^2},
 \qquad
 m=\frac{1}{2}(\rp^2+a^2)
 \left(1+\frac{\rp^2}{\ell^2}\right),
 \qquad
 P=-\frac{\Lambda}{8\pi}
 =\frac{3}{4\pi\ell^2}.
 \label{eq:extended-definitions}
\end{equation}
The enthalpy $M$, angular momentum, entropy, temperature, and angular
velocity are~\cite{GibbonsPerryPope:2005,CveticEtAl:2011}
\begin{align}
 M&=\frac{\pi m(2+\Xi_\ell)}{4\Xi_\ell^2},
 &
 J&=\frac{\pi ma}{2\Xi_\ell^2},
 \label{eq:extended-MJ}\\
 S&=\frac{\pi^2\rp(\rp^2+a^2)}{2\Xi_\ell},
 &
 T&=\frac{\rp\left[1+(2\rp^2+a^2)/\ell^2\right]}
 {2\pi(\rp^2+a^2)},
 \label{eq:extended-ST}\\
 \Omega&=\frac{a(1+\rp^2/\ell^2)}{\rp^2+a^2}.
 \label{eq:extended-Omega}
\end{align}
Writing the horizon area as $\mathcal A=4S$, the thermodynamic volume is
\begin{equation}
 \boxed{
 V=\frac{\rp\mathcal A}{4}
 \left[
 1+\frac{a^2(1+\rp^2/\ell^2)}
 {3\Xi_\ell\rp^2}
 \right]
 =\frac{\rp\mathcal A}{4}+\frac{2\pi}{3}aJ .
 }
 \label{eq:extended-volume}
\end{equation}
The rotational term is essential: the thermodynamic volume is not simply
the geometric volume $\rp\mathcal A/4$.  Direct differentiation of
Eqs.~\eqref{eq:extended-definitions}--\eqref{eq:extended-volume} gives
\begin{equation}
 \boxed{
 \dd M=T\,\dd S+\Omega\,\dd J+V\,\dd P,
 \qquad
 2M=3(TS+\Omega J)-2PV .
 }
 \label{eq:extended-first-law}
\end{equation}
The sign of $V\,\dd P$ is positive because $M$ is the enthalpy.  These
relations give a parametric representation of $M(S,J,P)$; fixing $J$ and
$P$ selects a one-dimensional family in the $(\rp,a)$ parameter space.
\subsubsection{Real Gibbs branches and critical point}

\begin{figure}[tbp]
 \centering
 \includegraphics[width=\textwidth]{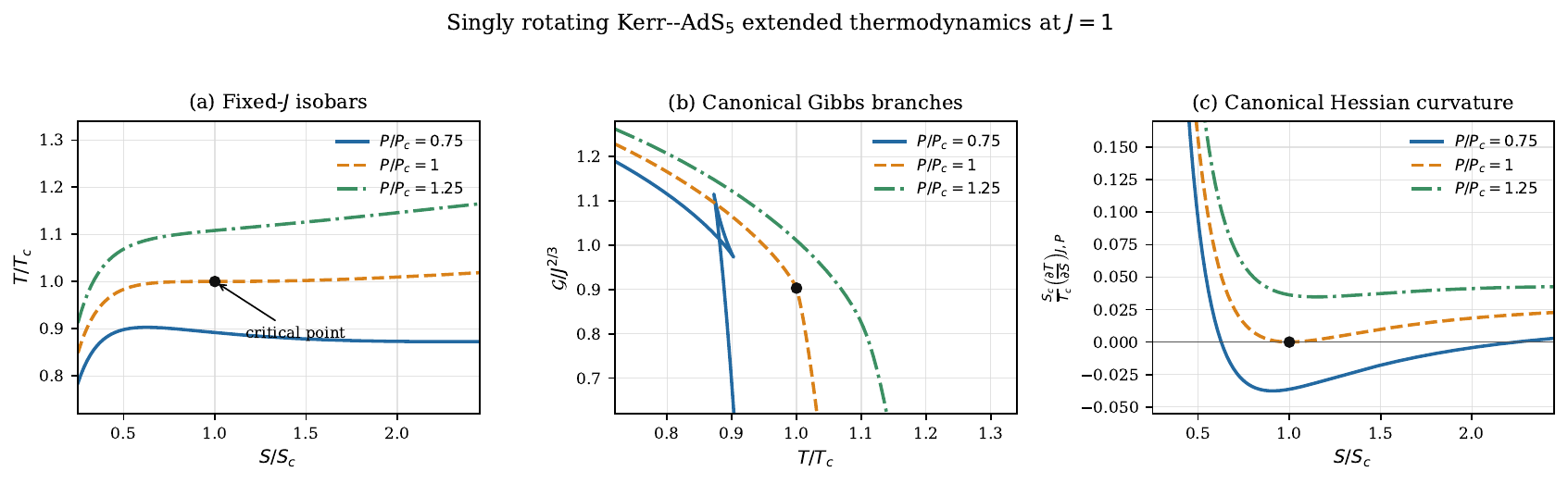}
 \caption{Fixed-$J$ extended thermodynamics of singly rotating
 Kerr--AdS$_5$ at $J=1$.  (a) For $P<P_c$ the isobar has two turning
 points; they merge at $(S_c,T_c)$ and disappear for $P>P_c$.
 (b) The parametric Gibbs branches exhibit the small/large black hole
 swallowtail below $P_c$: its stable-branch crossing is first order.
 Its endpoint at $(P_c,T_c)$ is continuous, with vanishing latent heat
 and a divergent equilibrium response. Above $P_c$ there is a smooth
 single branch.  (c) The dimensionless canonical slope
 $\kappa_{J,P}=(S_c/T_c)(\partial T/\partial S)_{J,P}$ has two simple zeros
 below $P_c$, a double zero at $P_c$, and no zero above $P_c$.}
 \label{fig:extended-thermodynamics}
\end{figure}

On a fixed-$(J,P)$ slice, the on-shell canonical exponent is
$-\beta\mathcal G$, where
\begin{equation}
 \mathcal G=M-TS,
 \qquad
 \dd\mathcal G=-S\,\dd T+V\,\dd P+\Omega\,\dd J.
 \label{eq:canonical-Gibbs}
\end{equation}
In terms of $(\rp,a,\ell)$ it has the compact parametric form
\begin{equation}
 \mathcal G
 =
 \frac{\pi}{8\Xi_\ell^2}
 \left[
 \rp^2+3a^2
 -\frac{(\rp^2-a^2)^2}{\ell^2}
 +\frac{3a^2\rp^4+a^4\rp^2}{\ell^4}
 \right].
 \label{eq:extended-Gibbs-parametric}
\end{equation}
The fixed-$J$ branch is obtained by solving the second equation in
Eq.~\eqref{eq:extended-MJ} for $a(\rp;J,P)$ and substituting it into
$T$, $S$, and $\mathcal G$.  Its folds and their merger are determined by
Eqs.~\eqref{eq:canonical-spinodal} and
\eqref{eq:canonical-criticality}.  For singly rotating Kerr--AdS$_5$, these
conditions give the critical scales~\cite{WeiChengLiu:2016}:
\begin{align}
 P_c&=0.0294179734\,J^{-2/3},
 &
 T_c&=0.156283679\,J^{-1/3},
 \nonumber\\
 S_c&=25.4041441\,J,
 &
 V_c&=43.8741110\,J^{4/3}.
 \label{eq:canonical-critical-values}
\end{align}
For $J=1$, the corresponding solution parameters are
\begin{equation}
 \rp^{\,c}=1.701767978,
 \qquad
 a_c=0.306605018,
 \qquad
 \ell_c=2.848717083.
 \label{eq:canonical-critical-parameters}
\end{equation}
Figure~\ref{fig:extended-thermodynamics} shows representative subcritical,
critical, and supercritical branches and the corresponding canonical slope.

For $P<P_c$, the two folds bound the locally unstable intermediate branch
and separate it from the locally stable small and large black hole
branches. The equilibrium potential is the lower Gibbs branch. Its
transition order is determined by its derivatives, rather than by the
existence of turning points alone.

\paragraph{First-order coexistence below the critical pressure.}
At fixed $(J,P)$, let $T_{\mathrm{tr}}(P,J)$ be the coexistence temperature;
$T_c$ is reserved for the critical endpoint. The two stable phases obey
\begin{equation}
 \mathcal G_{\mathrm{s}}(T_{\mathrm{tr}},P,J)
 =\mathcal G_{\mathrm{l}}(T_{\mathrm{tr}},P,J),
 \qquad \Delta S=S_{\mathrm{l}}-S_{\mathrm{s}}>0.
 \label{eq:canonical-coexistence-entropy}
\end{equation}
Equation~\eqref{eq:canonical-Gibbs} then yields
\begin{equation}
 \boxed{
 L_{J,P}=T_{\mathrm{tr}}\Delta S
 =-T_{\mathrm{tr}}\Delta\!\left[
 \left(\frac{\partial\mathcal G}{\partial T}\right)_{J,P}\right]
 =\Delta M>0.
 }
 \label{eq:canonical-latent-heat}
\end{equation}
The equality with $\Delta M$ holds because $M$ is enthalpy and both $J$
and $P$ are fixed; no rotational work term is present in this ensemble.
Thus the Gibbs potential is continuous at its stable-branch crossing,
while its first derivative jumps. This is the first-order small/large
black hole transition. Its spinodals are separate limits of metastability,
not additional coexistence transitions.

\begin{figure}[tbp]
 \centering
 \includegraphics[width=\textwidth]{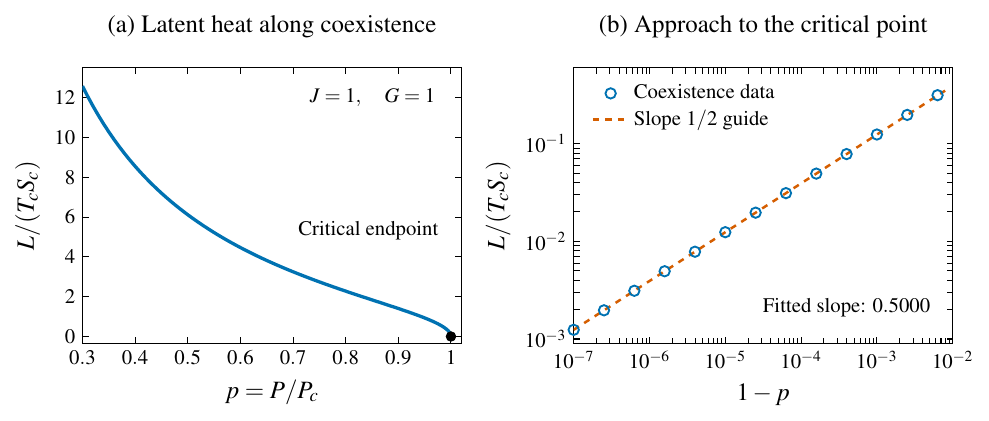}
 \caption{Canonical coexistence latent heat for singly rotating
 Kerr--AdS$_5$, with $J=1$ and $G=1$.
 (a) $L/(T_cS_c)$ along the equal-temperature, equal-Gibbs-potential
 small/large black hole coexistence line, for $0.3\leq p=P/P_c<1$.
 The black point at $(1,0)$ is the critical limiting value.
 (b) Near-critical data on logarithmic axes; open circles sample the
 computed coexistence curve and the dashed line is a slope-$1/2$ guide.
 The fitted exponent is $0.50001$ over
 $10^{-7}\leq1-p\leq10^{-4}$. The critical point itself is excluded
 from the logarithmic panel. The plot concerns the fixed-$J$ transition,
 not the grand canonical Hawking--Page latent heat.}
 \label{fig:canonical-latent-heat}
\end{figure}

Figure~\ref{fig:canonical-latent-heat} gives a direct numerical test of
this coexistence interpretation. At $J=1$, we solve the fixed angular momentum
constraint together with equal temperature and equal Gibbs potential on two
distinct physical branches. Both branches have positive $C_{J,P}$, and the
computed heat $T_{\mathrm{tr}}(S_{\mathrm{l}}-S_{\mathrm{s}})$ agrees
with $M_{\mathrm{l}}-M_{\mathrm{s}}$. The normalized latent heat is
positive below $P_c$ and approaches zero at the endpoint. In the window
$10^{-7}\leq1-P/P_c\leq10^{-4}$, a logarithmic fit gives exponent
$0.50001$, consistent with the generic square-root disappearance of the
coexistence entropy jump. This coexistence-path exponent is distinct from
the $2/3$ critical-isobar response exponent in
Eq.~\eqref{eq:canonical-critical-response-scaling}.

\paragraph{The three-halves law at a canonical spinodal.}
The free-energy splitting of a merging stable/unstable pair provides a
separate test of the fold geometry. Fix $J$ and $P<P_c$, away from the
critical endpoint, and let
$T_{SS,\mathrm{sp}}=(\partial^2T/\partial S^2)_{J,P,\mathrm{sp}}\ne0$.
For $x=S-S_{\mathrm{sp}}$ and $\theta=T-T_{\mathrm{sp}}$, the local
fixed-temperature potential is
\begin{equation}
 M(S,J,P)-TS=\mathcal G_{\mathrm{sp}}-S_{\mathrm{sp}}\theta
 -\theta x+\frac{T_{SS,\mathrm{sp}}}{6}x^3+O(x^4).
 \label{eq:canonical-fold-G-expansion}
\end{equation}
On the side $\theta/T_{SS,\mathrm{sp}}>0$, its two stationary branches
have $x_\pm\sim\pm\sqrt{2\theta/T_{SS,\mathrm{sp}}}$.
Their Gibbs gap, evaluated at the same $(T,J,P)$, is therefore
\begin{gather}
 \boxed{\Delta\mathcal G_{\mathrm{fold}}
 =\mathcal G_{\mathrm{u}}-\mathcal G_{\mathrm{m}}
 =K_{\mathrm{sp}}|T-T_{\mathrm{sp}}|^{3/2}
 \bigl[1+O(\delta)\bigr],}
 \label{eq:canonical-fold-three-halves}
 \\
 K_{\mathrm{sp}}=\frac{4\sqrt{2}}{3\sqrt{|T_{SS,\mathrm{sp}}|}},
 \qquad \delta=\frac{|T-T_{\mathrm{sp}}|}{T_{\mathrm{sp}}}.
 \nonumber
\end{gather}
Here $\mathrm{u}$ labels the unstable branch and $\mathrm{m}$ its locally
stable, metastable partner near the spinodal. The common regular term
$-S_{\mathrm{sp}}\theta$ cancels in this difference. Thus a plot of an
individual unsubtracted Gibbs potential would not isolate the $3/2$ law.
This gap is distinct from the free-energy difference of the two stable
phases at coexistence, which vanishes exactly.

\begin{figure}[tbp]
 \centering
 \includegraphics[width=\textwidth]{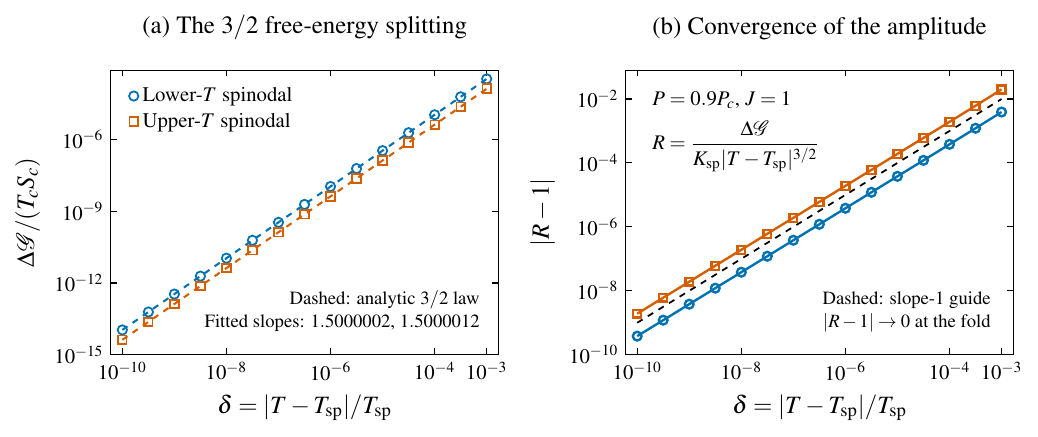}
 \caption{Three-halves Gibbs splitting at the two canonical spinodals,
 for $P/P_c=0.9$, $J=1$ and $G=1$.
 (a) The positive gap between the unstable and metastable branches at the
 same temperature, pressure and angular momentum. Open circles and squares
 sample the lower- and upper-temperature spinodals; dashed curves use the
 analytic coefficient in Eq.~\eqref{eq:canonical-fold-three-halves}.
 (b) The relative amplitude discrepancy $|R-1|$ on logarithmic axes, where
 $R=\Delta\mathcal G_{\mathrm{fold}}/[K_{\mathrm{sp}}|T-T_{\mathrm{sp}}|^{3/2}]$.
 Both curves approach zero, independently checking the analytic amplitude.
 The black dashed curve is the slope-one guide $10\delta$; its normalization
 is chosen only for visibility. Colors and markers identify the same
 spinodals in both panels. The two temperatures are
 $T_{\mathrm{sp}}=0.1488571492$ and $0.1499855330$.
 The approach is from $T>T_{\mathrm{sp}}$ at the lower fold and
 $T<T_{\mathrm{sp}}$ at the upper fold.}
 \label{fig:canonical-three-halves}
\end{figure}

Figure~\ref{fig:canonical-three-halves} tests both the exponent and the
coefficient using the full Kerr--AdS$_5$ thermodynamics at $P=0.9P_c$ and
$J=1$. We compute 57 same-temperature branch pairs at each spinodal, verify
opposite signs of $C_{J,P}$, and fit over $10^{-10}\leq\delta\leq10^{-6}$.
The fitted slopes are $1.5000002$ and $1.5000012$. Define the amplitude ratio
$R=\Delta\mathcal G_{\mathrm{fold}}/[K_{\mathrm{sp}}|T-T_{\mathrm{sp}}|^{3/2}]$.
Panel (b) resolves the small difference $|R-1|$ on logarithmic axes: it
vanishes linearly with $\delta$, consistent with the first relative
correction in Eq.~\eqref{eq:canonical-fold-three-halves}. The $3/2$ law describes an isolated spinodal; its amplitude
is not uniform as $P\to P_c$, where $T_{SS,\mathrm{sp}}\to0$ and the cusp
expansion replaces the fold expansion. It neither classifies the spinodal
as an equilibrium transition nor establishes a sequence of zeros of the partition function.

\paragraph{Continuous critical endpoint.}
As $P\to P_c$ along coexistence, the two phases merge:
$\Delta S\to0$, $\Delta V\to0$, and $L_{J,P}\to0$.
Zero latent heat alone is insufficient to identify the transition order.
Here the additional diagnostic is the singular equilibrium response
\begin{equation}
 C_{J,P}=-T\left(\frac{\partial^2\mathcal G_{\mathrm{eq}}}
 {\partial T^2}\right)_{J,P}
 =\frac{T}{(\partial T/\partial S)_{J,P}}.
 \label{eq:canonical-response-order}
\end{equation}
At the critical point the first two entropy derivatives of $T$ vanish,
while $T_{SSS,c}>0$. Along the critical isobar $P=P_c$, put
$\tau=T-T_c$ and $B_c=T_{SSS,c}/6>0$. Then
\begin{align}
 \tau&=B_c(S-S_c)^3+O\!\left((S-S_c)^4\right),
 \nonumber\\
 S-S_c&\sim\operatorname{sgn}(\tau)(|\tau|/B_c)^{1/3},
 \qquad
 C_{J,P_c}\sim\frac{T_c}{3B_c^{1/3}}|\tau|^{-2/3}.
 \label{eq:canonical-critical-response-scaling}
\end{align}
The entropy is continuous but the second temperature derivative of the
equilibrium Gibbs potential diverges. This is the continuous, conventionally
second-order endpoint of the first-order coexistence line
\cite{WeiChengLiu:2016}. The exponent $2/3$ refers specifically to approach
along $P=P_c$; it is not the generic fold exponent $1/2$ or a
fixed-volume heat-capacity exponent. The local quartic/Pearcey form
encodes this cusp geometry. Establishing complex zeros of the partition function
still requires the canonical contour, as explained in
Subsection~\ref{sec:canonical-cusp}.

For $P>P_c$ the real equilibrium branch is smooth and there is no
small/large black hole phase transition. A projected complex spinodal may
indicate a crossover, but cannot define an additional real transition.
These classifications refer to the classical equilibrium description, or
its thermodynamic/semiclassical limit; they do not assert a real-axis
nonanalyticity of a regular finite-parameter reduced integral.

\subsubsection{Pressure--volume representation}

The van der Waals structure is manifest in the reduced pressure--volume
plane.  Introduce
\begin{equation}
 p=\frac{P}{P_c},
 \qquad
 \vartheta=\frac{T}{T_c},
 \qquad
 \nu=\frac{V}{V_c}.
 \label{eq:reduced-PVT}
\end{equation}
At fixed $J$, the criticality conditions are equivalently
\begin{equation}
 \left(\frac{\partial P}{\partial V}\right)_{T,J}=0,
 \qquad
 \left(\frac{\partial^2P}{\partial V^2}\right)_{T,J}=0.
 \label{eq:PV-criticality}
\end{equation}

\begin{figure}[tbp]
 \centering
 \includegraphics[width=0.92\textwidth]{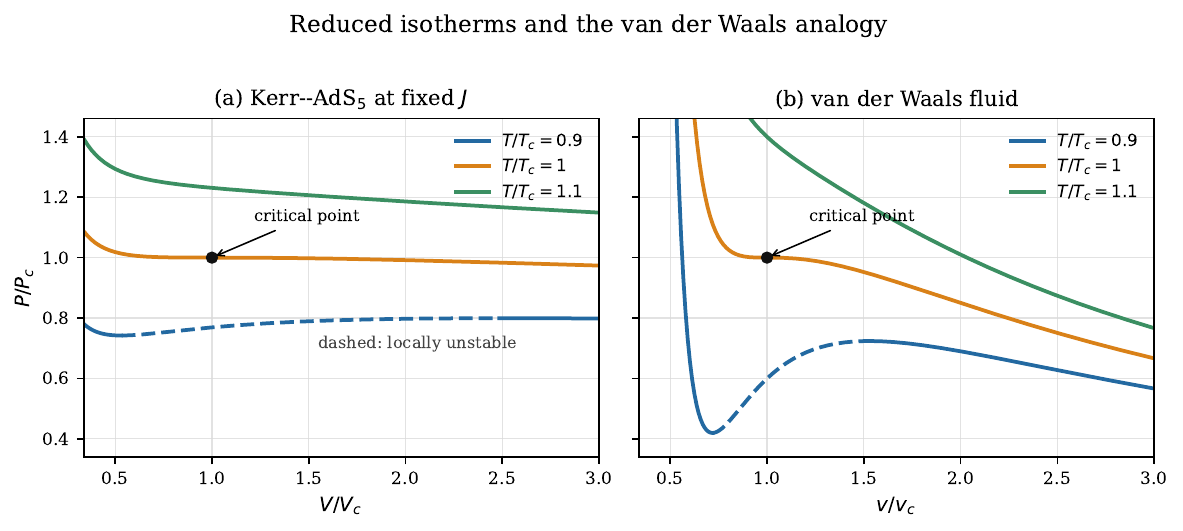}
 \caption{Reduced pressure--volume isotherms.  (a) Singly rotating
 Kerr--AdS$_5$ at fixed $J=1$, obtained from the exact parametric
 thermodynamic relations.  (b) The standard van der Waals fluid.  Solid
 segments have negative slope; dashed segments are the locally unstable
 positive-slope portions of the subcritical isotherm.  Both families pass
 through the critical point $(\nu,p)=(1,1)$.}
 \label{fig:vdw-comparison}
\end{figure}

Figure~\ref{fig:vdw-comparison} compares the resulting Kerr--AdS$_5$
isotherms with the reduced van der Waals equation
\begin{equation}
 p_{\mathrm{vdW}}
 =
 \frac{8\vartheta}{3\nu-1}
 -\frac{3}{\nu^2}.
 \label{eq:vdw-reduced}
\end{equation}
For $\vartheta<1$, both systems contain an oscillatory segment with
$(\partial P/\partial V)_{T,J}>0$.  This segment is locally unstable.
The Maxwell construction fixes the coexistence pressure, whereas its two
turning points are the spinodals.  At $\vartheta=1$ the turning points merge
into the critical inflection, and for $\vartheta>1$ the isotherm is
monotone.  The detailed equations of state differ, but the fold and cusp
organization is the same.

\subsubsection{Complex continuation of the spinodals}

Although no real spinodal remains for $P>P_c$, the fixed-$J$ fold equation
can be continued analytically in $\rp$ and $a$.  This continuation probes
the complex saddle geometry of the cusp; it does not by itself define a new
equilibrium phase or a zero of the partition function.  For this continuation,
define the reduced complex temperature and radius
\begin{equation}
 t=\frac{T}{T_c}, \qquad z=\frac{r_+}{r_c},
 \label{eq:complex-reduced-temperature}
\end{equation}
so that $t=\vartheta$ on the real thermodynamic slice.

\begin{figure}[tbp]
 \centering
 \includegraphics[width=0.49\textwidth,height=5cm,keepaspectratio]{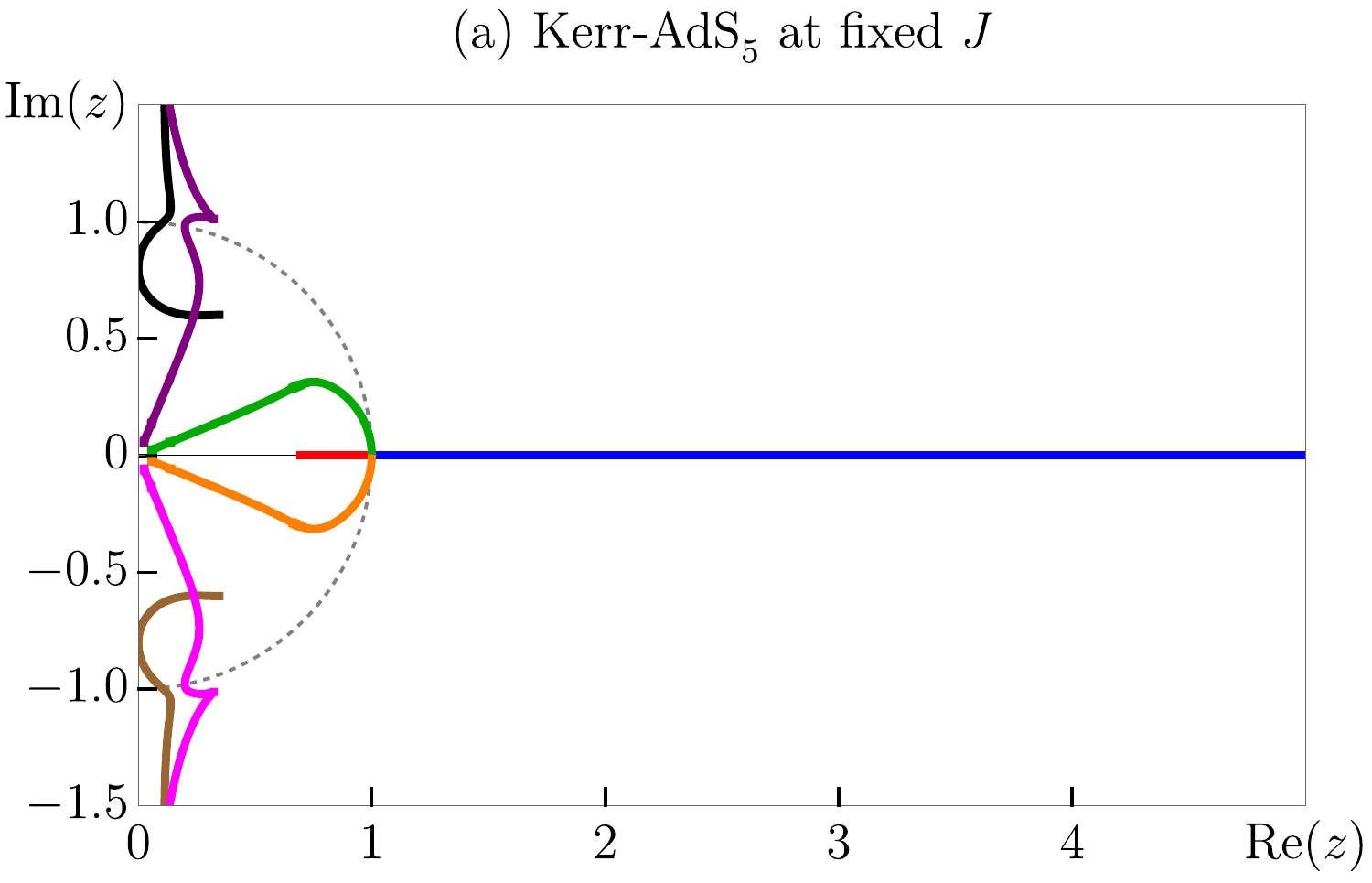}
 \includegraphics[width=0.49\textwidth,height=5cm,keepaspectratio]{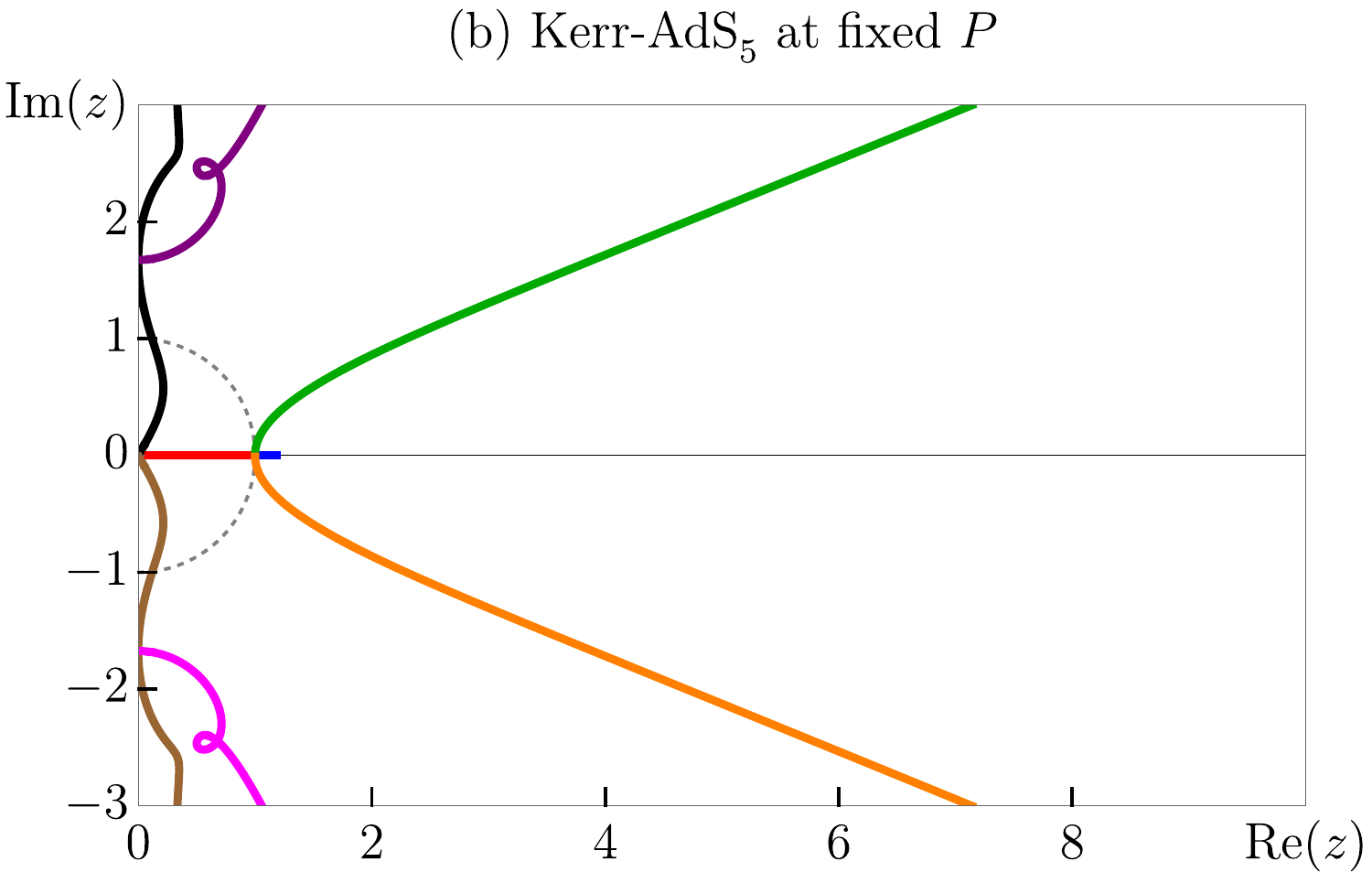}
 \caption{Solution trajectories in the reduced complex radius plane. (a) Fixed $J$ with varying pressure. (b) Fixed $P$ with varying angular momentum. In both panels, the red and blue branches lie on the real axis and meet at $z=1$, where the green and orange branches extend into the upper and lower half-planes, respectively. The purple and black curves show additional branches in the upper half-plane, with their magenta and brown counterparts in the lower half-plane. The gray dotted curves mark the unit circle $|z| = 1$.}
 \label{fig:complex-roots}
\end{figure}

Figure~\ref{fig:complex-roots} shows the trajectories of the spinodal solutions in the complex reduced-radius plane $z = \rp/r_c$. The curves are obtained by solving the angular-momentum constraint together with Eq.~\eqref{eq:canonical-spinodal} for complex $(\rp, a)$, while keeping $J$ and $P$ real. In panel (a), $J$ is fixed and $p = P/P_c$ varies. The red and blue branches represent the two real spinodal solutions for $p < 1$ and merge at $z = 1$ when $p = 1$. For $p > 1$, they continue as the green and orange branches in the upper and lower half-planes, forming a complex conjugate pair. In panel (b), $P$ is fixed and $j = J/J_c$ varies, where $J_c$ is the critical angular momentum at the chosen pressure. The corresponding real branches occur for $j < 1$, merge at $j = 1$, and continue as a complex conjugate pair for $j > 1$, with the same color convention. In both panels, the purple and black curves show additional complex branches in the upper half-plane, with their magenta and brown counterparts in the lower half-plane. The gray dotted curves mark $|z| = 1$. The critical scales $r_c$ and $T_c$ refer to the critical point of the corresponding fixed-$J$ or fixed-$P$ slice.

\begin{figure}[tbp]
    \centering
    \includegraphics[width=0.49\textwidth,height=5cm,keepaspectratio]{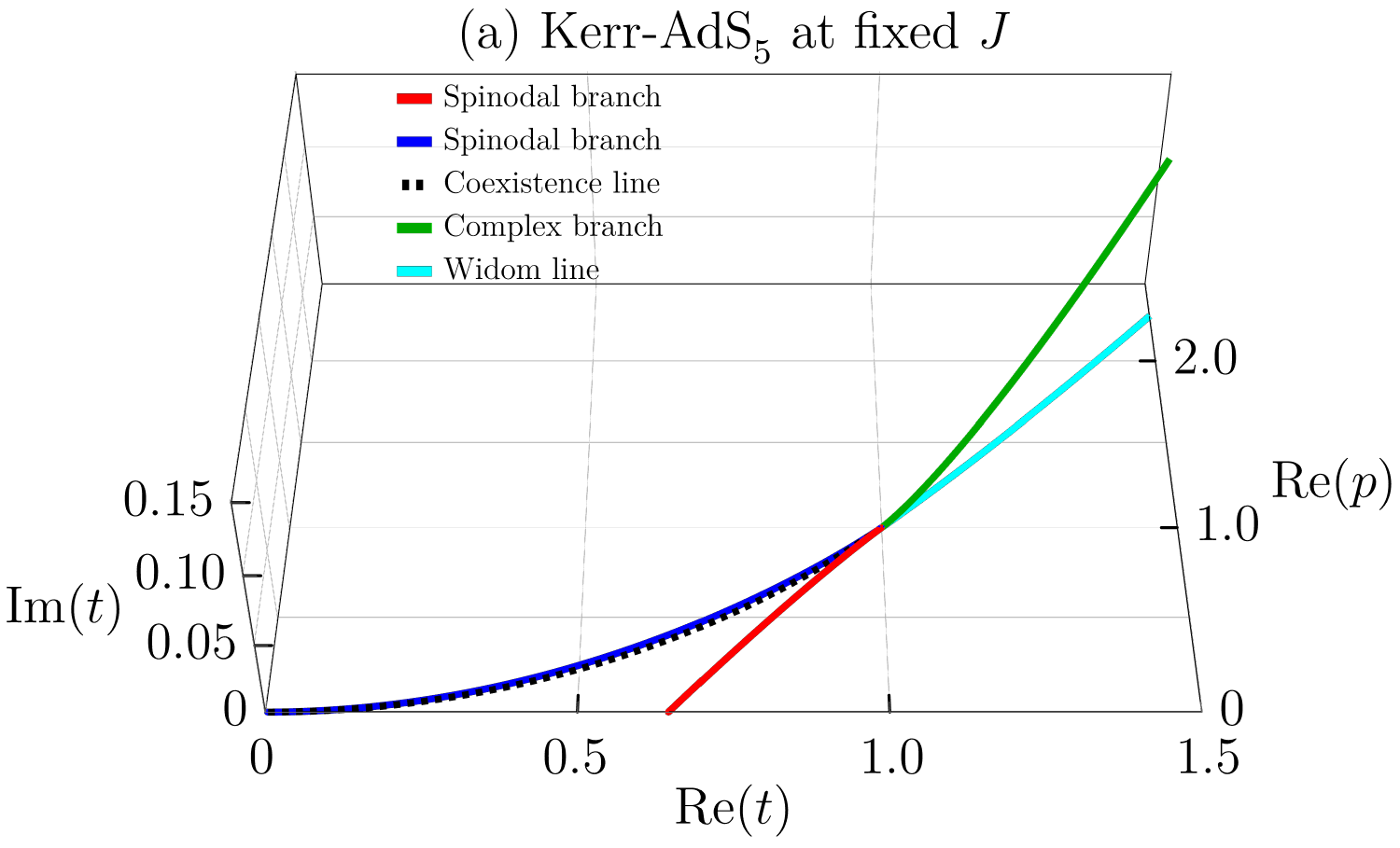}
    \includegraphics[width=0.49\textwidth,height=5cm,keepaspectratio]{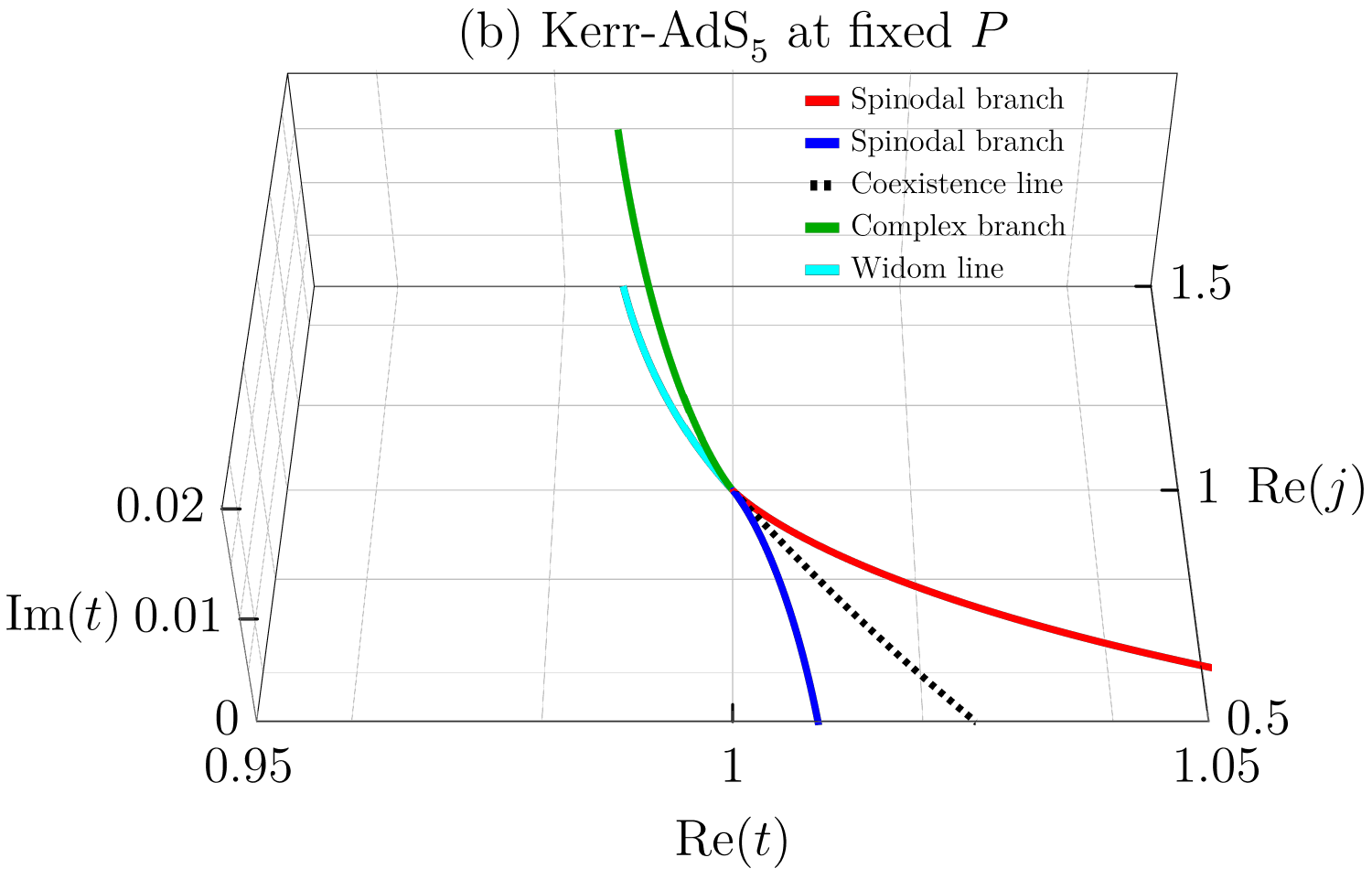}
 \caption{Complex temperature diagrams of singly rotating Kerr--AdS$_5$. (a) Fixed $J$, with reduced pressure $p = P/P_c$. (b) Fixed $P$, with reduced angular momentum $j = J/J_c$. In both panels, the red and blue curves show the real spinodal branches, and the black dotted curves denote the coexistence lines on the real plane. The green curves show the complex continuations with $\operatorname{Im}(t) > 0$, while the cyan curves are their projections onto this plane. The curves meet at $(t, p) = (1, 1)$ in (a) and $(t, j) = (1, 1)$ in (b).}
 \label{fig:complex-phase}
\end{figure}

Figure~\ref{fig:complex-phase} shows the corresponding complex-temperature diagrams, with $t = T/T_c$. In panel (a), the red and blue curves represent the real spinodal branches at fixed $J$ for $p < 1$, while the black dotted curve shows the coexistence line on the real plane. These three curves meet at $(t, p) = (1, 1)$. For $p > 1$, the green curve shows the continuation with $\operatorname{Im}(t) > 0$, and the cyan curve is its projection onto the real $(\operatorname{Re}(t),p)$ plane. In panel (b), the same colors distinguish the real spinodal branches and coexistence line at fixed $P$ for $j < 1$, which meet at $(t, j) = (1, 1)$. For $j > 1$, the green curve shows the upper complex branch, and the cyan curve is its projection onto the real $(\operatorname{Re}(t),j)$ plane. Following the projection used in Ref.~\cite{Xu:2025jrk}, the cyan curves are labeled Widom lines; this term refers to the projected complex spinodals.

\FloatBarrier
\section{Discussion and conclusions}
\label{sec:discussion}
The central result of this analysis is that the small/large Kerr--AdS$_5$
saddle merger and the accumulation of Fisher zeros describe different
features of the grand canonical partition function. The merger is a local
Hessian degeneracy, whereas the zeros studied here arise from interference
between thermal AdS and the large black hole on a specified reduced contour.
Their distinction persists throughout $|\Omega|<1$. The fixed-$J$ extended
ensemble has a separate critical structure: two canonical folds merge into
a cusp. Together, these results show why a complex continuation of the
thermodynamic saddle diagram must be supplemented by contour and fluctuation
data before it becomes a prediction for zeros of the partition function.

The exact determinant factorization in Eq.~\eqref{eq:detH-factorized}
locates the grand canonical fold at $2\rp^2-a^2-1=0$. At a generic point,
one Takagi value vanishes while the other remains nonzero, and rotation
mixes the radial and angular components of the soft direction. The
nonvanishing cubic term establishes the local Airy structure. By contrast,
both reduced rates remain nonzero on the Hawking--Page line
(Figure~\ref{fig:stability-exponents}). Hawking--Page coexistence therefore
involves an equality of saddle actions without a local merger of the
competing saddles. The strict ordering
$\beta_{\mathrm{HP}}(\Omega)<\beta_{\mathrm{sp}}(\Omega)$ leaves a window
in which the large black hole is metastable and thermal AdS has lower
classical action. Their common limit at $(|\Omega|,\beta)=(1,\pi)$ is
excluded from the physical open domain and does not provide a finite
critical endpoint of the Hawking--Page line.

This distinction also clarifies the meaning of the local rates. For a
complex symmetric Hessian they are Takagi singular values, not ordinary
complex eigenvalues; for the real family they reduce to the magnitudes of
the signed Hessian eigenvalues. A positive Takagi value consequently does
not distinguish a thermodynamically stable direction from an unstable one.
Its numerical magnitude depends on the chosen collective coordinates and
flow metric, while the loss of Hessian rank at a saddle is preserved by
nonsingular coordinate changes. The rates calculated here characterize
Picard--Lefschetz flow in the reduced $(\rp,a)$ space and should not be
identified with Lorentzian dynamical instabilities or real-time chaos.

 The contour prescription determines which local saddle sectors can
interfere. We adopt the regulated decomposition of Ref.~\cite{Singhi:2025},
including the half-weight of the radially extended thermal thimble. Each
lateral decomposition contains the unstable small black hole sector. Its
coefficient-weighted leading Gaussian contribution cancels only when the
two complete lateral expansions are combined in the conjugation-symmetric
real-$G$ prescription. The resulting effective sum,
Eq.~\eqref{eq:Kerr-symmetric-sum}, is a leading saddle statement, not an
exact termwise cancellation of thimbles on the Stokes surface. The absence
of a selected Airy-zero family asserted here is restricted to that
prescription and the order retained; it is not an all-orders statement or
a conclusion for arbitrary integration cycles. A uniform evaluation of
the complete selected contour through the fold is needed to test this
conclusion within the coalescence window. The Airy uniformization
remains necessary for a contour-supported contribution of the coalescing
pair.

The selected leading competition is instead between thermal AdS and the
large Kerr saddle. The direct integral with a flat measure and the independent
reduced Gaussian calculation support the same Hawking--Page accumulation
pattern, with conjugate zeros and leading imaginary offsets proportional
to $2n+1$, governed by the latent charge $L_\Omega=\Delta(E-\Omega J)$. The asymptotic relation
\eqref{eq:Kerr-Fisher-zero-asymptotic} separates this action-controlled
spacing from the displacement due to the prefactor ratio $R_\Omega$.
Since $L_\Omega=O(G^{-1})$, the near-transition imaginary spacing is
$O(G)$, as expected for this first-order saddle competition. It differs
from the $O(G^{2/3})$ source scale of a contour-supported Airy edge.
Agreement with quadrature is a test of the semiclassical approximation,
not an equality between the direct integral and its Gaussian truncation:
the thermal endpoint already has the $O(\sqrt G)$ relative correction in
Eq.~\eqref{eq:thermal-boundary-expansion}. A heat-capacity divergence or a vanishing Takagi value therefore identifies a local degeneracy but is insufficient to locate the zeros of the selected partition function.

 The fixed-$J$ critical point has a different origin. In the extended
ensemble, two entropy folds meet where the quadratic and cubic terms in
the canonical exponent vanish. A nonzero quartic term then yields the
local Pearcey approximation, with the two-variable scaling window
$X=O(\varepsilon^{1/2})$ and $Y=O(\varepsilon^{3/4})$ of
Eq.~\eqref{eq:canonical-Pearcey-uniform}. Here $\varepsilon$ is the
semiclassical parameter defined in the canonical normal form; relating
it to Newton's constant requires a specified semiclassical scaling of the
geometrized thermodynamic variables. The real Gibbs swallowtail and the
complex continuation of the spinodals expose this cusp geometry, but they
do not determine the canonical integration cycle or its zero distribution.
In particular, the projected continuation described as a Widom-type line
is not, by itself, a new thermodynamic transition. A canonical Fisher zero
calculation requires the contour-selected Pearcey solution, or the
corresponding saddle weights and phases away from the cusp. It cannot be
inferred by transferring the grand canonical contour prescription to a
different ensemble.

KSW allowability supplies an independent condition on the spacetime
metric. For the real quasi-Euclidean family, a strict asymptotic phase
margin is equivalent to $|\Omega|<1$, or $\rp^2>|a|$. Both the fold and
the Hawking--Page locus lie strictly inside this domain at $|a|<1$.
The thermodynamic soft mode thus occurs without a loss of the strict KSW
margin. At $\rp^2=|a|$, pointwise allowability at finite exterior radius
must be distinguished from saturation of the margin at conformal infinity.
These results neither select a thimble nor establish allowability for a
genuinely complex saddle: the latter requires a specified real
middle-dimensional spacetime cycle and a pointwise phase test on that
cycle.

\paragraph{Scope and further work.}

The exact results concern the stated two-variable mini-superspace exponent:
its Hessian factorization, real-saddle spectra, and local fold coefficients.
The Airy and Pearcey expressions are uniform local semiclassical
approximations under the stated nondegeneracy and regular-measure
assumptions. The numerical zeros belong to the flat-measure reduced
integral, and the grand canonical contour data are imported from
Ref.~\cite{Singhi:2025}. A full gravitational calculation must also include
the functional fluctuation determinants and modes omitted by the
truncation. Such contributions can modify prefactors, phases, and finite-$G$
zero locations. If they remain subleading, introduce no additional critical
mode, and preserve the relevant saddle sectors, the distinction between a
local fold and first-order thermal AdS/black hole interference remains
valid.

The most direct next step is to determine the connecting Kerr flows and
intersection numbers over the complex source domain. The rotating phase
networks shown here concern the analytically continued small/large pair;
they omit the third algebraic saddle and do not constitute a complete
Stokes diagram. Combining a global flow analysis with fluctuation
determinants and KSW tests on specified complex spacetime cycles would
establish which candidate phase boundaries survive in the gravitational
integral. An independent construction of the fixed-$J$ contour would then
allow the canonical Pearcey zeros to be calculated and compared with the
grand canonical Fisher sequence. Two independent angular momenta provide a
further test of this organization in a larger control space.
A natural extension of this work is the study of complex spinodals and supercritical crossovers in Kerr--Newman--AdS$_4$ black holes. Although the real spinodal branches terminate at criticality, their analytic continuation into complex thermodynamic parameter space may retain information about the underlying saddle structure. In forthcoming work, we will investigate whether these complex spinodals organize the supercritical crossover and its Widom-line structure, providing a direct connection between complex black hole saddles and real thermodynamic response~\cite{Thakur:ComplexSpinodals}.

Lower-dimensional models offer complementary settings in which to test
the same separation of local geometry and global contour selection.
Deformed JT gravity permits comparison with matrix model
constructions~\cite{WittenJT:2020,WittenMatrixJT:2020,JohnsonRosso:2021},
while black hole/winding condensate saddles motivate related questions in
Horowitz--Polchinski systems~\cite{HorowitzPolchinski:1997,
IshibashiEtAl:2025,LeeLeeThakur:2026LargeD}. The criterion carried over from
the present calculation is precise: the local catastrophe determines the
appropriate uniform approximation, while the zero distribution is fixed by
the complete contour-selected expression, including its phases and
prefactors.

\section*{Acknowledgments}
BHL is supported by the National Research Foundation of Korea (NRF) grants  (RS-2020-NR049598, RS-2024-00339204, and RS-2026-25473640) and Overseas Visiting Fellow Program of Shanghai University. BHL thanks Asia Pacific Center for Theoretical Physics (APCTP) and Korea Institute for Advanced Study (KIAS) for the hospitality during his visit, where a part of this project was done. HL is supported by Basic Science Research Program through the National Research Foundation of Korea (NRF) funded by the Ministry of Education (NRF-2022R1I1A2063176) and the Dongguk University Research Fund of 2026. ST thanks Center for Quantum Spacetime (CQUeST), Sogang University for support for this work. 

OpenAI Codex was used for language editing and LaTeX consistency checks.

\appendix

\section{Takagi values and local Picard--Lefschetz flow}
\label{app:takagi-examples}

This appendix reviews two elementary examples of the local Takagi description
of Picard--Lefschetz flow.  They fix the conventions used in
Subsection~\ref{sec:PL} and illustrate why the local flow rates are Takagi
singular values rather than ordinary eigenvalues of a complex symmetric
Hessian.

\subsection{One complex variable}

Let a holomorphic exponent have the local expansion
\begin{equation}
 \mathcal I(Z)
 =\mathcal I_\sigma
 +\frac{1}{2}\mathcal I''_\sigma(Z-Z_\sigma)^2+\cdots .
 \label{eq:one-dimensional-local-exponent}
\end{equation}
For the upward flow $\dd Z/\dd u=\overline{\partial_Z\mathcal I}$, write
\begin{equation}
 \mathcal I''_\sigma=A+iB,
 \qquad
 Z-Z_\sigma=x+iy.
\end{equation}
The linearized flow is then
\begin{equation}
 \frac{\dd}{\dd u}
 \begin{pmatrix}x\\ y\end{pmatrix}
 =
 \begin{pmatrix} A&-B\\ -B&-A\end{pmatrix}
 \begin{pmatrix}x\\ y\end{pmatrix},
 \label{eq:one-dimensional-doubled-flow}
\end{equation}
whose two real eigenvalues are
\begin{equation}
 \kappa_\pm=\pm\sqrt{A^2+B^2}
 =\pm|\mathcal I''_\sigma|.
 \label{eq:one-dimensional-flow-rates}
\end{equation}
Thus the positive local PL exponent is
\begin{equation}
 \boxed{\lambda_{\rm PL}=|\mathcal I''_\sigma|}.
 \label{eq:one-dimensional-PL-rate}
\end{equation}
The phase of $\mathcal I''_\sigma$ does not give an imaginary flow rate.
Instead it fixes the orientation of the local thimble.  Indeed, with
\begin{equation}
 \mathcal I''_\sigma=|\mathcal I''_\sigma|e^{i\theta_\sigma},
 \qquad
 Z-Z_\sigma=\rho e^{i\varphi},
\end{equation}
the local constant-phase directions satisfy
\begin{equation}
 \varphi=-\frac{\theta_\sigma}{2}+\frac{k\pi}{2},
 \qquad k\in\mathbb Z.
 \label{eq:one-dimensional-thimble-directions}
\end{equation}
Even and odd $k$ give the two alternating tangent lines of the downward and
upward cycles; which line is the thimble is fixed by the sign of the real
quadratic term and by the chosen flow convention.  Thus the Hessian phase
fixes the local orientation data, while its modulus fixes the flow rate.

\subsection{A two-dimensional toy Hessian}

Consider the complex symmetric Hessian
\begin{equation}
 H_{\rm toy}=
 \begin{pmatrix}1+i&0\\0&2-i\end{pmatrix}.
 \label{eq:toy-Hessian}
\end{equation}
Its ordinary eigenvalues are $1+i$ and $2-i$, which are complex and hence
cannot be interpreted as local growth or decay rates.  By contrast,
\begin{equation}
 H_{\rm toy}^\dagger H_{\rm toy}
 =\begin{pmatrix}2&0\\0&5\end{pmatrix},
 \qquad
 \lambda_{\rm PL,1}=\sqrt{5},\quad
 \lambda_{\rm PL,2}=\sqrt{2}.
 \label{eq:toy-Takagi-values}
\end{equation}
For example, before ordering the singular values this factorization is
\begin{equation}
 H_{\rm toy}=U_{\rm toy}
 \begin{pmatrix}\sqrt{2}&0\\0&\sqrt{5}\end{pmatrix}
 U_{\rm toy}^{T},
 \qquad
 U_{\rm toy}=\diag\!\left(
 e^{i\pi/8},e^{-\frac{i}{2}\arctan(1/2)}
 \right).
 \label{eq:toy-Takagi-factorization}
\end{equation}
The two positive numbers in Eq.~\eqref{eq:toy-Takagi-values} are the local
PL exponents.  The corresponding phase information is instead contained in
the Takagi frame and in
\begin{equation}
 \arg\det H_{\rm toy}=\arg(3+i).
 \label{eq:toy-determinant-phase}
\end{equation}
This elementary split between positive local rates and orientation data is
precisely the one used for the Kerr Hessian.  Away from a real Kerr saddle,
the entries mix the radial and rotational coordinates, but the Takagi
construction and its interpretation are unchanged.

\section{Derivation of the Hessian}
\label{app:hessian}

Write
\begin{equation}
 \Phi=\frac{\pi}{8G}\,g,
 \qquad
 g=4\pi\rp f_1-\beta v f_2,
 \qquad
 f_1=\frac{u}{\XiA},
 \qquad
 f_2=\frac{uw}{\XiA^2}.
\end{equation}
The useful derivatives are
\begin{align}
 \partial_{\rp}f_1&=\frac{2\rp}{\XiA},
 &
 \partial_{\rp}^2f_1&=\frac{2}{\XiA},
 \\
 \partial_af_1&=\frac{2av}{\XiA^2},
 &
 \partial_a^2f_1&=\frac{2v(1+3a^2)}{\XiA^3}.
\end{align}
The saddle conditions imply
\begin{equation}
 \beta w=\frac{2\pi\XiA(3\rp^2+a^2)}{\rp s},
 \qquad
 a+\Omega=\frac{as}{u}.
 \label{eq:saddle-identities}
\end{equation}
Substituting these identities only after differentiating gives
Eqs.~\eqref{eq:H11}--\eqref{eq:H12}.

For example,
\begin{align}
 \left.\partial_{\rp}^2g\right|_{\mathrm{saddle}}
 &=\frac{4\pi}{\rp\XiA s}
 \left[6\rp^2s-(3\rp^2+a^2)(6\rp^2+a^2+1)\right]
 \\
 &=\frac{4\pi N_1}{\rp\XiA s}.
\end{align}
The mixed derivative reduces to
\begin{equation}
 \left.\partial_{\rp}\partial_ag\right|_{\mathrm{saddle}}
 =\frac{16\pi av(1-\rp^2)}{\XiA^2s}.
\end{equation}
The remaining diagonal derivative is
\begin{equation}
 \left.\partial_a^2g\right|_{\mathrm{saddle}}
 =\frac{2\pi vN_2}{\rp\XiA^3s},
\end{equation}
which gives Eq.~\eqref{eq:H22}.  Combining the three entries yields
\begin{equation}
 N_1N_2-32\rp^2a^2v(1-\rp^2)^2
 =
 2(a^2+3)u^2v(2\rp^2-a^2-1).
\end{equation}
This polynomial identity establishes the factorization
\eqref{eq:D-factorized} without dividing by the fold factor.

\section{Explicit derivation of the fold and Airy formulas}
\label{app:fold-Airy}

This appendix derives
Eqs.~\eqref{eq:fold-normal-form}--\eqref{eq:one-loop-scaling} without
assuming their powers of $G$ or $\delta\lambda$.  It is useful first to
remove the universal factor of $G^{-1}$ by defining
\begin{equation}
 \varphi(z;\beta,\Omega)=G\Phi(z;\beta,\Omega),
 \qquad z=(\rp,a).
 \label{eq:reduced-varphi}
\end{equation}
The function $\varphi$ is independent of $G$ when the dimensionless sources
are held fixed.

\subsection{Kerr soft and hard coordinates}

At a point on the rotating fold,
\begin{equation}
 \rp^{\mathrm{sp}}=\sqrt{\frac{1+a^2}{2}},
 \qquad
 \Omega_{\mathrm{sp}}=\frac{a(a^2+3)}{1+3a^2},
 \qquad
 \beta_{\mathrm{sp}}
 =\frac{\pi(1+3a^2)}
 {\sqrt{2}(1+a^2)^{3/2}},
 \label{eq:fold-source-values-app}
\end{equation}
define the normalized vectors
\begin{equation}
 \widehat e_0
 =\frac{1}{\sqrt{1+\kappa^2}}
 \begin{pmatrix}1\\ \kappa\end{pmatrix},
 \qquad
 \widehat e_1
 =\frac{1}{\sqrt{1+\kappa^2}}
 \begin{pmatrix}-\kappa\\ 1\end{pmatrix},
 \qquad
 \kappa=\frac{4\sqrt{2}\,a\sqrt{1+a^2}}{a^2+3}.
 \label{eq:fold-orthonormal-frame}
\end{equation}
Equation~\eqref{eq:Kerr-soft-vector} implies that $\widehat e_0$ is the
null direction of the Hessian, while $\widehat e_1$ is its nondegenerate
direction.  We introduce local coordinates by
\begin{equation}
 z-z_{\mathrm{sp}}=p\,\widehat e_1+q\,\widehat e_0,
 \qquad
 \partial_1=\widehat e_1^{\,i}\partial_i,
 \qquad
 \partial_0=\widehat e_0^{\,i}\partial_i .
 \label{eq:fold-pq-coordinates}
\end{equation}
For a fixed-$\Omega$ slice take
$\delta\lambda=\beta-\beta_{\mathrm{sp}}$.  More generally,
$\delta\lambda$ may be any analytic source coordinate transverse to the
fold.

The three coefficients entering the fold normal form are the following
derivatives of the reduced exponent:
\begin{equation}
 \boxed{
 C=-\left.\partial_1^2\varphi\right|_{\mathrm{sp}},
 \qquad
 A=\left.\partial_{\lambda}\partial_0\varphi\right|_{\mathrm{sp}},
 \qquad
 B=\frac12\left.\partial_0^3\varphi\right|_{\mathrm{sp}}.
 }
 \label{eq:ABC-derivatives}
\end{equation}
The minus sign makes $C>0$ on the real hard steepest-descent direction.  In
particular, Eq.~\eqref{eq:Kerr-fold-rates} gives the fully explicit Kerr
result
\begin{equation}
 C
 =G\lambda_{\mathrm{PL},\max}^{\mathrm{sp}}
 =\frac{\pi^2}{2}\widehat\lambda_{\max}^{\mathrm{sp}}
 =\frac{\pi^2\sqrt{2}(1+3a^2)(9+11a^2)}
 {32(1-a^2)(1+a^2)^{3/2}}.
 \label{eq:fold-C-explicit}
\end{equation}
The conditions $A\neq0$ and $B\neq0$ say, respectively, that the chosen
source crosses the fold transversely and that the degeneracy is a fold
rather than a higher catastrophe.

To see the normal form directly, Taylor expand $\varphi$ at the fold.
Stationarity removes the terms linear in $p$ and $q$,
$\partial_0^2\varphi=0$ removes the $q^2$ term, and the orthogonal
eigenbasis removes the quadratic $pq$ term.  A parameter-dependent Morse
change in the hard coordinate absorbs the remaining $p$-dependent
couplings into an analytic regular part.  The singular part is therefore
\begin{equation}
 \varphi_{\mathrm{loc}}
 =\varphi_{\mathrm{reg}}(\delta\lambda)
 -\frac{C}{2}p^2
 {}+A\,\delta\lambda\,q+\frac{B}{3}q^3
 {}+O\!\left(q^4,\delta\lambda q^2,\delta\lambda^2q\right).
 \label{eq:fold-Taylor-explicit}
\end{equation}
Dividing by $G$ and absorbing
$\varphi_{\mathrm{reg}}(\delta\lambda)/G$ into the definition
\begin{equation}
 \Phi_{\mathrm{reg}}(\delta\lambda)
 =\frac{\varphi_{\mathrm{reg}}(\delta\lambda)}{G},
 \qquad
 \Phi_{\mathrm{reg}}(0)=\Phi_{\mathrm{sp}},
 \label{eq:fold-regular-exponent}
\end{equation}
gives Eq.~\eqref{eq:fold-normal-form}.  The regular exponent is analytic in
the source displacement but is not generally constant.  At fixed $\Omega$,
for example, stationarity gives
\begin{equation}
 \left.\partial_{\delta\lambda}\Phi_{\mathrm{reg}}
 \right|_{0}
 =-\frac{\pi(3+a^2)(3+5a^2)}{32G(1-a^2)}\neq0.
 \label{eq:fold-regular-slope}
\end{equation}
This nonvanishing analytic factor does not move the local Airy zeros.  As a
simple check, at the Schwarzschild
fold $(\rp,a)=(1/\sqrt2,0)$ and fixed $\Omega=0$,
\begin{equation}
 A=-\frac{3\pi}{2\sqrt2},
 \qquad
 B=-\frac{3\pi^2}{4},
 \qquad
 C=\frac{9\pi^2\sqrt2}{32},
 \label{eq:ABC-Schwarzschild}
\end{equation}
so all three generic-fold conditions are manifestly satisfied.

\subsection{Gaussian--Airy integral}

Let $\mu_{\mathrm{sp}}$ include the smooth measure and the regular Jacobian
of the coordinate change.  To leading uniform order, the local contribution
is
\begin{equation}
 Z_{\mathrm{loc}}
 \simeq
 \mu_{\mathrm{sp}}e^{\Phi_{\mathrm{reg}}(\delta\lambda)}
 \int_{\Gamma_p}\dd p\,
 e^{-Cp^2/(2G)}
 \int_{\Gamma_q}\dd q\,
 \exp\!\left[
 \frac{A\delta\lambda q+Bq^3/3}{G}
 \right].
 \label{eq:fold-local-integral}
\end{equation}
The hard thimble gives
\begin{equation}
 \int_{\Gamma_p}\dd p\,e^{-Cp^2/(2G)}
 =\eta_p\sqrt{\frac{2\pi G}{C}},
 \label{eq:fold-hard-Gaussian}
\end{equation}
where $\eta_p$ is the contour-orientation phase.  In the soft integral set
\begin{equation}
 q=\left(\frac{G}{B}\right)^{1/3}t,
 \qquad
 c=-\frac{A}{B^{1/3}},
 \label{eq:fold-Airy-rescaling}
\end{equation}
with the branch of $B^{1/3}$ fixed by the soft thimble.  Using the standard
Airy-contour identity for an elementary Airy thimble,
\begin{equation}
 \operatorname{Ai}(x)
 =\frac{1}{2\pi i}
 \int_{\Gamma_{\mathrm A}}\dd t\,
 \exp\!\left(\frac{t^3}{3}-xt\right)
 \label{eq:Airy-contour-identity}
\end{equation}
then yields
\begin{equation}
 \int_{\Gamma_q}\dd q\,
 e^{(A\delta\lambda q+Bq^3/3)/G}
 =
 \eta_q\,2\pi i\,G^{1/3}B^{-1/3}
 \operatorname{Ai}\!\left(
 c\,\frac{\delta\lambda}{G^{2/3}}
 \right).
 \label{eq:fold-soft-Airy}
\end{equation}
Here $\eta_q$ records the orientation of the chosen elementary thimble.  A
general local contour gives a linear combination of rotated Airy solutions,
which cannot be encoded by a single orientation phase.  For the elementary
thimble, multiplying Eqs.~\eqref{eq:fold-hard-Gaussian} and
\eqref{eq:fold-soft-Airy} gives
\begin{equation}
 Z_{\mathrm{loc}}
 \simeq
 \mathcal C\,e^{\Phi_{\mathrm{reg}}(\delta\lambda)}G^{1/2+1/3}
 \operatorname{Ai}\!\left(
 c\,\frac{\delta\lambda}{G^{2/3}}
 \right),
 \qquad
 \mathcal C
 =2\pi i\,\eta_p\eta_q\mu_{\mathrm{sp}}
 \sqrt{\frac{2\pi}{C}}\,B^{-1/3}.
 \label{eq:fold-Airy-prefactor}
\end{equation}
This proves both the $G^{5/6}$ power and the Airy argument in
Eq.~\eqref{eq:Airy}.  If the selected local cycle is this elementary Airy
thimble, its zeros obey $\delta\lambda_n=a_nG^{2/3}/c$, where $a_n$ is an
Airy zero; for a general linear combination their locations depend on its
two contour coefficients.  As explained in the main text, the unstable
sector is absent from the leading conjugation-symmetric lateral combination.
The Airy relation is therefore a local fold statement rather than the Kerr
Fisher zero sequence obtained in that prescription.

\subsection{Soft rate and one-loop divergence}

The two critical points of the cubic normal form satisfy
\begin{equation}
 0=\partial_q\varphi_{\mathrm{loc}}
 =A\delta\lambda+Bq^2,
 \qquad
 q_\pm=\pm\sqrt{-\frac{A}{B}\delta\lambda}.
 \label{eq:fold-saddle-splitting}
\end{equation}
At either critical point, the soft entry of the Hessian of $\Phi$ is
\begin{equation}
 \left.\partial_q^2\Phi\right|_{q_\pm}
 =\frac{2Bq_\pm}{G}
 {}+O\!\left(\frac{\delta\lambda}{G}\right).
 \label{eq:fold-soft-Hessian}
\end{equation}
Consequently its Takagi magnitude is
\begin{equation}
 \boxed{
 \lambda_{\mathrm{PL},\min}
 =\frac{2|AB|^{1/2}}{G}
 |\delta\lambda|^{1/2}
 \left[1+O\!\left(|\delta\lambda|^{1/2}\right)\right],
 }
 \label{eq:fold-soft-rate-explicit}
\end{equation}
which proves Eq.~\eqref{eq:soft-rate-scaling}.  The hard rate remains
\begin{equation}
 \lambda_{\mathrm{PL},\max}
 =\frac{C}{G}
 \left[1+O\!\left(|\delta\lambda|^{1/2}\right)\right].
 \label{eq:fold-hard-rate-explicit}
\end{equation}
Since the modulus of the Hessian determinant is the product of the two
Takagi values, Eqs.~\eqref{eq:fold-soft-rate-explicit} and
\eqref{eq:fold-hard-rate-explicit} give
\begin{equation}
 |\det H|^{-1/2}
 =
 \boxed{
 \frac{G}{\sqrt{2C}\,|AB|^{1/4}}\,
 |\delta\lambda|^{-1/4}
 \left[1+O\!\left(|\delta\lambda|^{1/2}\right)\right].
 }
 \label{eq:fold-determinant-explicit}
\end{equation}
This is the coefficient-level form of
Eq.~\eqref{eq:one-loop-scaling}.  Its divergence is not a divergence of the
partition function: it is precisely the failure of treating the two
coalescing saddles as separate Gaussians, repaired by the finite uniform
integral in Eq.~\eqref{eq:fold-Airy-prefactor}.

\section{Jacobian and thermodynamic response}
\label{app:jacobian}

The Jacobian of the saddle map is
\begin{equation}
 \mathcal{J}_{\mathrm{map}}
 =\det\frac{\partial(\beta,\Omega)}{\partial(\rp,a)}.
\end{equation}
Direct differentiation of Eq.~\eqref{eq:saddle-equations} shows that
\begin{equation}
 \mathcal{J}_{\mathrm{map}}
 =-\frac{2\pi(1+\rp^2)}{\rp^2(2\rp^2+a^2+1)^2}
 \left(2\rp^2-a^2-1\right).
 \label{eq:saddle-map-Jacobian-explicit}
\end{equation}
Consequently, in the physical real domain $\rp>0$,
\begin{equation}
 \mathcal{J}_{\mathrm{map}}=0
 \quad\Longleftrightarrow\quad
 2\rp^2-a^2-1=0.
\end{equation}
The same factor appears in the denominator of $C_\Omega$ and in $\det H$,
confirming that all three diagnostics locate the same physical fold.

The heat capacity follows from the Jacobian identity
\begin{equation}
 \left(\frac{\partial S}{\partial T}\right)_\Omega
 =\frac{\partial(S,\Omega)/\partial(\rp,a)}
 {\partial(T,\Omega)/\partial(\rp,a)},
\end{equation}
which yields Eq.~\eqref{eq:C-Omega}.

\section{Checks in different regimes}
\label{app:checks}

For $a=0$, the on-shell sources are
\begin{equation}
 \beta=\frac{2\pi\rp}{2\rp^2+1},
 \qquad \Omega=0,
\end{equation}
and Eq.~\eqref{eq:H-Schwarzschild} is diagonal.  The following substitutions
therefore check both the signs of the Hessian and the ordering of its Takagi
values.  We assume $G>0$ throughout.

\paragraph{Large Schwarzschild black hole.}
At $(\rp,a)=(2,0)$, for which $\beta=4\pi/9$,
\begin{equation}
 H_{11}=-\frac{7\pi^2}{3G},
 \qquad
 H_{22}=-\frac{25\pi^2}{9G},
 \qquad
 H_{12}=0.
\end{equation}
Consequently,
\begin{equation}
 \det H=\frac{175\pi^4}{27G^2}>0,
 \qquad
 \lPL{1}=\frac{25\pi^2}{9G},
 \qquad
 \lPL{2}=\frac{7\pi^2}{3G}.
\end{equation}

\paragraph{Small Schwarzschild black hole.}
At $(\rp,a)=(1/2,0)$, for which $\beta=2\pi/3$,
\begin{equation}
 H_{11}=\frac{\pi^2}{4G},
 \qquad
 H_{22}=-\frac{25\pi^2}{96G},
 \qquad
 H_{12}=0.
\end{equation}
Thus
\begin{equation}
 \det H=-\frac{25\pi^4}{384G^2}<0,
 \qquad
 \lPL{1}=\frac{25\pi^2}{96G},
 \qquad
 \lPL{2}=\frac{\pi^2}{4G},
\end{equation}
and $\phiM=\pi/2$ modulo $\pi$.

\paragraph{Schwarzschild fold.}
At $(\rp,a)=(1/\sqrt{2},0)$, for which $\beta=\pi/\sqrt{2}$,
\begin{equation}
 H_{11}=0,
 \qquad
 H_{22}=-\frac{9\pi^2\sqrt{2}}{32G}.
\end{equation}
Hence $\lPL{1}=9\pi^2\sqrt{2}/(32G)$ and $\lPL{2}=0$.  The Gaussian
determinant phase is undefined at the fold, as expected when the isolated
saddle approximation fails.

\paragraph{Hawking--Page point.}
At $(\rp,a)=(1,0)$, for which $\beta=2\pi/3$,
\begin{equation}
 \Phi_{\mathrm{BH}}=0,
 \qquad
 H_{11}=-\frac{\pi^2}{2G},
 \qquad
 H_{22}=-\frac{2\pi^2}{3G}.
\end{equation}
Here
\begin{equation}
 \det H=\frac{\pi^4}{3G^2}>0,
 \qquad
 \lPL{1}=\frac{2\pi^2}{3G},
 \qquad
 \lPL{2}=\frac{\pi^2}{2G}.
\end{equation}
The Hawking--Page saddle is therefore nondegenerate despite satisfying the
anti-Stokes condition $\Phi_{\mathrm{BH}}=\Phi_{\mathrm{th}}=0$.

\bibliographystyle{JHEP}
\bibliography{kads5ref}

\end{document}